\RequirePackage{fix-cm}
\documentclass[smallcondensed]{svjour3}     
\smartqed  

\def\arcdeg{\hbox{$^{\circ}$}}

\usepackage{soul}
\usepackage{graphicx}
\usepackage{journal_macros}
\usepackage{natbib}
\usepackage{amsmath}
\usepackage{amssymb}
\usepackage{xcolor}
\usepackage{rotating}
\usepackage[percent]{overpic}
\usepackage[bookmarks=false, 
     pdfnewwindow=true, 
     colorlinks=true,   
     linkcolor=magenta, 
     citecolor=blue,    
     filecolor=cyan,    
     urlcolor=purple,   
	final=true]{hyperref}
\newcommand*{\doi}[1]{\href{http://dx.doi.org/#1}{DOI: #1}}

\journalname{Space Science Reviews}

\begin{document}

\title{Understanding the Origins of Problem Geomagnetic Storms Associated With ``Stealth" Coronal Mass Ejections}

\titlerunning{Problem Geomagnetic Storms and Stealth CMEs}

\author{Nariaki~V.~Nitta \and
        Tamitha~Mulligan \and
        Emilia~K.~J.~Kilpua \and
        Benjamin~J.~Lynch \and
        Marilena~Mierla \and
        Jennifer~O'Kane \and
        Paolo~Pagano \and
        Erika~Palmerio \and
        Jens~Pomoell \and
        Ian~G.~Richardson \and
        Luciano~Rodriguez \and
        Alexis~P.~Rouillard \and
        Suvadip~Sinha \and
        Nandita~Srivastava \and
        Dana-Camelia~Talpeanu \and
        Stephanie~L.~Yardley \and
        Andrei~N.~Zhukov
}

\authorrunning{Nitta et al.} 

\institute{ Nariaki V. Nitta \at
            Lockheed Martin Solar and Astrophysics Laboratory, Palo Alto, CA 94304, USA \\
            \email{nitta@lmsal.com}         
            \and
            Tamitha Mulligan \at
            Space Sciences Department, The Aerospace Corporation, Los Angeles, CA 94305, USA
            \and
            Emilia K. J. Kilpua \and Erika Palmerio \and Jens Pomoell \at
            Department of Physics, University of Helsinki, FI-00014 Helsinki, Finland
            \and
            Benjamin J. Lynch \and Erika Palmerio \at
            Space Sciences Laboratory, University of California--Berkeley, Berkeley, CA 94720, USA \and
            Marilena Mierla \and Luciano Rodriguez \and Dana-Camelia Talpeanu \and Andrei N. Zhukov \at
            Solar--Terrestrial Centre of Excellence---SIDC, Royal Observatory of Belgium, 1180 Brussels, Belgium \and
            Marilena Mierla \at
            Institute of Geodynamics of the Romanian Academy, 020032 Bucharest-37, Romania \and
            Jennifer O{\textquoteright}Kane \and Stephanie L. Yardley \at
            Mullard Space Science Laboratory, University College London, Holmbury St. Mary, Dorking, Surrey RH5 6NT, UK \and
            Paolo Pagano \and Stephanie L. Yardley \at
            School of Mathematics and Statistics, University of St Andrews, North Haugh, St Andrews, Fife KY16 9SS, UK \and
            Paolo Pagano \at Dipartimento di Fisica \& Chimica, Universit{\`a} di Palermo, I-90134 Palermo, Italy \and
            Paolo Pagano \at INAF--Osservatorio Astronomico di Palermo, I-90134 Palermo, Italy \and
            Erika Palmerio \at
            Cooperative Programs for the Advancement of Earth System Science, University Corporation for Atmospheric Research, Boulder, CO 80301, USA \and
            Ian G. Richardson \at
            Department of Astronomy, University of Maryland, College Park, MD 20742, USA \and 
            Ian G. Richardson \at
            Heliophysics Science Division, NASA Goddard Space Flight Center, Greenbelt, MD 20771, USA \and
            Alexis P. Rouillard \at
            IRAP, Universit{\'e} Toulouse III---Paul Sabatier, CNRS, CNES, 31400 Toulouse, France \and
            Suvadip Sinha \and Nandita Srivastava \at
            Centre of Excellence in Space Sciences India, Indian Institute of Science Education and Research Kolkata, Mohanpur 741246, India \and
            Nandita Srivastava \at
            Udaipur Solar Observatory, Physical Research Laboratory, Udaipur 313001, India \and
            Dana-Camelia Talpeanu \at
            Centre for Mathematical Plasma Astrophysics (CmPA), KU Leuven, 3001 Leuven, Belgium \and
            Andrei N. Zhukov \at
            Skobeltsyn Institute of Nuclear Physics, Moscow State University, 119991 Moscow, Russia
}

\date{Received: date / Accepted: date}

\maketitle

\begin{abstract}
Geomagnetic storms are an important aspect of space weather and can result in significant impacts on space- and ground-based assets.  The majority of strong storms are associated with the passage of interplanetary coronal mass ejections (ICMEs) in the near-Earth environment.  In many cases, these ICMEs can be traced back unambiguously to a specific coronal mass ejection (CME) and solar activity on the frontside of the Sun. Hence, predicting the arrival of ICMEs at Earth from routine observations of CMEs and solar activity currently makes a major contribution to the forecasting of geomagnetic storms. However, it is clear that some ICMEs, which may also cause enhanced geomagnetic activity, cannot be traced back to an observed CME, or, if the CME is identified, its origin may be elusive or ambiguous in coronal images. Such CMEs have been termed ``stealth CMEs". In this review, we focus on these ``problem" geomagnetic storms in the sense that the solar/CME precursors are enigmatic and stealthy. We start by reviewing evidence for stealth CMEs discussed in past studies. We then identify several moderate to strong geomagnetic storms (minimum Dst $< -50$~nT) in solar cycle 24 for which the related solar sources and/or CMEs are unclear and apparently stealthy. We discuss the solar and in situ circumstances of these events and identify several scenarios that may account for their elusive solar signatures. These range from observational limitations (e.g., a coronagraph near Earth may not detect an incoming CME if it is diffuse and not wide enough) to the possibility that there is a class of mass ejections from the Sun that have only weak or hard-to-observe coronal signatures.  In particular, some of these sources are only clearly revealed by considering the evolution of coronal structures over longer time intervals than is usually considered.  
We also review a variety of numerical modelling approaches that attempt to advance our understanding of the origins and consequences of stealthy solar eruptions with geoeffective potential. Specifically, we discuss magnetofrictional modelling of the energisation of stealth CME source regions and magnetohydrodynamic modelling of the physical processes that generate stealth CME or CME-like eruptions, typically from higher altitudes in the solar corona than CMEs from active regions or extended filament channels.
\end{abstract}


\section{Introduction}
\label{sec:intro}

Geomagnetic storms are an important aspect of space weather with significant impacts on Earth-based modern technological infrastructures and on space-based assets \citep[e.g.,][]{Riley2018}. Forecasting such storms in advance remains, however, a complex problem. Geomagnetic storms usually occur during passages of interplanetary coronal mass ejections (ICMEs) and corotating interaction regions (CIRs) at Earth \citep[e.g.,][]{Gosling1991, Tsurutani1997, Gonzalez1999, Richardson2001,Zhang2007,Echer2008}. 
This is due to the fact that  these structures often contain sustained intervals of strong southward (negative $B_z$) interplanetary magnetic field (IMF). It is well established that such sustained southward-B$_\text{Z}$ intervals are extremely geoeffective and drive geomagnetic storms \citep[e.g.][]{Tsurutani92}, in particular when combined with elevated solar wind speeds \citep[for a discussion on solar wind--magnetosphere coupling, see e.g.][]{Newell2007}. However, it is currently  a major challenge to predict the orientation of the IMF before it comes in contact with the magnetosphere \citep[e.g.,][]{Riley2017}.

In the case of ICME-driven storms, the development of space-based coronagraphs, in particular those on \emph{Skylab} \citep{Macqueen1974}, \emph{Solwind} \citep{Sheeley1980} and the \emph{Solar Maximum Mission} \citep{House1981} during the 1970s and 1980s, and more recently, the \emph{Large Angle Spectroscopic Coronagraph} \citep[LASCO;][]{Brueckner1995} on board the \emph{Solar and Heliospheric Observatory} \citep[SOHO;][]{Domingo1995}, revolutionised the understanding of storms and their solar drivers by clearly demonstrating the association between ICMEs and coronal mass ejections (CMEs) at the Sun. In the ideal case, a storm associated with the arrival of an ICME is preceded a few days before by the observation of a CME in coronagraph images and associated activity on the frontside solar disc. In particular, a CME propagating directly towards the observing spacecraft may appear as a symmetrical ``halo"  surrounding the occulting disc of the coronagraph (note that a halo CME may also arise from activity on the far side of the Sun). Thus, the observation of a frontside symmetrical halo CME by a coronagraph on board a spacecraft along the Sun--Earth line is a strong indication that the CME is heading towards the Earth.
However, a spatially-extended CME ejected somewhat away from the Sun--Earth line may also arrive at Earth. Such a CME may be asymmetrical or even appear as a ``partial halo", surrounding only part of the occulting disc. The angular width of a full halo CME, irrespective of its symmetry, is defined to be 360$\arcdeg$.  There is no universal definition of a partial halo CME, but typically, the angular width is at least 100$\arcdeg$\,--\,140$\arcdeg$. 
Cases where a geomagnetic storm is clearly associated with specific frontside solar activity and CMEs are discussed for example by \citet{Webb2000}, \citet{Schwenn2005}, \citet{Bothmer2007}, \citet{Zhang2007}, and \citet{Scolini2018}.

A natural extension of space-based coronagraphs has been the development of heliospheric white-light imagers such as the \emph{Solar Mass Ejection Imager} \citep[SMEI;][]{Eyles2003} and the \emph{Heliospheric Imager} \citep[HI;][]{Eyles2009}, part of the \emph{Sun Earth Connection Coronal and Heliospheric Investigation} \citep[SECCHI;][]{HowardR2008} suite on the twin \emph{Solar Terrestrial Relations Observatory} \citep[STEREO;][]{Kaiser2008} spacecraft. A comprehensive survey of CME height--time tracks and the associated kinematics of CME propagation through the inner heliosphere between the Sun and Earth, has been performed by the ``Heliospheric Catalouguing Analysis and Techniques Services'' \citep[HELCATS;][]{Harrison2018, Barnes2019, Barnes2020} project.

In reality, not all ICMEs or their related geomagnetic storms can be traced back to CMEs that are expected to be Earth-directed based on solar and coronal observations \citep[for reviews of the in situ signatures of ICMEs, see e.g.][and references therein]{Zurbuchen2006,Kilpua2017a}. For example, around a third of the near-Earth ICMEs identified by \citet{Cane2003} and around 20\% of those identified by \citet{Schwenn2005} were not preceded by (partial or full) halo CMEs observed by LASCO.  Furthermore, \citet{Zhang2007} could not identify the solar sources of nine ($\approx$12\%) of the 77 intense (minimum Dst $<-100$~nT) geomagnetic storms in 1996 to 2005 that were not attributable to CIRs alone. This was despite a careful review by these authors of coronal images from, e.g., the \emph{Extreme-ultraviolet Imaging Telescope} \citep[EIT;][]{Delaboudiniere1995} on SOHO. Nevertheless, at least an associated partial halo CME was idenfied for each storm. The lack of identifiable solar sources for these nine storms is problematic in various ways:  Without knowledge of the solar source, the selected halo CME could have originated from the far side of the Sun, meaning that the ICME and geomagnetic storm could have been caused by another CME that was not observed by LASCO as a halo CME. The presence of such CMEs could also bring into question the validity of the ICME\,--\,CME pairs found for other events. Moreover, without identifying the solar origin of the CME, it is impossible to understand or predict the southward IMF at Earth based on observations of the magnetic configuration of the erupting solar source region \citep[e.g.,][]{Savani2015,Palmerio2018}. These ambiguities highlight two main issues, namely: (1) some ICMEs do not appear to have a corresponding Earth-directed CME at the Sun, and (2) some CMEs observed by a coronagraph are not associated with eruptive signatures on the solar disc (or signatures may in fact be present but appear elusive).

To address these unresolved issues,  a team was formed at the International Space Science Institute (ISSI) focussed on the topic of ``Understanding the Origins of Problem Geomagnetic Storms''.   This review is based on the activities of the team that included two in-person meetings at ISSI in 2018\,--\,2019.  It is organised as follows: In Section~\ref{sec:stealthcmes}, we briefly review observational evidence for the existence of stealth CMEs in the literature and summarise their known properties based on earlier studies. We then contrast these properties with observations of a more ``textbook'' example of an ICME\,--\,CME pair.  This will highlight the challenges in associating an ICME with a specific CME, and forecasting when the ICME will encounter Earth and its geomagnetic effects. In  Section~\ref{sec:observations}, we discuss several selected geomagnetic storms for which the solar sources of the solar wind structures that generate the storms appear to be problematic and stealthy. These events  illustrate the broad range of observational constraints and diversity of  structures that contribute to (and expand) the classification of an ICME\,--\,CME pair as ``stealthy".  We summarise the circumstances of these events, including why the CMEs are stealthy and why their associated ICMEs are geoeffective.  A key question is whether their ``stealthiness'' is simply due to limited observations, or whether these CMEs are physically distinct from ``normal'' CMEs. We also highlight an event that illustrates another aspect of such problematic storms, namely when the interpretation of the in situ structures driving the storm is uncertain. In this case, it is unclear whether the storm is driven by an ICME embedded in a CIR, or by structure within the CIR, which then would account for the absence of associated solar signatures.    
In Section~\ref{sec:modelling}, we review recent modelling efforts of the source region energisation, eruption mechanism(s), and the ambiguous low-coronal signatures associated with stealth CME eruptions. 
In Section~\ref{sec:discussion}, we discuss why certain geomagnetic storms cannot be predicted with presently available tools, and outline how future observational programs and modelling efforts should improve the predictability of these events. 


\begin{figure}
\includegraphics[width=1.0\textwidth]{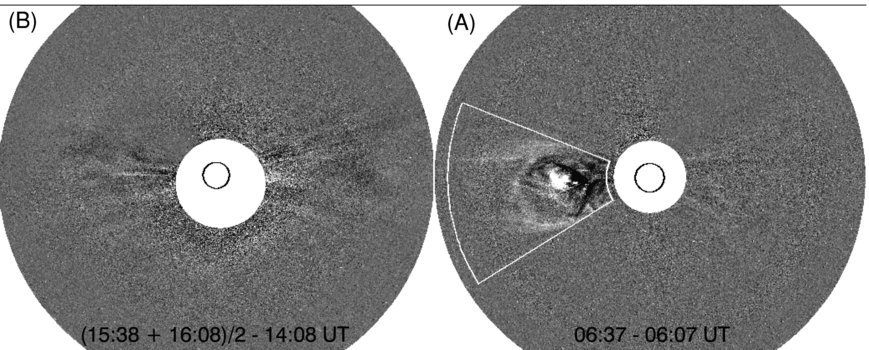}
\caption{STEREO-A (right) and STEREO-B (left) SECCHI/COR2 observations of the 2 June 2008 CME \citep[from][]{Robbrecht2009}. The images are running differences of total brightness images, and two images (at the times indicated) are combined in the STEREO-B observations to enhance the contrast. STEREO-B only observed a faint halo CME whereas STEREO-A, 53$^\circ$ to the west clearly observed the CME side-on before it was visible in the STEREO-B coronagraph field of view. \citep[The white sector in the STEREO-A image indicates the region used by][to calculate the CME mass.]{Robbrecht2009}}
\label{fig:robbrecht} 
\end{figure}

\section{Evidence for Stealth CMEs: An Observational Perspective} \label{sec:stealthcmes}

In Section~\ref{sec:intro}, we discussed how some geomagnetic storms can be problematic because the solar origin of the ICME causing the storm is difficult to observe or possibly to define. Unfortunately, problems identifying the solar signatures of ICMEs and the associated CMEs occur frequently, and there are many examples reported in the literature. In this section, we review existing studies that focus on events where the on-disc low coronal signatures (LCSs) of CMEs are apparently missing. Classic LCSs that can be traced back to a CME eruption include post-eruption arcades, coronal dimmings, extreme ultra-violet (EUV) waves, flare ribbons, and ejection of filament material \citep[e.g.,][]{Hudson2001,Zhukov2007,Nitta2014}. Events that lack such signatures are now commonly referred to as ``stealth CMEs", a term that became more widely accepted after \citet{Robbrecht2009} presented an outstanding example of a CME ``leaving no trace behind on the solar disc''. As such, this paper is an excellent place to begin reviewing the characteristics of this class of solar eruptions. 

Figure~\ref{fig:robbrecht} shows observations of the \citet{Robbrecht2009} event. The CME was observed on 1\,--\,2 June 2008 by the COR2 coronagraphs that form part of SECCHI on the twin STEREO spacecraft. At that time, STEREO-A (-B) was located 29$\arcdeg$ west (25$\arcdeg$ east) of the Sun--Earth line. Figure~\ref{fig:robbrecht} shows COR2 difference images of the CME as viewed at STEREO-B (left) and STEREO-A (right). The STEREO-A image shows what looks like a streamer-blowout CME. It had a slowly accelerating speed profile reaching only $\sim$200~km~s$^{-1}$ at 15\,$R_{\odot}$. The same CME was so diffuse that it was detected by COR2-B only by taking longer than usual temporal separations in running-difference images. The fact that it was a halo CME indicates that it was directed towards STEREO-B. Given the ${\sim}55^{\circ}$ of longitudinal separation between the two STEREO spacecraft, \citet{Lynch2010} were able to track the complete Sun-to-1~AU propagation of this CME via the COR and HI cameras on board STEREO-A. As anticipated, an ICME with a flux rope magnetic field structure was observed in situ at STEREO-B on 6\,--\,7 June 2008 \citep{Mostl2009,Lynch2010,Nieves-Chinchilla2011}. 

Regarding the origin of the CME observed by the COR2 coronagraphs and HI imagers on STEREO-A and -B, a slowly rising prominence was identified off the east limb in EUV images from the SECCHI Extreme-ultraviolet Imager \citep[EUVI;][]{Wuelser2004} on STEREO-A (EUVI-A). However, images from EUVI-B show none of the typical LCSs of CMEs in the apparently corresponding region as viewed against the solar disc.  \citet{Robbrecht2009} attributed the lack of dimmings to the idea that the eruption might have started at a greater height than normal CMEs, where there was not much mass to be evacuated.  The impact of this study was to make the community aware of the existence of Earth-directed CMEs with weak or no on-disc signatures that are extremely difficult to identify in near-Earth coronagraph imagery. 

Following the event in 2008 reported by \citet{Robbrecht2009}, several stealth CMEs were found in observations from 2009, during solar minimum conditions \citep{Ma2010, Kilpua2014}. At this time, STEREO-A and -B were positioned in near quadrature about the Sun--Earth line, making it possible to simultaneously compare limb and disc views of CMEs that were frontside events from the Earth's perspective. Due to the additional views of the limb provided by the STEREO spacecraft, these studies vastly increased our knowledge of the diversity of stealth CME characteristics. When observed as limb events in coronagraphs, these stealth CMEs without on-disc LCSs were generally slow and narrow. In addition, some were diffuse and lacked a clear front. These events apparently originated in quiescent regions, as active regions are rarely present at solar minimum.

Studies of stealth CMEs have continued to advance in the Solar Dynamics Observatory \citep[SDO;][]{Pesnell2012} era through the analysis of images from the \emph{Atmospheric Imaging Assembly} \citep[AIA;][]{Lemen2012}, which is a vast improvement over EIT and EUVI in terms of sensitivity, spatial resolution, cadence, and temperature coverage. For example, \citet{Vourlidas2011} and \citet{Pevtsov2012} associated stealth CMEs with erupting filaments or filament channels. \citet{OKane2019} and \citet{OKane2021a} studied the initiation and magnetic environment of stealth CMEs from active regions. \citet{DHuys2014} performed a comprehensive statistical study of eruptions in 2012 without LCSs in either limb or disc views. These studies confirmed that stealth CMEs usually present distinct characteristics when compared to more typical ``textbook'' CMEs. For example, they are usually rather slow and narrow, and tend to originate from the vicinity of coronal holes or open field regions.

One of the remaining open scientific questions about stealth CMEs is whether or not they represent a different class of CMEs that cannot be explained in terms of the standard model of eruptive flares \citep[e.g.,][]{Svestka1992}. However, \citet{HowardT2013} cautioned that the apparent lack of LCSs associated with these elusive events may be due to the limited sensitivity and temperature response of the observing instruments.  Moreover, there are limitations in the observational geometry of available spacecraft, based on their location relative to the observed stealth CME. The implication of these two limitations is beginning to become apparent. The limited sensitivity and temperature coverage may be partly addressed with improved imaging processing techniques.  
However, the issue of whether or not a spacecraft is positioned at the optimal location for observing these weak eruptive events is less trivial to solve. In fact, as will be shown in Section~\ref{sec:observations}, the relative location of the observing spacecraft (or suite of instruments) may actually determine whether one of these weak eruptive events is unambiguously observed, or if the failure to detect the event allows it to fall into the category of stealth CMEs.

\begin{figure}
\includegraphics[width=1.0\textwidth]{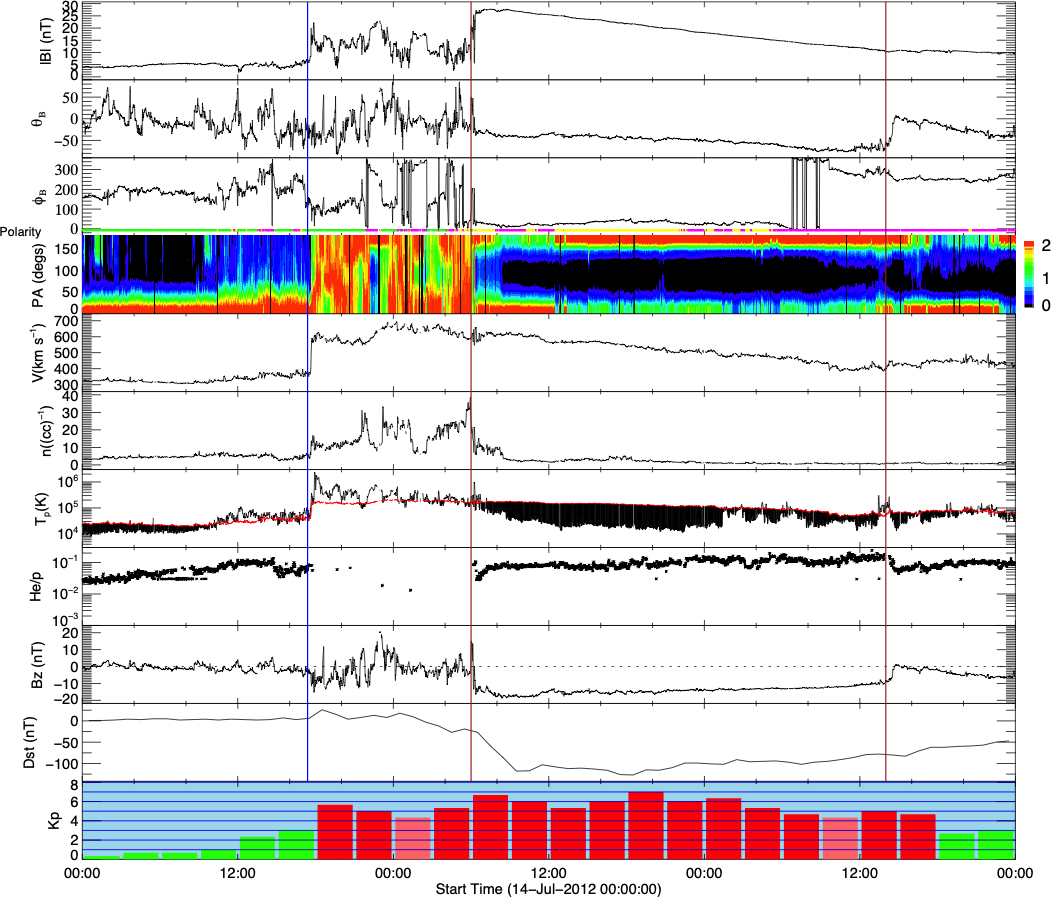}
\caption{In situ solar wind magnetic field and plasma observations at L1 (made by \emph{Wind}) and geomagnetic response for the ``textbook'' ICME\,--\,CME event initiated by a solar eruption on 12 July 2012.  The panels show, from the top: the magnetic field magnitude; the latitude and longitude angles of the magnetic field in the GSE coordinate system with the field polarity indicated below the longitude angle (green: positive/away, pink: negative/toward, yellow: uncertain); the pitch angle distribution of 165~eV electrons; the proton bulk speed, density, and kinetic temperature (with the ``expected temperature" \citep{Richardson1995} indicated in red; black shading indicates where the observed temperature is less than the expected temperature, a frequent characteristic of ICMEs); solar wind alpha to proton ratio, and the Z-component (in GSE) of the magnetic field.   The bottom two panels show the Dst and Kp geomagnetic indices. The blue vertical line shows the shock. The ICME interval (from the Wind ICME Catalog) is bounded by vertical brown lines and has the enhanced, slowly-rotating magnetic field characteristic of a magnetic cloud. }
\label{fig:sw4textbook} 
\end{figure}

\begin{figure}
\includegraphics[width=1.0\textwidth]{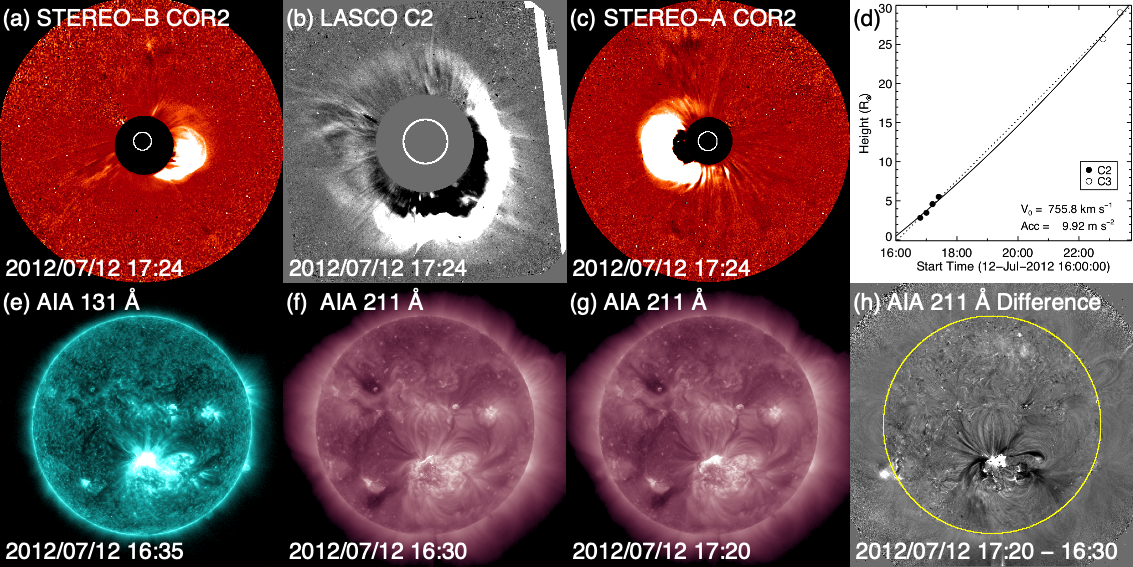}
\caption{Coronagraph (upper row) and low coronal (lower row) observations of the 12 July 2012 solar event.  (a) and (c) STEREO-B (-A) COR2 difference images. (b) LASCO C2 difference image.  (d) Heliocentric height vs. time plot, from LASCO C2 and C3 data with linear (quadratic) fits shown as the dotted (solid) line. (e) AIA 131~{\AA} image at the flare peak. (f) and (g) AIA 211~{\AA} images before and after the eruption.  (h) The difference image of (f) and (g).}
\label{fig:images4textbook} 
\end{figure}

In addition to the intriguing solar physics question of whether or not stealth CMEs represent a unique class of weak eruptive events, we must also consider their potential for geoeffectiveness \citep{Kilpua2014}. In a recent study, \citet{Nitta2017} collected examples of CMEs without clear LCSs that seemed to be associated with near-Earth ICMEs but were not necessarily classified as ``stealth CMEs''. Most of these events resulted in geomagnetic storms, with the largest occurring when the ICME driving the storm was compressed in a CIR ahead of a high-speed solar wind stream. This study emphasised that events that fail to show LCSs with routine data reduction pose an additional challenge to creating accurate space weather forecasts and geomagnetic storm predictions. Events that exhibit ``super-stealthiness'', namely the lack of any Earth-directed CME observations (either on-disc or in coronagraphs), are also of major concern. In the next section, we discuss observations of the ICME\,--\,CME pairs associated with these kinds of events and their relation to problem geomagnetic storms occurring during solar cycle 24. Rather than attempting to quantitatively define the stealth CME category, we will compare and contrast our selected events with a ``textbook'' ICME\,--\,CME pair, in which links between the ICME and CME, and between the CME and LCSs are solidly established, as illustrated in Figures~\ref{fig:sw4textbook} and \ref{fig:images4textbook}.

Figure~\ref{fig:sw4textbook} shows in situ observations from the \textit{Wind} spacecraft \citep{Wilson2021} of a clear ICME that consists of a ``magnetic cloud" (in the interval between the vertical brown lines) characterised by an enhanced magnetic field magnitude and a smooth rotation of the magnetic field through a large angle, suggestive of a flux-rope-like magnetic field structure, and a low proton temperature  \citep[as discussed by e.g.,][]{Burlaga1981,Klein1982,Lepping1990}. This follows a shock (vertical blue line) and a sheath region between the shock and magnetic cloud.  The extended interval of strong southward magnetic fields (negative $B_z$) in the ICME was responsible for an intense geomagnetic storm on 15 July 2012 (minimum Dst $= -139$~nT, maximum Kp $= 7-$).  

During the few days before the arrival of the ICME, the symmetrical halo CME observed by LASCO on 12 July shown in Figure~\ref{fig:images4textbook}(b) was the only possible progenitor. At this time, the STEREO spacecraft were positioned $\approx$120$\arcdeg$ from the Sun--Earth line (see Figure~\ref{fig:stereoloc}(a)), which is a favourable configuration for observing Earth-directed CMEs as limb events. The CME was seen moving to the right in COR2-B images (Figure~\ref{fig:images4textbook}(a)) and moving to the left in COR2-A images (Figure~\ref{fig:images4textbook}(c)), confirming that it was Earth-directed.
Figure~\ref{fig:images4textbook}(d) shows the LASCO CME height--time profile, indicating that the linear speed of the CME was $\approx$800~km~s$^{-1}$ as projected onto the plane of the sky (POS); similar STEREO observations suggest a speed of more than 1000~km~s$^{-1}$ when seen as a limb CME moving with a small angle to the POS. These speeds are comparable to the 1~AU transit speed inferred, for example, from the arrival of the shock ($\approx$850~km~s$^{-1}$), further supporting the CME\,--\,ICME association.   The CME was associated with a flare with an X1.4 peak soft X-ray intensity observed by the GOES spacecraft. The flare is also clearly seen in the AIA image in Figure~\ref{fig:images4textbook}(e) within a large active region (AR~11520) in the southern hemisphere near central meridian.  The eruption left unmistakable coronal dimmings, the dark regions  near the active region in Figure~\ref{fig:images4textbook}(h), which is the difference between the AIA observations separated in time by 50 minutes in Figures~\ref{fig:images4textbook}(f) and (g). Due to the clear connection between the ICME and CME, and between the CME and its corresponding LCSs, this example is considered a ``textbook'' case where the links between the solar eruption and the ICME driving a geomagnetic storm are solid \citep{Webb2017}.


\section{Identification of Problem Geomagnetic Storms in the SDO/STEREO Era}
\label{sec:observations}

In this section, we discuss several problem geomagnetic storms, the associated in situ solar wind structures that drive the storms, and the preceding solar observations that occurred in solar cycle 24 during the SDO era (since May 2010).  In particular, we will take advantage of full-disc EUV images from AIA on board SDO that are a significant improvement over those from previous instruments in terms of sensitivity, resolution (both spatial and temporal), and temperature coverage. These images are expected to reveal weaker on-disc LCSs of CMEs than previously observed.  We also use the capability of the STEREO spacecraft to observe CMEs from locations far from the Sun--Earth line, increasing the probability of unambiguously identifying the Earth-directed CMEs associated with specific ICMEs (cf. the 1\,--\,2 June 2008 CME in Figure~\ref{fig:robbrecht}). However, both STEREO spacecraft drift away from Earth by $\approx$22$\arcdeg$ a year, so the observational geometry changes with time and is not always favourable for viewing Earth-directed CMEs (see Figure~\ref{fig:stereoloc}). This is a critical factor for interpreting certain events, as will be discussed later in this section.  In addition, contact with STEREO-B was lost on 1 October 2014.

\subsection{Event Selection}
\label{sec:observations:event_selection}

The starting point for the event selection for this study is the Richardson and Cane near-Earth ICME catalogue\footnote{\url{http://www.srl.caltech.edu/ACE/ASC/DATA/level3/icmetable2.htm}}, updated from that in \citet{Cane2003} and \citet{Richardson2010}. Since the focus is on problem geomagnetic storms, we began by selecting those ICMEs associated at least with moderate to strong storms, i.e., with minimum Dst $< -50$~nT.  We first removed all ICMEs for which the corresponding CME was already indicated in the catalogue and for which the solar source was confirmed in contemporaneous reports and/or later publications.  We then examined LASCO data for each of the remaining cases and removed those events for which a single wide (angular width $>$120$\arcdeg$) CME from the front side of the Sun could be clearly identified.\footnote{Note that the current on-line version of the Richardson and Cane ICME catalogue includes additional CME associations that were identified as a result of this study.} We also excluded cases where the geomagnetic storm was driven by an ICME consisting of more than one structure (such as more than one magnetic field rotation suggestive of multiple flux ropes), or by the interaction of multiple ICMEs. These storms most likely had complex solar sources and the in situ structures may have evolved during transit from the Sun.  Such complex cases are also unlikely to be good candidates for assessing whether stealth CMEs may be involved in producing geomagnetic storms. 

The 16 ICMEs remaining, for which either the isolation of the associated CME is problematic in LASCO imagery or the LCSs are elusive in AIA imagery, are listed in Table~\ref{tab:tab_1}, which includes the basic information on the geomagnetic storm and the ICME driver.  For these events, we carefully examined solar and coronagraph observations in an attempt to isolate the CME associated with the ICME and the related LCSs, as described in Sections~\ref{sec:observations:cme} and \ref{sec:observations:lcs}.  The results of this search are given in Table~\ref{tab:tab_2}; the incomplete table reflects the difficulty of this task. In Sections~\ref{sec:observations:event_24}\,--\,\ref{sec:observations:event_15}, we discuss the six events indicated by asterisks in  Tables~\ref{tab:tab_1} and \ref{tab:tab_2}, summarise the identification and characteristics of the ICME\,--\,CME pairs, and contrast these  with those of the textbook event presented in Figures~\ref{fig:sw4textbook} and \ref{fig:images4textbook}.

\begin{table}
\renewcommand{\arraystretch}{1.0}
\setlength{\tabcolsep}{.02in}
\caption{List of Selected ICMEs With Dst $\le -50$~nT Geomagnetic Storms }
\label{tab:tab_1}
\begin{tabular}{ccrlcccccc}

\hline
\\
 & \multicolumn{3}{c}{Geomagnetic Storms} & & \multicolumn{4}{c}{ICMEs} \\
\cline{2-4} \cline{6-10} \\
 1 & 2 & 3 & 4 & & 5 & 6 & 7 & 8 & 9 \\
 ID & Time & Dst  & Kp & & Disturbance  & Start &  End & Speed        &  ICME \\
    & (UT) &(nT)  &    & & (UT)   & (UT)  & (UT) & (km~s$^{-1}$)  & Signatures \\
\\
\hline
 1  & 2011/02/04 22:00 & -63 & 6- & & 02/04 13:00 & 02/04 13:00 & 02/04 20:00 & 470/430 & 201101 \\
 2  & 2012/02/15 17:00 & -67 & 5+ & & 02/14 07:00 & 02/14 21:00 & 02/16 06:00 & 400/370 & 110000\\
 3  & 2012/09/03 11:00 & -69 & 6- & & 09/01 06:00 & 09/01 07:00 & 09/03 15:00 & 310/330 & 111000\\
 4  & 2012/10/13 07:00 & -90 & 6- & & 10/12 19:00 & 10/12 22:00 & 10/13 10:00 & 530/490 & 210001\\
 5  & 2013/01/17 22:00 & -53 & 4o & &   ---              & 01/17 16:00 & 01/18 12:00 & 400/390 & 211001\\
 6  & 2013/06/07 05:00 & -73 & 6- & & 06/06 02:55 & 06/06 14:00 & 06/08 00:00 & 510/430 & 211000\\
 7  & 2013/06/30 01:00 & -98 & 6+ & & 06/27 13:51$^{s}$ & 06/28 02:00 & 06/29 12:00 & 450/390 & 211001\\
 8*  & 2013/11/09 08:00 & -81 & 5o & &   ---              & 11/08 22:00 & 11/09 07:00 & 460/420 & 210001\\
 9*  & 2014/04/12 09:00 & -81 & 5- & &   ---              & 04/11 06:00 & 04/12 20:00 & 380/350 & 211000\\
10  & 2014/04/30 09:00 & -64 & 4o & &   ---             & 04/29 20:00 & 04/30 21:00 & 320/310 & 211001\\
11  & 2015/01/04 17:00 & -62 & 5+ & &   ---              & 01/03 14:00 & 01/04 16:00 & 460/430 & 111001\\
12*  & 2015/01/07 12:00 & -99 & 6+ & & 01/07 05:39$^{s}$ & 01/07 06:00&01/07 21:00&470/450 & 211100\\
13*  & 2015/05/11 04:00 & -51 & 4- & &   ---              & 05/10 12:00&05/11 02:00&400/370 & 210000\\
14*  & 2016/10/14 00:00 &-104 & 6+ & & 10/12 21:15$^{s}$ & 10/13 06:00&10/14 16:00&460/390 & 211101\\
15  & 2017/05/28 08:00 &-125 & 7o & & 05/27 14:41$^{s}$ & 05/27 22:00&05/29 14:00&390/360 & 201010\\
16*  & 2018/08/26 07:00 &-174 & 7+ & &  08/25 01:05$^{s}$              & 08/25 12:00&08/26 10:00&440/410 & 210011\\

\hline
\end{tabular}
Column: (1) Event ID; (2) Time (year/month/day hour:min (UT)) of storm Dst minimum; (3) Minimum Dst index (nT) from the Kyoto Dst server (\url{http://wdc.kugi.kyoto-u.ac.jp/dstdir/} with final values up to 2014; provisional in 2015\,--\,2016: and quicklook from  2017 onwards); (4) Maximum Kp index from the official Kp index page at the German Research Centre for Geosciences (\url{https://www.gfz-potsdam.de/en/kp-index/});  (5) Disturbance arrival time -- shock/bow wave if present (indicated by letter ``s''); (6 and 7) ICME start and end times to the nearest ${\sim}1$ hour; (8) Maximum solar wind speed in sheath or ICME (km~s$^{-1}$)/mean speed in ICME; (9) in situ ICME signatures. The first number (i) indicates (1) that the ICME contains a flux rope structure or (2) that the ICME is a magnetic cloud according to the stricter definition of \citet{Burlaga1981}. The remaining five numbers designate the presence (1) or absence (0) of (ii) abnormally low solar wind proton temperature \citep[cf.][]{Richardson1995}, (iii) bidirectional solar wind electron heat flux (from ACE/SWEPAM, WIND/3DP), (iv) composition/charge state signatures (from ACE/SWICS), (v) enhanced solar wind He/p ratio), and (vi) a HSS following the ICME ejecta. *: Events discussed in Sections~\ref{sec:observations:event_24}\,--\,\ref{sec:observations:event_34}.
    
\end{table}


\subsection{Isolation of the CMEs associated with the selected ICMEs}
\label{sec:observations:cme}

The in situ ICME speed and arrival time can provide an estimate of the time when the associated CME left the Sun (assuming, probably unrealistically, that the ICME travelled at a constant speed) and hence indicate the approximate time of the solar eruption responsible for the ICME.  However, since stealthy CMEs tend to be slow, and slow CMEs tend to be accelerated by the ambient solar wind flow, the speed of the background solar wind through which they propagate may significantly impact their transit speed and thus the inferred eruption time at the Sun. Hence, we searched for evidence of a related solar eruption  around the back-extrapolated launch time inferred from in situ observations, using this to guide rather than constrain the solar event time.

\begin{figure}
\includegraphics[width=0.99\linewidth]{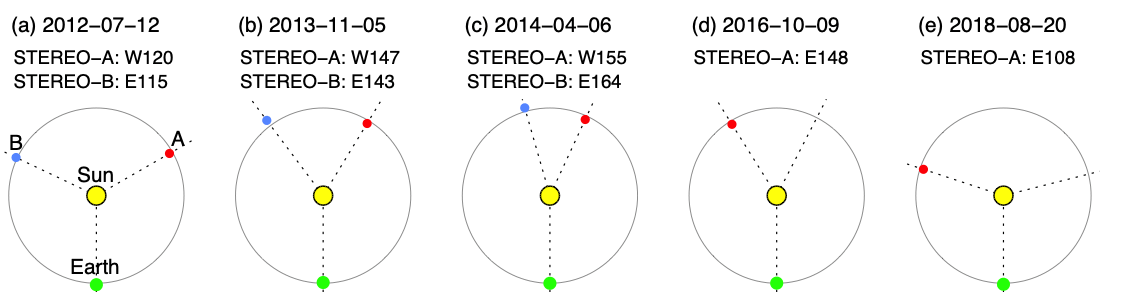}
\caption{Configuration of Earth and the STEREO spacecraft close to the ecliptic plane at five different times. Note that contact with STEREO-B was lost on 1~October~2014 (between (c) and (d)) when it was $161^\circ$ east of Earth. }
\label{fig:stereoloc} 
\end{figure}

The results to be discussed below clearly confirm the value of STEREO observations when identifying the Earth-directed CME associated with a specific ICME.  In particular, these observations allow us to (1) observe CMEs that are not bright or wide enough to be detected by LASCO, and (2) determine whether they are indeed directed towards Earth. Figure~\ref{fig:stereoloc} shows the relative locations of STEREO and Earth at five representative times. As already noted, at the time of the ``textbook event" in Figure~\ref{fig:images4textbook}, STEREO-A and STEREO-B were $\approx$120$\arcdeg$ west and east of the Sun--Earth line, respectively (Figure~\ref{fig:stereoloc}(a)), and favourably positioned to observe faint/narrow CMEs that might be Earth-directed. However, this became more difficult as the STEREO spacecraft drifted farther away from the Sun--Earth line (Figure~\ref{fig:stereoloc}(b, c)).  In the configuration shown in Figure~\ref{fig:stereoloc}(c), the STEREO spacecraft have similar lines of sight (but in the opposite direction) to that from Earth. Nevertheless, EUVI observations of the far side of the Sun are still useful for confirming the far-side origin of CMEs when AIA images show no LCSs. Following superior conjunction in 2015, the location of STEREO-A again facilitated the detection of Earth-directed CMEs (see Figures~\ref{fig:stereoloc}(d) and (e)). Unfortunately, as already noted, contact was lost with STEREO-B on 1 October 2014 when it was 161$\arcdeg$ east of the Sun--Earth line.  

\subsection{Low Coronal Signatures}
\label{sec:observations:lcs}

Once an Earth-directed CME is isolated, we extensively analyzed AIA images to find its LCSs, such as coronal dimmings, post-eruption arcades (PEAs), and separating flare ribbons, around the time of CME liftoff, deduced by extrapolating the observed height--time profile of the CME back to $\approx$1\,$R_{\odot}$. This extrapolation is not always straightforward, in particular if the CME very gradually accelerates close to the Sun, and the time at which this acceleration starts is hard to determine exactly. 

Coronal dimmings are one of the most reliable LCSs of CMEs \citep[e.g.,][]{Thompson2000}, even though some CMEs do not show them. Coronal dimmings are often detected in so-called ``base-difference'' images \citep{Zhukov2004, Attrill2010}, in which changes of brightness are recorded with respect to a pre-event image, in contrast to ``running-difference'' images generated by differencing successive images. Running-difference images are commonly used for finding transient changes such as coronal waves, but base-difference images are less widely used. One reason may be the greater need to compensate for differential rotation as the time difference becomes longer.  Even when differential rotation is compensated for, spurious effects can become more serious for longer time differences. This is partly because the rotation rate at the photosphere is usually assumed, whereas emission detected in each pixel is contributed by coronal plasma at a range of heights that may rotate faster or slower than the photosphere. Another caveat when interpreting difference images with long time separations is that they are more likely to include variations that are spatially or temporally unrelated to the CME. In order to find dimming in less bright regions, the logarithmic base ratio may be used \citep{Dissauer2018}.

It turns out that none of the events to be discussed here were detected as dimming events with automated algorithms \citep[e.g., Solar Demon\footnote{\url{http://solardemon.oma.be}};][]{Kraaikamp2015}, even when base-difference images were used. One reason is that these algorithms may use criteria based on the rate of (negative) change in brightness (E. Kraaikamp, private communication 2020).  As reported by \citet{Nitta2017}, stealthy events tend to start very slowly, with an initial low CME speed and acceleration.  This is expected to result in a very small rate of dimming development \citep[see][]{JMason2016, Dissauer2019} that may be insufficient to trigger an automated algorithm. Therefore, we also examine non-differenced intensity images to identify dimmings \citep[cf.][]{Krista2013}.

Another type of LCSs  are the large-scale (i.e., active-region scale) brightenings that may grow with time, and be interpreted as PEAs or flare ribbons.  Unlike normal CMEs for which the LCSs are unquestionable (Figure~\ref{fig:images4textbook}), stealthy CMEs tend to show only weak and slowly evolving associated dimmings and brightenings, which may be difficult to distinguish from other unrelated activities at different times and locations. The temporal coincidence of dimmings and brightenings with a CME, even though the CME liftoff time may be uncertain by several hours, increases our confidence that these weak signatures are associated with the CME. Another test is whether the quadrant of the source region on the solar disc is consistent with the position angle of the most prominent part of the CME in coronagraph data.

In addition to these approaches, the members of the ISSI team (see Introduction) have started using other techniques, such as the Multiscale Gaussian Normalization technique \citep{Morgan2014} to enhance lower level signals of the low coronal signatures of CMEs \citep{Alzate2017,OKane2019}, and the Graduated Cylindrical Shell model \citep{Thernisien2006,Thernisien2009,Thernisien2011} that gives the 3D trajectory of a CME that may be traced back to the initiation site. The usefulness of these tools for understanding stealthy CMEs is evaluated in a separate paper \citep{Palmerio2021a}. 

\begin{table}
\caption{CMEs and Low Coronal Signatures Associated With the Events in Table~\ref{tab:tab_1}}
\label{tab:tab_2}
\begin{tabular}{llrcccc}

\hline
\\
 & \multicolumn{3}{c}{CMEs} & & \multicolumn{2}{c}{Sources}\\
\cline{2-4} \cline{6-7} \\
 1 & 2 & 3 & 4 & & 5 & 6 \\
 ID & Time & Speed & Width/PA & & Coronal & Source \\
    & (UT) & (km~s$^{-1}$)  & (Degrees) & & Signatures & Location \\
\\ 
\hline
 1 & 2011/01/30 12:36            & 120      & 264/184    &&  DE     &  S25E40 \\
 2 & 2012/02/10 10:48            & ---      & ---        &&  BDE    &  S16W03\\
 3 & 2012/08/26 07:24            & 87       &  44/164    &&  BD    &  S27W05\\
 4 & 2012/10/09 02:36            & ---      & 138/246    &&  BD    &  S26W10\\
 5 & 2013/01/13 12:00            & 229      & 176/384    &&  B     &  N13E21\\
 6 & 2013/06/02 20:00            & 222      &  87/93     &&  BDEF   &  N10W18\\
 7 & 2013/06/23 22:36            & 174      & 174/284    &&  BDE    &  N25W05\\
 8* & 2013/11/05 07:24           & ---      &  30/275    &&  D     &  S10W20\\
 9* &    ---                     & ---      & ---        &&  ---   &  ---   \\
10 &    ---                      & ---      & ---        &&  ---    &  ---   \\
11 &    ---                      & ---      & ---        &&  ---   &  ---   \\
12* & 2015/01/03 03:12           & 163      & 153/118    &&  DE     &  S25E06\\
13a* & 2015/05/06 00:12           & 221      &  55/252    &&  BDF   &  S35W05\\
13b* & 2015/05/06 16:48           & 479      & 100/263    &&  BDF   &  S08W14\\
14* & 2016/10/09 01:25           & 179      & 360        &&  DE     &  S05E17\\
15 & 2017/05/23 05:36            & 259      & 243/259    &&  BDF   &  N12W13\\
16* & 2018/08/20 21:24           & 126      & 120/252    &&  BDEF   &  N20E03 \\

\hline
\end{tabular}

Column: (1) Event ID; (2) Time of the associated CME, if present, observed by LASCO C2; 
(3) CME linear speed; (4) CME angular width/position angle measured anti-clockwise from the north pole; (5) On-disc signatures if present (B=brightening; D=dimming; E=enlargement of a coronal hole; F=filament); (6) Source location. *: Events discussed in Sections~\ref{sec:observations:event_24}\,--\,\ref{sec:observations:event_15}. The characteristics of the CMEs (in columns 3 and 4) are taken from the CDAW LASCO CME catalogue \citep[\url{https://cdaw.gsfc.nasa.gov/CME_list/};][]{Yashiro2004,Gopalswamy2009}.


\end{table}


\subsection{A Common Stealth Event: Event 12}
\label{sec:observations:event_24}

\begin{figure}
\includegraphics[width=0.99\textwidth]{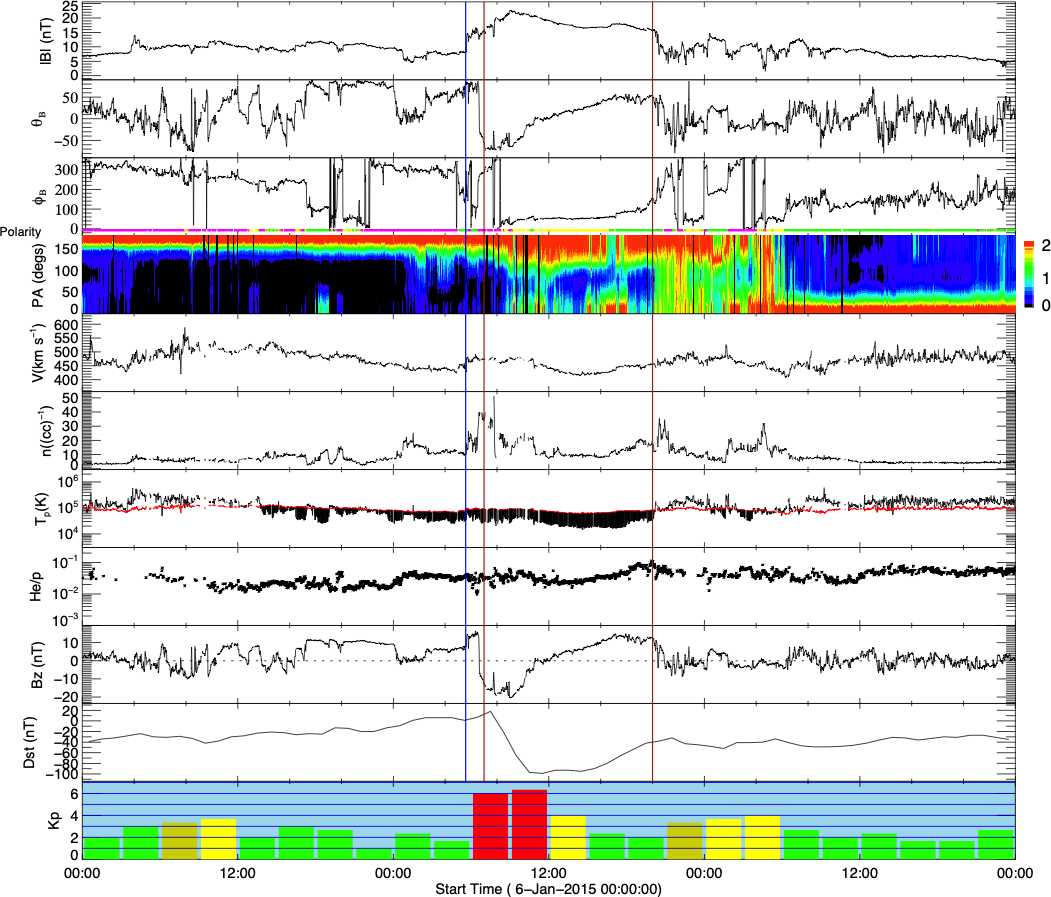}
\caption{In situ solar wind magnetic field and plasma observations at L1 and geomagnetic response (Dst and Kp indices, last two rows) for Event 12 (storm of 7 January 2015) in the same format as Figure~\ref{fig:sw4textbook}.}
\label{fig:sw4event24} 
\end{figure}

\begin{figure}
    \includegraphics[width=\textwidth]{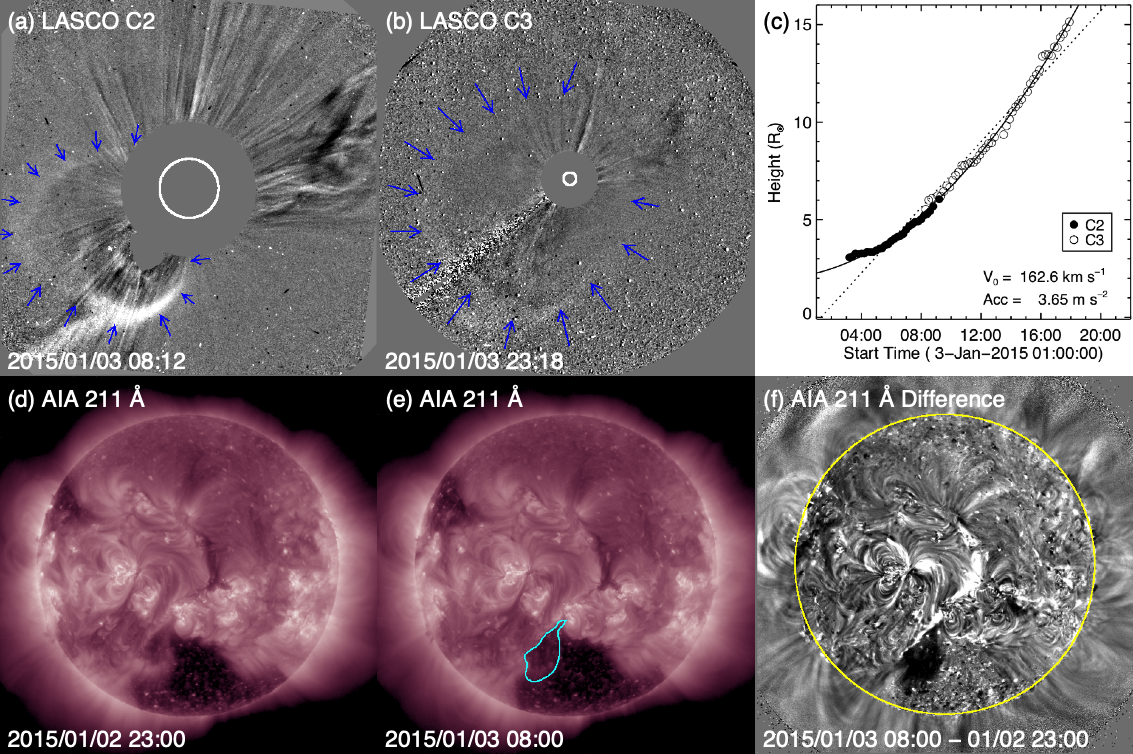}
    \caption{Coronagraph (upper row) and low coronal (lower row) data showing the slow and diffuse CME on 3 January 2015 (Event 12). (a), (b) LASCO C2 and C3 difference images. The CME front is indicated with arrows. (c) Heliocentric height vs. time plot, from LASCO C2 and C3 data with linear (quadratic) fits shown as the dotted (solid) line.  (d) and (e) AIA 211~\AA\ images before and after the eruption.  (f) The difference image of (d) and (e) which reveals the brightening and dimming associated with the eruption. The region outlined in (e) shows an enlarging of the southern polar coronal hole. }
    \label{fig:images4event24}
\end{figure}

We begin our individual case descriptions of the selected events in Table~\ref{tab:tab_1} with an event that was associated with a type of stealth CME that is quite commonly observed. Although \citet{Nitta2017} considered this particular stealth CME event, on 3 January, 2015, to be more challenging to understand than others in their study, we consider it to be the simplest of the selected events to contrast with the textbook event, and therefore an excellent example with which to start. Figure~\ref{fig:sw4event24}, in the same format as Figure~\ref{fig:sw4textbook}, summarises the related in situ solar wind magnetic field and plasma observations from \textit{Wind} along with the Kp and Dst indices. This particular stealthy CME caused a strong geomagnetic storm on 7 January 2015 with a minimum Dst of $-99$~nT at 11~UT and maximum Kp of 6+ in the interval of 9\,--\,12~UT. This storm and the related in situ and solar observations have been discussed by \citet{Cid2016}, \citet{Nitta2017}, \citet{Palacios2017}, and \citet{Yardley2021b}. Similar to the textbook ICME shown in Figure~\ref{fig:sw4textbook}, the in situ signatures of this ICME are clearly observed. The geomagnetic storm is driven by the leading region  of strong southward magnetic field within the reasonably slow ($\sim$450~km~s$^{-1}$) ICME (bounded by the vertical brown lines) which includes a magnetic cloud \citep{Burlaga1981}, characterised by an enhanced magnetic field strength that rotates through a large angle, and depressed proton temperatures. This is preceded by a narrow sheath with a weak or developing shock at the leading edge (vertical blue line).

Comparing directly with Figure~\ref{fig:sw4textbook}, the similar in situ signatures of this particular ICME give no obvious indication that it is associated with a stealthy CME. The major difference is in the lower solar wind speeds associated with this ICME, which can be used to identify the associated CME. Assuming that the ICME propagated from the Sun to 1~AU at a constant speed of 450~km~s$^{-1}$ (based on the in situ ICME speed) suggests a transit time of $\approx$93 hours and hence an origin at the Sun around 13~UT on 3 January 2015. The only candidate CME observed within a reasonable window of time is the partial halo CME first observed by LASCO C2 at 03:12~UT on 3 January with a speed of 163~km~s$^{-1}$, a width of 153$\arcdeg$, and a central position angle (measured from north and anti-clockwise) of 118$\arcdeg$ (see Figure~\ref{fig:images4event24} (a)\,--\,(c)). The low CME speed is not inconsistent with the assumed 1~AU transit speed since Figure~\ref{fig:sw4event24} suggests that the ICME was carried out by the ambient solar wind flow.

Why is this CME considered to be stealthy? Focusing on the on-disc source regions for this event, analyses of the SDO/AIA passbands are not particularly revealing, as compared to the textbook event. The LCSs of this CME are not readily isolated using standard methods of on-disc detection, whether intensity, running-difference, or even base-difference images are used, unless images taken over a very long time span are examined. As reported in \citet{Nitta2017}, the source location is determined to be at S25E06 from evidence of a slow coronal dimming revealed in long-duration observations of the AIA 211~{\AA} images and in base-difference imaging taken 9 hours apart. The signatures in the intensity images (before and after eruption) and in the base-difference image are shown in Figure~\ref{fig:images4event24}(d)\,--\,(e) and (f), respectively. However, the methods employed by \citet{Nitta2017} to reveal these source regions are not commonly used, and are not required, for example, to elucidate the on-disc source region of the CME in the textbook case shown in Figure~\ref{fig:images4textbook}. The contrasts between the on-disc signatures of the textbook CME and the stealthy CME are striking. For example, in Figures~\ref{fig:images4textbook}(e) and (g), the source region of the textbook CME eruption is clearly observed in the raw imagery. A rapid intensification in the flaring region, including flare ribbons and PEAs, clearly defines the source regions. In the base-difference image in Figure~\ref{fig:images4textbook}(h), not only is the flaring region well-defined, but the dimming region is also well-developed. By contrast, in Figures~\ref{fig:images4event24}(d)\,--\,(f), there is no obvious flaring region, and the brightenings possibly related to the CME are only revealed through the analysis of long-duration base-difference imagery. The dimming region is peculiar as well. Not only does it require over 9 hours to manifest, it also results in the slow widening of the southern polar coronal hole. The perimeter of the coronal hole region enlarged by the stealthy CME eruption is outlined in Figure~\ref{fig:images4event24}(e) and is visible as the dimming region in Figure~\ref{fig:images4event24}(f).

Unfortunately, at the time of this event, contact had already been lost with STEREO-B and STEREO-A was at 172$\arcdeg$ west of Earth on the far side of the Sun, during which time no useful data were downlinked. The  lack  of  STEREO data makes it difficult to confirm that the CME observed by LASCO was in fact directed towards Earth. As \cite{Nitta2017} noted, it is in principle possible that this CME came from the far side.  Thus, this event illustrates one aspect of investigating apparently stealthy CMEs: In order to identify Earth-directed CMEs, and to locate the source regions of stealthy events, multiple viewpoints of the eruption are highly desirable. Comparison of Figures~\ref{fig:images4event24}(a) and (b) with Figures~\ref{fig:images4textbook}(a)\,--\,(c) illustrates why multipoint measurements of stealthy CMEs are so critical. In Figure~\ref{fig:images4textbook}(b), the halo CME is obvious and well-defined, and the STEREO observations in Figure~\ref{fig:images4textbook}(a) and (c) clearly confirm that it is Earthward-directed. However, in Figures~\ref{fig:images4event24}(a) and (b), there is no clear halo CME, but the diffuse partial halo CME has an appearance of consisting of multiple parts.
Thus, without coronagraph imagery from STEREO, particularly in quadrature, the unambiguous interpretation of the stealthy CME on 3 January 2015 is severely hindered. We will return to this limiting aspect when discussing subsequent examples of stealthy CMEs.

Although, as previously mentioned, the signatures of this stealth CME are more visible than those of some other examples covered in this review, it nevertheless establishes several critical concepts. The first is that it is possible to unambiguously observe a stealth CME in both on-disc signatures and in coronagraphs, but this may require non-standard techniques such as long-time difference imagery. One can surmise that this is because these CMEs erupt in a radically different way from typical CMEs. The second critical concept is that multipoint observations play a vital role in accurate stealth CME detection. In this case, neither STEREO-A nor STEREO-B coronagraph observations were available, which made proper identification of the source region more difficult. In the following examples, we will discuss in detail how STEREO coronagraph observations in near-quadrature with Earth are necessary to pinpoint the proper stealth CME candidate, especially when the stealth CME is being ``camouflaged'' by other, near-simultaneous CME eruptions seen in LASCO imagery.


\subsection{Multiple Candidates, Including a Prominent but Unrelated Event: Event 13}
\label{sec:observations:event_26}

\begin{figure}
\includegraphics[width=0.99\textwidth]{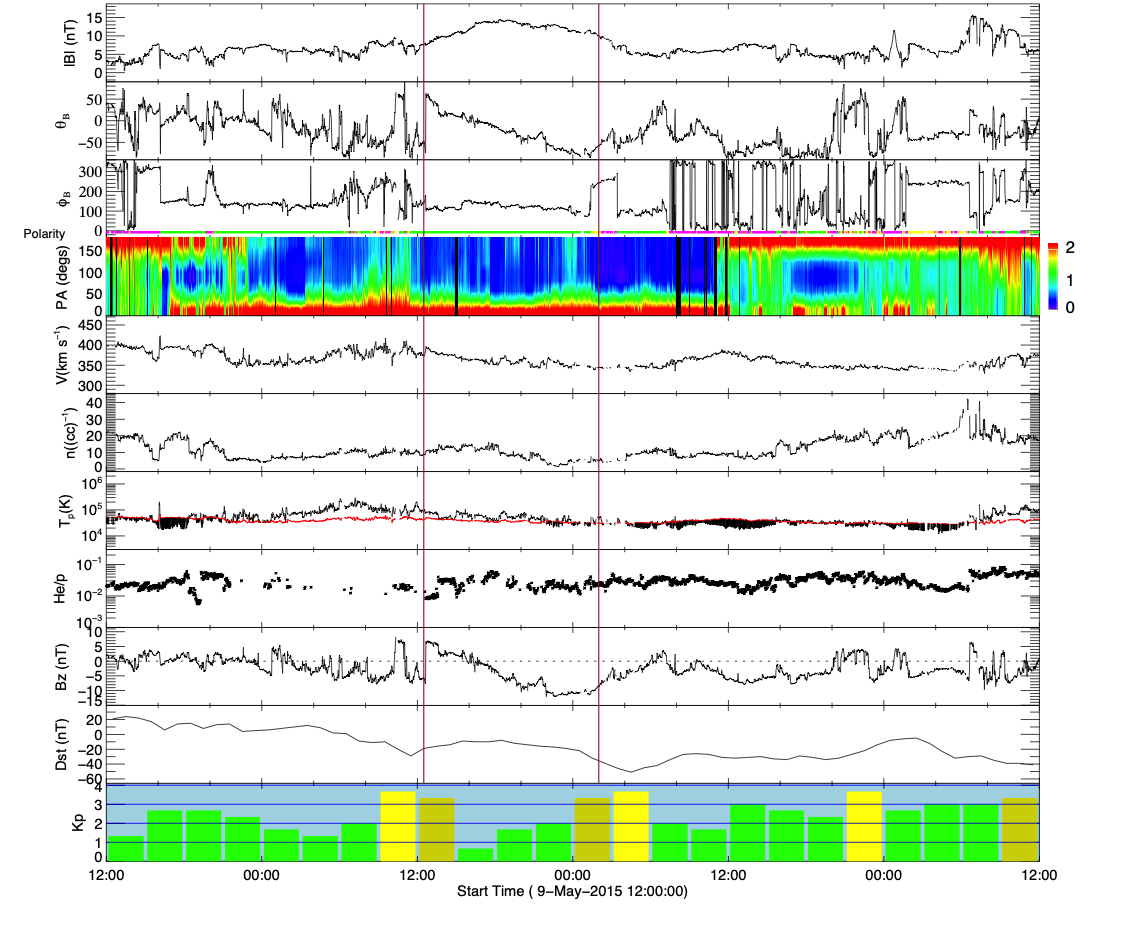}
\caption{In situ observations at L1 and geomagnetic responses (last two rows) for Event~13 (storm of 11 May 2015) in the same format as Figure~\ref{fig:sw4textbook} except that there was no shock.  }
\label{fig:sw4event26} 
\end{figure}

\begin{figure}
    \centering
    \includegraphics[width=0.99\textwidth]{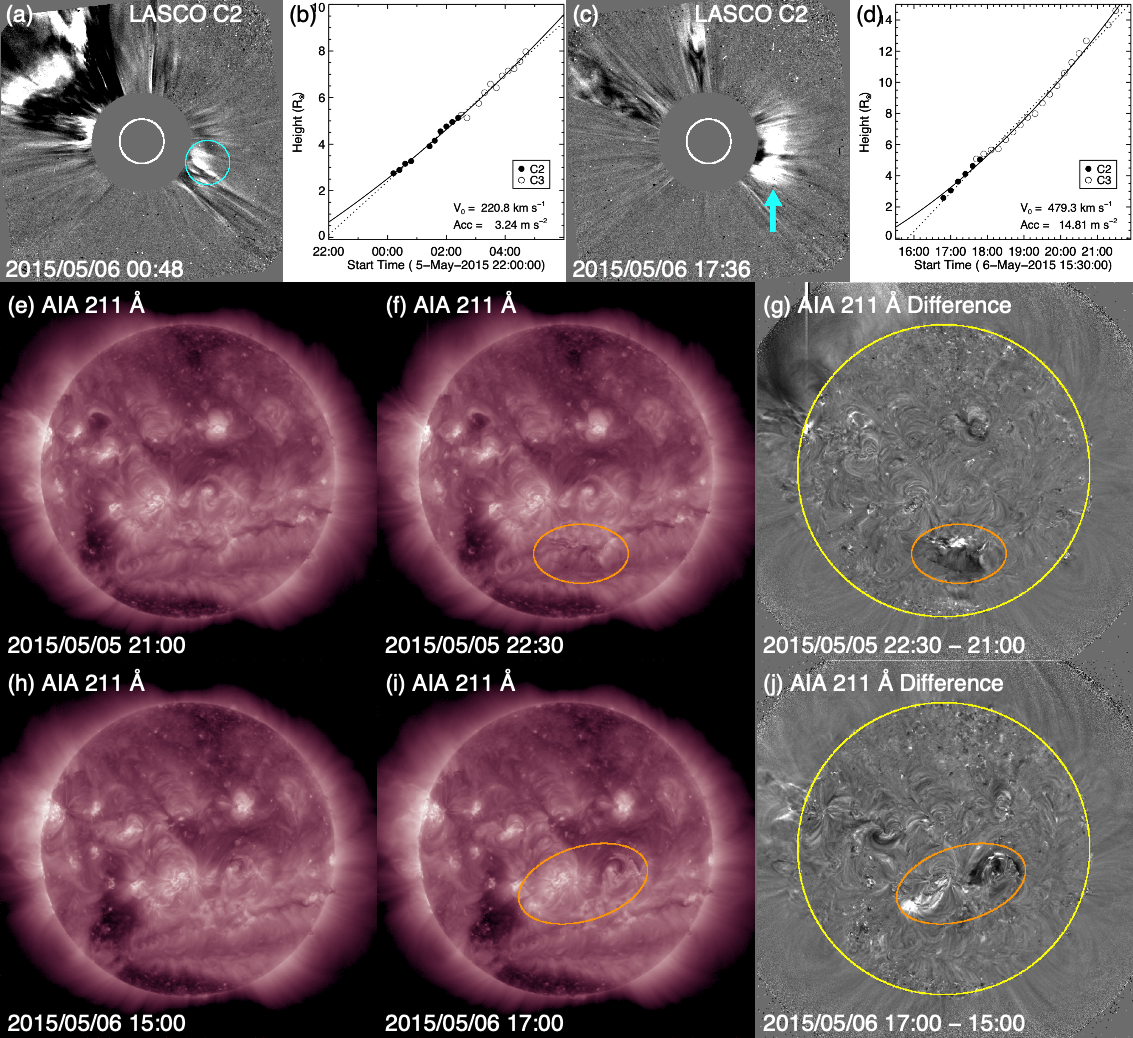}
    \caption{Event 13: (a) and (c) Two possible CMEs that could have been responsible for the ICME that started on 10~May~2015.  Their heliocentric height vs. time plots are shown in (b) and (d).  (e) and (f) AIA 211~{\AA} images before and after the filament eruption responsible for the first CME.  (f) The difference image of (d) and (e).  The source region for the CME is encircled in (e) and (f). (h)\,--\,(j) Same as (e)\,--\,(g) for the second CME.}
    \label{fig:images4event26}
\end{figure}

Another cause of problem geomagnetic storms can arise during a period of heightened solar activity that makes it difficult to properly identify the CME that is responsible for the storm. Such a CME by itself may not necessarily be stealthy. Event 13 is such a case, where we could not convincingly single out the CME associated with the ICME that caused the problem geomagnetic storm on 11 May 2015. Wind and geomagnetic data for this ICME and storm are shown in Figure~\ref{fig:sw4event26}. This weak storm, with the minimum Dst index of only $-51$~nT at $\sim$04:00~UT, was driven by a slow ($\sim$370~km~s$^{-1}$) ICME in which the magnetic field rotated from north to south, preceded by a small initial increase in geomagnetic activity contributed by southward fields in the sheath. Comparing this ICME to the textbook event in Figure~\ref{fig:sw4textbook}, the ICME plasma signatures are less evident. The slow ICME did not drive a shock nor are there clearly depressed proton temperatures or bidirectional suprathermal electron flows in the ICME. However, both ICMEs include the enhanced and smoothly rotating magnetic fields characteristic of a magnetic cloud.

\begin{table}
\caption{Event 13: List of Partial/Full Halo CMEs}
\label{tab:cmes_201505}
\begin{tabular}{llcc}

\hline
\\
Time & Speed & Width/PA & Source \\
(UT) & (km~s$^{-1}$)  & (Degrees) & Location \\
\\ 
\hline
 2015/05/05 22:24            & 715     & 360/41        &  N15E79  \\
 2015/05/06 12:12*         & 738     & 132/53        &  N15E67  \\
 2015/05/06 19:00            & 308     & 199/149       &  Behind S limb  \\
\hline
\end{tabular}

*: The CME resulted from at least three separate eruptions on or close to the east limb. The location is referred to the M1.9 flare  associated with the widest ($\approx$50$\arcdeg$) component of the compound CME.
\end{table}

The in situ ICME speed (assumed to be constant from the Sun to 1~AU) suggests a solar origin time of around 4~UT on 6 May 2015. LASCO halo or partial halo CMEs around this time are listed in Table~\ref{tab:cmes_201505}. Figure~\ref{fig:images4event26}(a) shows a prominent CME at the northeast limb associated with an X2.7 flare at N15E79 starting at 22:05~UT on 5 May (first CME in Table~\ref{tab:cmes_201505}). Weak western extensions of this CME almost camouflaged a second, fainter CME above the west limb (encircled) that appeared in the C2 FOV at 00:12~UT on 6 May 2015. This CME was apparently due to the eruption of a quiet-Sun filament around S30W06 at the location indicated in  Figures~\ref{fig:images4event26}(f)\,--\,(g); Figure~\ref{fig:images4event26}(e) shows the reference image used to create the 1.5 hour-separation difference image in Figure~\ref{fig:images4event26}(g).  This was therefore not a stealthy CME. We can reasonably exclude the strong east limb CME associated with the X2.7 flare (at E79) as the origin of the ICME observed in near-Earth space, which then leaves the frontside filament eruption as a candidate, even though the CME was faint, slow (221~km~s$^{-1}$, see Figure~\ref{fig:images4event26}(b)), and too narrow (55$^\circ$) to be a partial halo CME.

However, it is not possible to determine with certainty whether this CME from the filament eruption was responsible for the geomagnetic storm in Figure~\ref{fig:sw4event26}, as a second CME appeared in LASCO imagery $\approx$17 hours later (indicated in Figure~\ref{fig:images4event26}(c)). This is also a viable candidate for the ICME on 11~May, requiring a 1~AU transit speed (${\approx}460$~km~s$^{-1}$) that is not inconsistent with the measured ICME speed. In solar disc imagery, we note two eruptions from the area  west of the coronal hole in the southeast quadrant (circled in Figure~\ref{fig:images4event26}(h)\,--\,(j)) that were likely sympathetic eruptions. They both featured coronal dimmings and post-eruption arcades.  The eruption in the western part of the area apparently accounted for the CME coming from the west, so this is again not a stealth CME; it is not clear whether the eruption in the eastern part of the area also contributed to the CME.  This CME was brighter and faster (479~km~s$^{-1}$) in LASCO imagery than the CME from the previous day, and almost as wide as a partial halo event (angular width of 100$\arcdeg$). 

This example showcases that it is not a trivial exercise to identify the correct CME\,--\,ICME connection in the absence of a prominent parent CME (and when all candidates are equally weak, originate from nearby regions on the Sun, and are not associated with a clear full halo CME). In this case, there are two CMEs that might have been associated with the geomagnetic storm. Unfortunately, instruments on STEREO-A, on the far side of the Sun close to superior conjunction, were turned off at this time. Therefore, STEREO-A was unable to help confirm whether these CMEs were directed towards Earth and help identify the CME most likely to be associated with the storm. Ideally, a spacecraft near quadrature to Earth might have observed a minor eruption from the Earth-facing limb and helped to resolve this problem.


\subsection{A ``Super Stealthy'' Event: Event 9}
\label{sec:observations:event_19}

\begin{figure}
\includegraphics[width=0.99\textwidth]{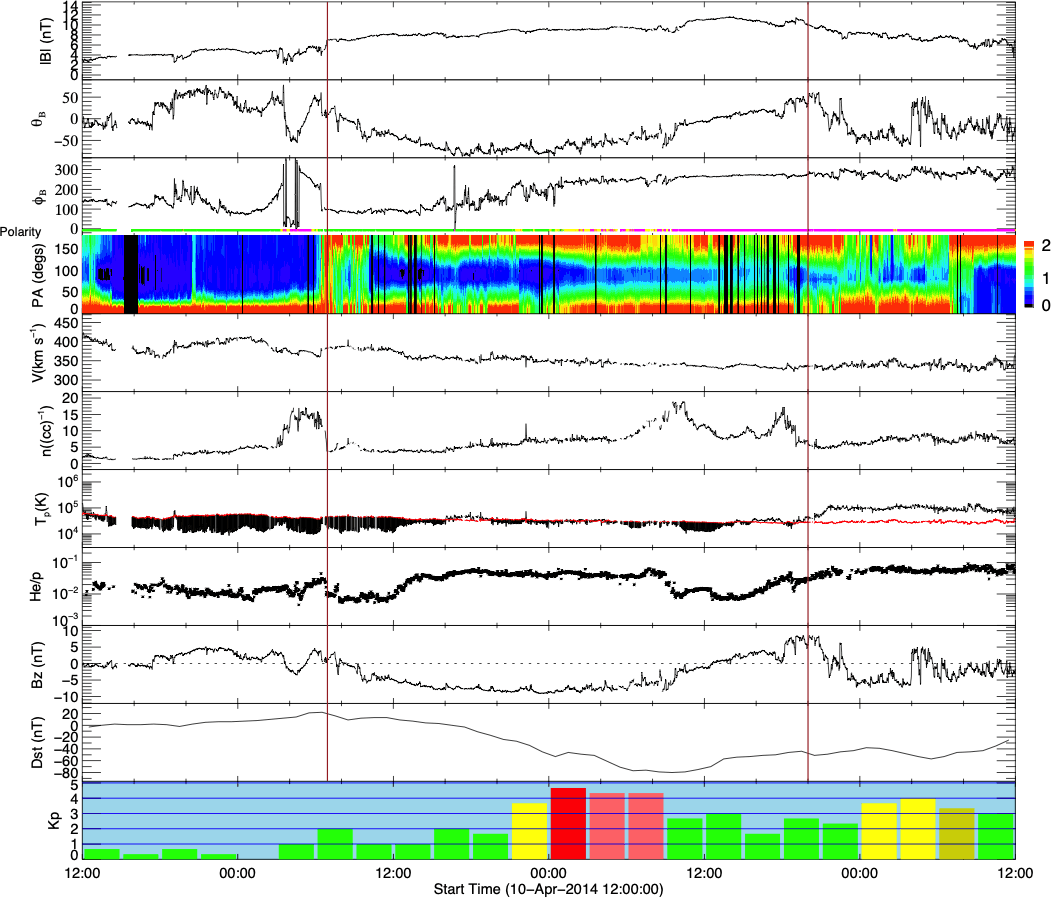}
\caption{In situ observations at L1 and geomagnetic response for Event 9 (storm of 12 April 2014) in the same format as Figure~\ref{fig:sw4textbook} except that there was no shock in this case. }
\label{fig:sw4event19} 
\end{figure}

The next event in our discussion is significant, not so much in the intensity of the geomagnetic storm it caused, but rather in the apparent lack of CME signatures in Earth-based observations, both in coronagraph data and on disc images. This event in particular highlights the necessity of multipoint observations in quadrature to be able to unambiguously detect the stealthiest of CME eruptions. 

The geomagnetic storm on 12 April 2014 was caused by the ICME shown in Figure~\ref{fig:sw4event19}. The Dst index during the storm reached a minimum of $-81$~nT at 9~UT and the Kp index registered a maximum of $5-$. Once again comparing the ICME causing this geomagnetic storm to the textbook case shown in Figure~\ref{fig:sw4textbook}, the in situ signatures of this event are also clearly evident and give no indication that it is associated with a stealthy CME. The geomagnetic storm is not particularly strong and was driven by the persistent southward component of the interplanetary magnetic field inside the ICME. In this case, no shock was present ahead of the ICME because the ICME speed was similar to that of the upstream solar wind. The density enhancement just ahead of the ICME may be associated with an encounter with the heliospheric plasma sheet (HPS) (e.g., \cite{Winterhalter1994}; see also Section~\ref{sec:observations:event_15}), based on the slight depression in magnetic field strength and localized variation in the field direction,  rather than evidence of a sheath region.

\begin{table}
\caption{Event 9: List of Partial Halo CMEs}
\label{tab:cmes_201404}
\begin{tabular}{llcc}

\hline
\\
Time & Speed & Width/PA & Source \\
(UT) & (km~s$^{-1}$)  & (Degrees) & Location \\
\\ 
\hline
 2014/04/05 00:12            & 585     & 149/106       &  S25E134 \\
 2014/04/05 06:24            & 798     & 203/218       &  S13W167 \\
 2014/04/06 01:48            & 526     & 132/121       &  S25E121 \\
 2014/04/06 09:36            & 464     & 182/309       &  N20W145 \\

\hline
\end{tabular}
\end{table}

Assuming Sun to 1~AU propagation at the in situ ICME speed (350~km~s$^{-1}$), we infer a solar origin time of around 07~UT on 6 April 2014.  Several LASCO partial halo CMEs occurring around this time are listed in Table \ref{tab:cmes_201404}.  All originated on the far side as confirmed by EUVI observations from the STEREO spacecraft (the spacecraft configuration is shown in Figure~\ref{fig:stereoloc}(c)), and therefore were likely not associated with this geomagnetic storm.

\begin{figure}[ht]
\includegraphics[width=1\textwidth]{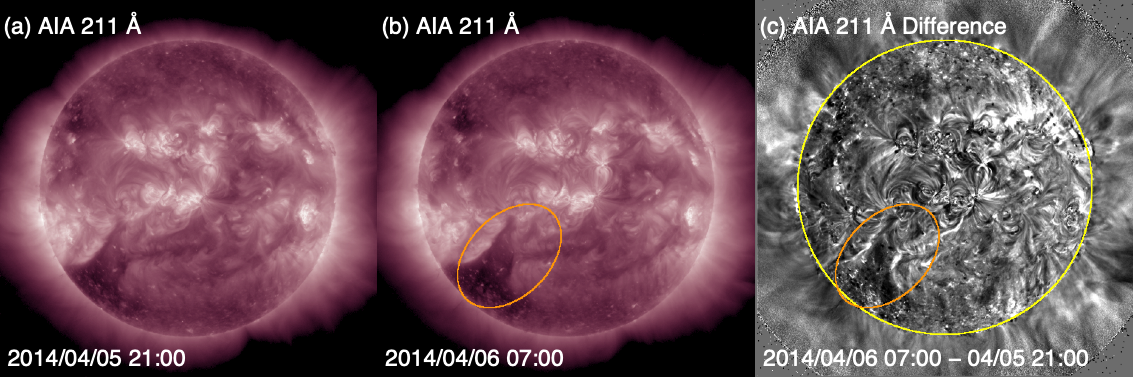}
\caption{Possible source region for the missing CME in Event 9. (a) and (b) AIA 211~\AA\ images before and after the assumed eruption.  (c) The difference image of (a) and (b).  A possible source region for the undetected CME is indicated by the ellipse in (b) and (c).   }
\label{fig:images4event19_1} 
\end{figure}

We could not identify any additional, unreported frontside CME that might have been associated with this ICME after examining movies of LASCO C2 images.  
Referring to Figure~\ref{fig:stereoloc}(c), observations from the STEREO spacecraft do not help much with identifying any potentially Earthward CME that is not evident in the LASCO observations. Both spacecraft were approaching superior conjunction of the far side of the Sun and hence had a similar line of sight as LASCO, making an unambiguous detection of any narrow and faint Earth-directed CMEs extremely difficult. 

We have however tried to find on-disc LCSs that might signal an Earth-directed CME. Experimenting with base-difference images in different time ranges, we initially identified a potential candidate source region for the CME. Figure~\ref{fig:images4event19_1} shows the results of this analysis using AIA 211~{\AA} images separated by 10~hours. The region indicated by the ellipse in the ``after'' image (Figure~\ref{fig:images4event19_1}(b)) indicates where the northernmost boundary of the southern coronal hole changes shape. The change in these boundaries is consistent with the bright ribbons revealed using base-difference image analysis (Figure~\ref{fig:images4event19_1}(c)). Although reminiscent of the LCSs for Event 12 presented in Section~\ref{sec:observations:event_24}, overall, the signatures are less convincing, especially since no frontside CME is found at southeastern position angles in LASCO during the rather wide time frame over which the coronal hole boundaries change shape. 
Such a CME would have been better observed by COR2 on STEREO-B than on STEREO-A, but there is no evidence of a CME in the STEREO-B observations.

\begin{figure}[!t]
\includegraphics[width=1\textwidth]{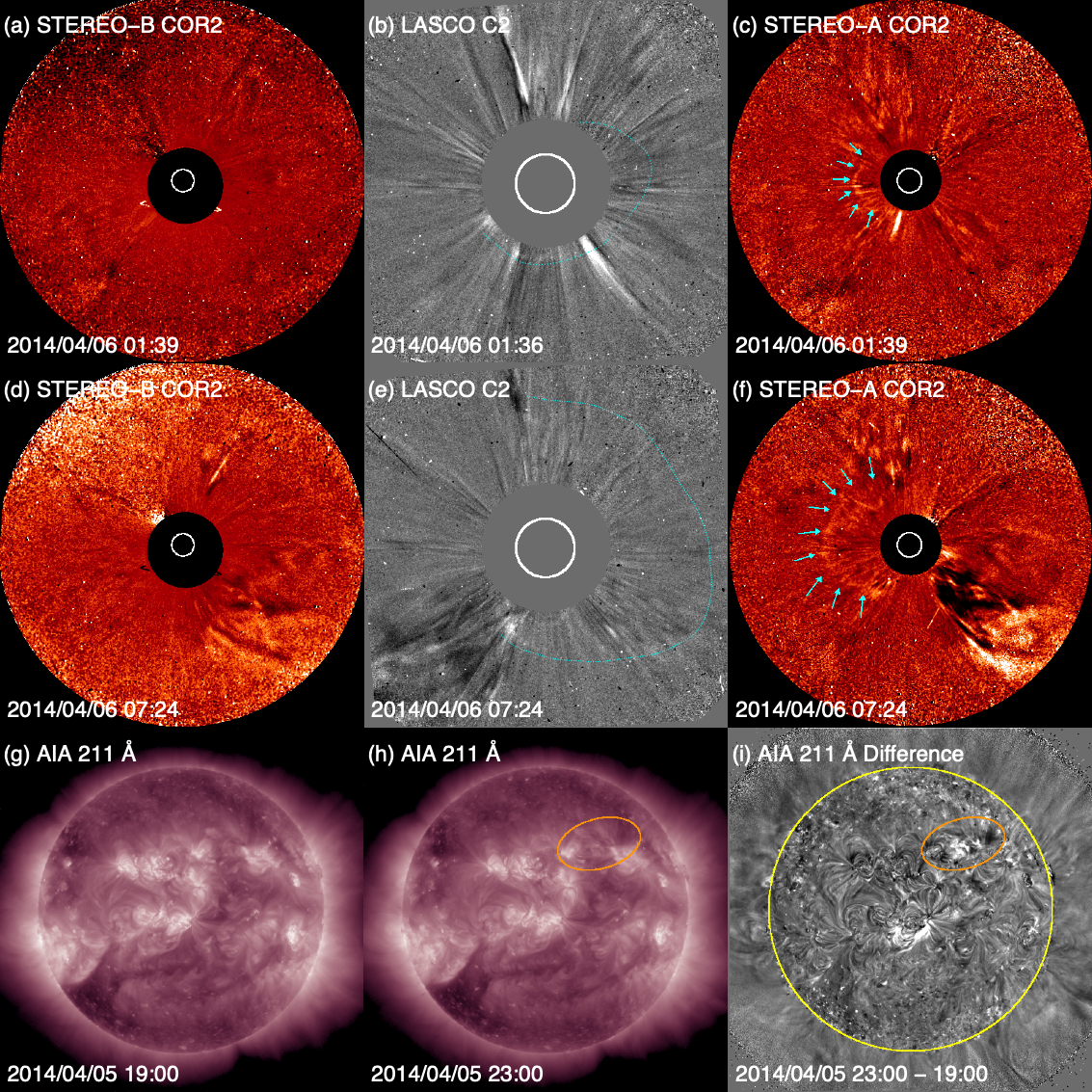}
\caption{Coronagraph (two upper rows) and low coronal (bottom row) observations for Event 9 in April 2014 where no Earth-directed CME was found with confidence.  (a) and (d) ((c) and (f)): COR2/STEREO-B (STEREO-A) difference images.  (b) and (e): LASCO C2 difference images.  (g) and (h) AIA 211~\AA\ images before and after the eruption.  (i) The difference image of (g) and (h).  The diffuse CME is pointed by arrows in cyan in (c) and (f), and by curves in cyan in (b) and (e).  A possible source region for this CME is indicated by an ellipse in brown in (h) and (i).   }
\label{fig:images4event19_2} 
\end{figure}

In a further effort to identify the CME associated with this storm, team members took a fresh look at coronagraph difference images using longer time separations than in the standard cadences. These attempts revealed a diffuse, semi-circular feature moving outward in COR2/STEREO-A images (indicated by the cyan arrows in Figures~\ref{fig:images4event19_2}(c) and (f)), although at a very low speed ($\approx$95~km~s$^{-1}$). None of the automated CME catalogues using COR2 data recognised this CME\footnote{CACTus: \url{http://sidc.oma.be/cactus/catalog.php}, SEEDS: \url{http://spaceweather.gmu.edu/seeds/secchi.php}, APL: \url{http://solar.jhuapl.edu/Data-Products/COR-CME-Catalog.php}}.  Moreover, this was barely observed by LASCO even in difference images with a longer separation, i.e., 36~minute, as opposed to the native 12~minute cadence; see Figures~\ref{fig:images4event19_2}(b) and (e), where a CME-like feature is traced out by the cyan curves. COR2 on STEREO-B (Figures~\ref{fig:images4event19_2}(a) and (d)) did not show anything. This leads us to expect that the source region was on the western hemisphere around 180$^\circ$ from the Sun--STEREO-B line.  We indeed note a minor eruption on the western hemisphere (Figures~\ref{fig:images4event19_2}(g)\,--\,(i)) hours before the diffuse CME-like feature appeared in images taken by COR2 on STEREO-A, but its onset time cannot be accurately determined because of a STEREO-A data gap between 5 April 17~UT and 6 April 00~UT. EUVI on STEREO-A detected no eruptive signatures. This indicates that the eruption responsible for the marginal CME would have been east of $\approx$W50, assuming that eruptions are not detectable if they occur $\gtrsim 15\arcdeg$ behind the east limb from STEREO-A.  However, the link between the eruption marked in Figures~\ref{fig:images4event19_2}(g)\,--\,(i)) and the CME may not be thoroughly convincing because, in view of the extremely slow speed in the COR2 field-of-view, the CME-like feature might have left the Sun much earlier than the eruption.

We suggest that the CME associated with the geomagnetic storm starting on 11 April 2014 should be classified as ``super stealthy''.  Due to the lack of a definite front-side source region, the CME-like feature in images from COR2 on STEREO-A may not be Earth-directed or linked to the ICME.  Assessment of this event was also not helped by the unfavourable spacecraft configuration. This event exemplifies one of the most extreme kinds of stealth CME and illustrates how challenging such events can be for the reliable prediction of geomagnetic storms. Events 10 and 11 (see Tables~\ref{tab:tab_1} and \ref{tab:tab_2}) may belong to an even more challenging category because, unlike Event 9, it was not possible  to identify any CME candidates, though the  unfavourable position of STEREO, or absence of STEREO observations due to the superior conjunction, respectively, are contributing factors.


\subsection{An Event Associated With a Large-scale CME: Event 14}
\label{sec:observations:event_30}

\begin{figure}
\includegraphics[width=0.99\textwidth,clip=]{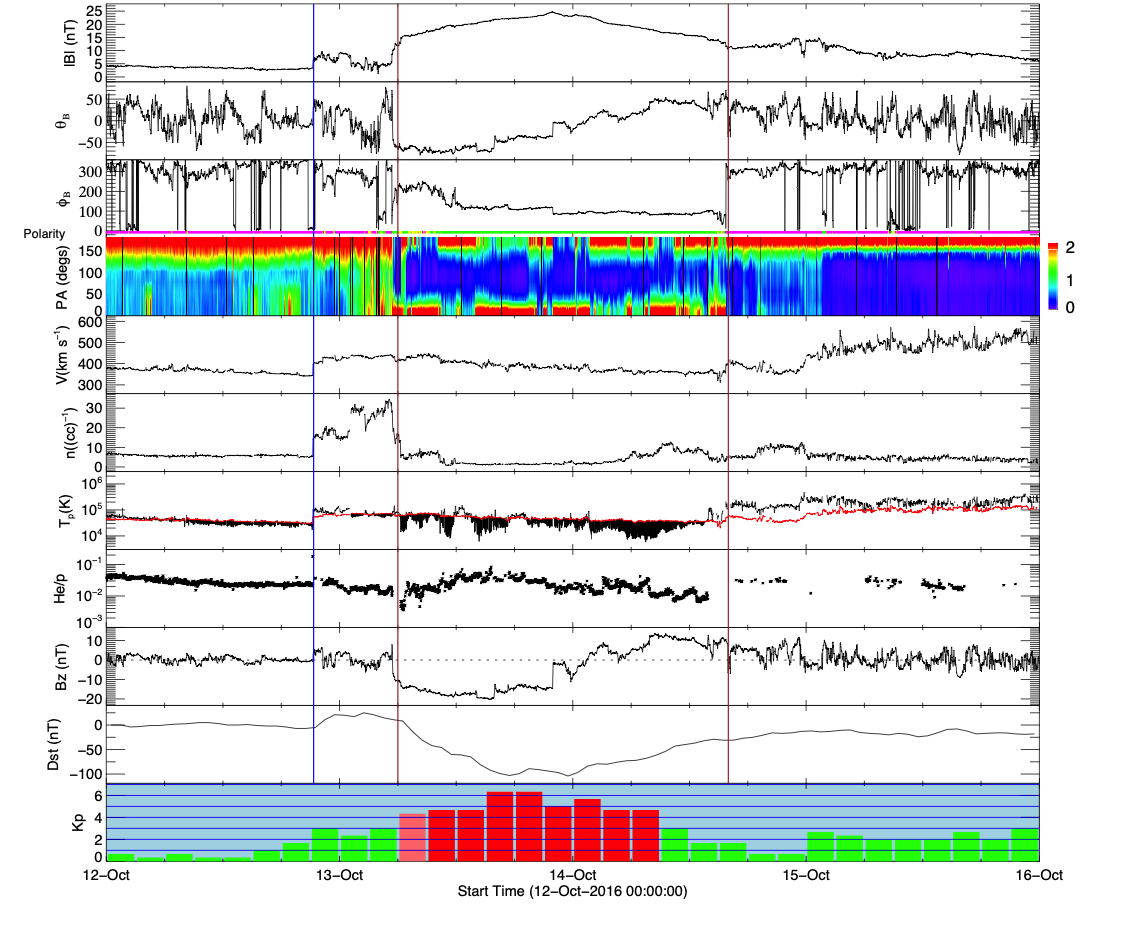}
\caption{In situ solar wind magnetic field and plasma observations at L1 and geomagnetic responses (last two rows) for Event 14 (storm of 13\,--\,14 October 2016) in the same format as Figure~\ref{fig:sw4textbook}. }
\label{fig:sw4event30} 
\end{figure}

In the next event, the associated CME appears to be of a much larger scale, and a small isolated filament eruption that occurred a few hours before the CME was first observed may not represent the CME source region. The possible LCSs are identified only after producing difference images with long time separations.  

A strong double-peaked geomagnetic storm occurred on 13 October 2016 with Dst minima of $-103$~nT and $-104$~nT at 17~UT and 23~UT, respectively, and a maximum Kp of 6+ during 15\,--\,21~UT (see the bottom two panels of Figure~\ref{fig:sw4event30}).  The associated near-Earth solar wind structures (Figure~\ref{fig:sw4event30}) consisted of a shock late on 12 October followed by a sheath region and a magnetic cloud in which the  magnetic field vector rotated smoothly for almost 1.5 days.  This resulted in an enhanced southward field in the first half of the MC that lasted for $\approx$15~hours and generated the storm. This particular ICME again is similar to the textbook ICME in Figure~\ref{fig:sw4textbook}. A difference is that it was followed by a high speed stream that may have enhanced the geoeffectiveness of the storm \citep{He2018}.

\begin{figure}
\includegraphics[width=1\textwidth]{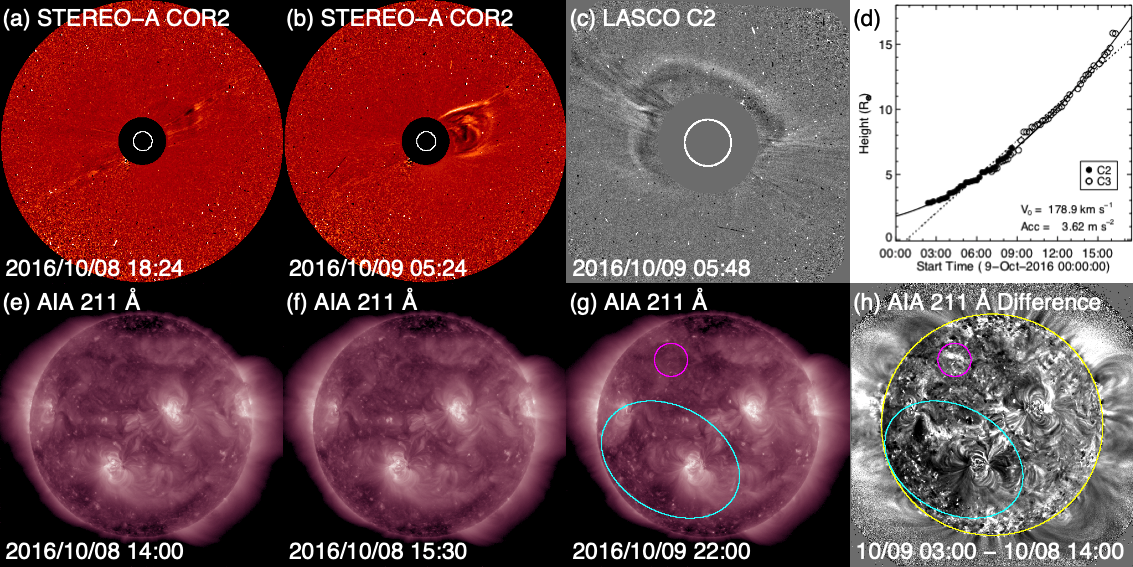}
\caption{Event 14: Coronagraph (upper row) and low coronal (lower row) information on the event in October 2016. (a) and (b) Difference images from COR2 on STEREO-A.  (c) LASCO C2 difference image. (d) Heliocentric height vs. time plot, from LASCO C2 and C3 data with linear (quadratic) fits shown as the dotted (solid) line. (e)\,--\,(g) AIA 211~\AA\ images before and after the eruption.  (h) The difference image of (f) and (g).  The possible source region for the CME is indicated by the cyan ellipse in (g) and (h). The location of the minor filament eruption seen in (f) is encircled in magenta in (g) and (h). }
\label{fig:images4event30} 
\end{figure}

Assuming a constant speed of 450~km~s$^{-1}$, based on the solar wind speed in the ICME observed by \textit{Wind}, the CME is expected to have left the Sun around 2~UT on 9~October 2016. There were two LASCO halo CMEs close to this time.  One was a partial halo at 08:36~UT on 10 October, but it was from the far side as revealed by EUVI on STEREO-A, and thus could not be the progenitor of the above ICME.  The other was a full halo CME, which was first detected at 02:24 UT on 9~October in the northeastern sector.  In a few hours, it surrounded a large part of the limb (see Figure~\ref{fig:images4event30}(c)). It was faint and very slow (179~km~s$^{-1}$) (Figure~\ref{fig:images4event30}(d)).  There were no obvious LCSs (including no soft X-ray enhancement in GOES data) but the side view offered by STEREO-A (Figure~\ref{fig:images4event30}(b)), located 148$\arcdeg$ east of the Sun--Earth line (Figure~\ref{fig:stereoloc}(d)), confirms that the CME was from the front side. 

As described by \cite{Nitta2017}, finding the on-disc LCSs for this CME is challenging. There was a minor filament eruption in the northern hemisphere during 15:00\,--\,16:00~UT on 8 October, as seen by comparing Figures~\ref{fig:images4event30}(e) and (f) with the area encircled in magenta in (g). But this activity seems too localised and too early to account for the extended CME.  Instead, it may have resulted in the blob-like outflow in Figure~\ref{fig:images4event30}(a) observed by STEREO-A. It could also have helped destabilize other nearby regions, causing them to erupt.  To search for more global changes, we tried various start and end times for base-difference images within the range between 10~UT on 8 October and 6~UT on 9 October. The most notable signature that almost always appears in these difference images is the dimming in the area within the cyan ellipses  in Figures~\ref{fig:images4event30}(g) and (h). Figure (h) also shows spurious artifacts typical of difference images with long temporal separations (in this case, 13 hours), such as the tendency of the western (eastern) hemisphere to become brighter (darker) due to solar rotation, combined with the variation of the area per pixel with distance from the disc center. Although this southern dimming region may be large enough to account for the extended CME, it is not clear how it was related to the body of the CME, which was directed more to the north rather than the south  (Figure~\ref{fig:images4event30}(b)).


\subsection{A Very Geoeffective Event: Event 16}
\label{sec:observations:event_34}

\begin{figure}
\includegraphics[width=0.99\textwidth]{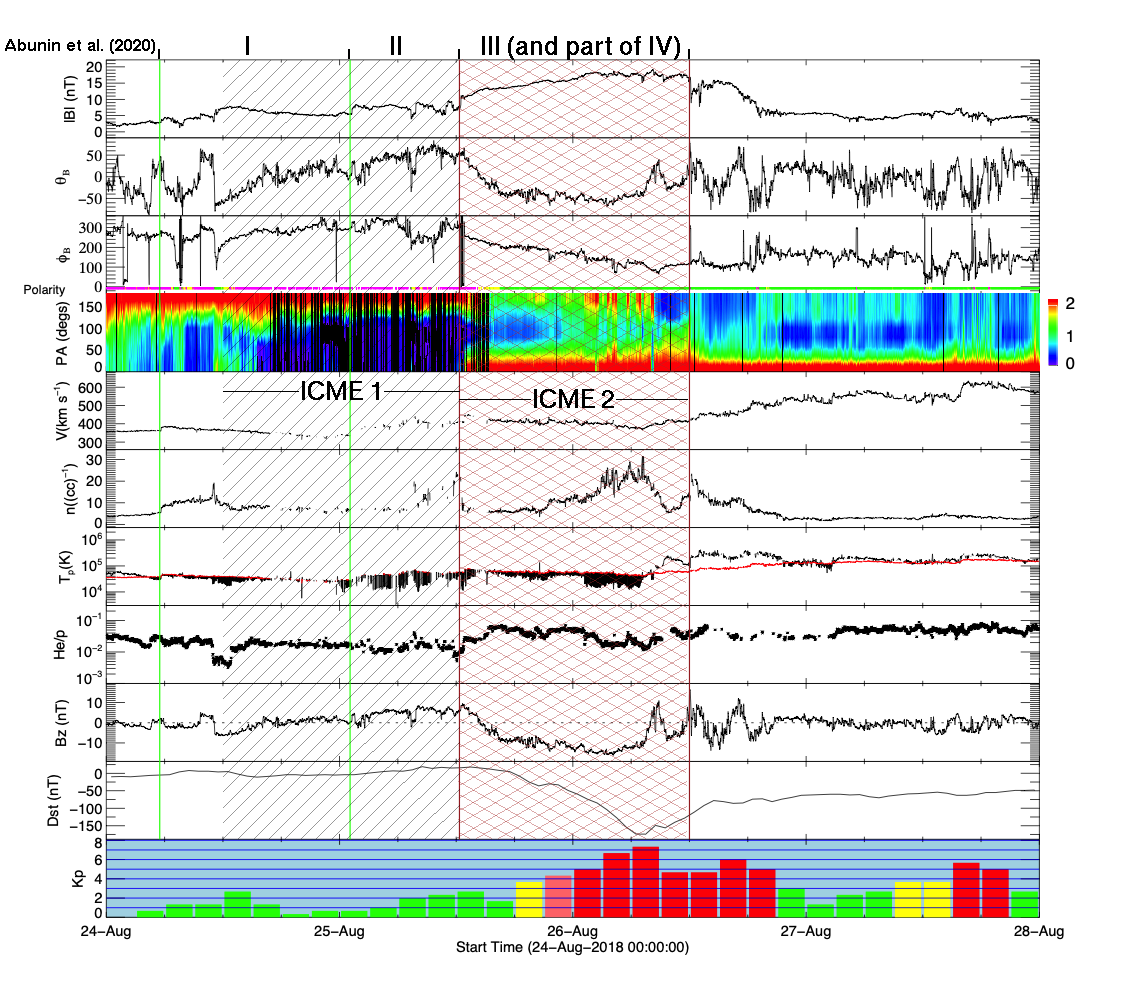}
\caption{In situ solar wind observations at L1 and geomagnetic response for Event 16 (storm of 26 August 2018) in the same format as Figure~\ref{fig:sw4textbook}. The regions (I to III) identified by \citet{Abunin2020} are indicated at the top of the figure (see the text for more details). Two ICMEs present are indicated by grey diagonal shading (ICME1) or brown cross hatching (ICME2; also bounded by vertical brown lines). Southward magnetic fields in the magnetic cloud of ICME2 generated the intense geomagnetic storm. The vertical green lines indicate probable weak shocks. }
\label{fig:sw4event34} 
\end{figure}

\begin{figure}
    \centering
    \includegraphics[width=0.99\textwidth]{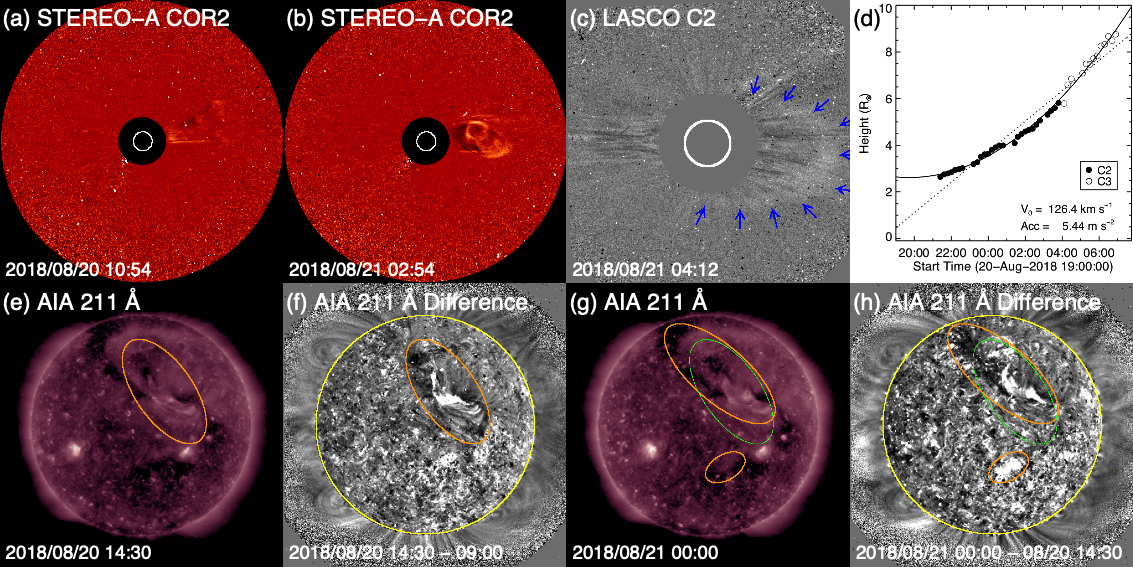}
    \caption{Event 16: Coronagraph (upper row) and low coronal (lower row) information for the two slow and diffuse CMEs on 20\,--\,21 August 2018  responsible for the solar wind structures in Figure~\ref{fig:sw4event34}. (a)\,--\,(b) COR2 (STEREO-A) difference images showing the two CMEs (c) LASCO C2 difference image showing the second CME. (d) Heliocentric height vs. time plot, from LASCO C2 and C3 data, for the second CME, with a linear (quadratic) fit shown as the dotted (solid) line.  (e) and (g) AIA 211~{\AA} images taken at two stages of the associated filament eruption.  (f) The difference image of (e) and an image taken 5.5 hours earlier. (h) The difference image of (g) and (e), taken 9.5 hours apart. The changes (dimmings and brightenings) likely associated with the CMEs as found in the difference images (f) and (h) are encircled in light brown, respectively, in (e)\,--\,(f) and (g)\,--\,(h). The area that contains changes in (e) and (f) is replicated in green in (g)\,--\,(h). }
    \label{fig:images4event34}
\end{figure}

This event is significant because it illustrates that even an intense geomagnetic storm can be attributed to a stealthy CME, providing a clear example of a ``problem'' geomagnetic storm. This storm, on 26~August 2018, reached a minimum Dst $=-174$~nT at 07~UT, with Kp = 7+ in the 6\,--\,9 UT interval, and was the third largest in solar cycle 24 in terms of the Dst index. The solar wind observations in Figure~\ref{fig:sw4event34} show that the storm occurred at a time when multiple structures were present upstream of a high-speed stream (HSS). We discuss these using the notation of \citet{Abunin2020}. 

The first transient, commencing at the first vertical green line and labelled ``region~I'' by \citet{Abunin2020}, leads off with a weak shock observed around 05:30~UT on 24 August 2018. After a brief sheath, this is followed by an ICME (ICME1), indicated by grey diagonal shading extending from around 12~UT on 24 August to 12~UT on 25 August (the first brown line).  ICME1 includes a modestly-enhanced magnetic field with a coherent rotation in direction suggestive of a flux-rope in a magnetic cloud. The depressed proton temperature is also consistent with an ICME, even though the suprathermal electron pitch angle distribution does not show bidirectional flows. The region between the second green and first brown lines,  labelled ``region~II" by \citet{Abunin2020}, commences with a weak disturbance, possibly a weak shock, at around 02~UT on 25 August that is propagating through ICME1. This region forms the sheath upstream of a slow ($\sim$370~km~s$^{-1}$) ICME (ICME2),  bounded by the brown lines and indicated by brown cross hatching, that includes ``region~III" and part of ``region~IV’’ of \citet{Abunin2020}. Bidirectional suprathermal electron flows are evident in ICME2, which also appears to be embedded in the heliospheric current sheet (HCS) as indicated by the reversal of the magnetic field azimuthal angle and suprathermal electron flow across the structure. The region of enhanced solar wind density ahead of ICME2 could be a signature of the HPS or alternatively result from ICME2 compressing the upstream solar wind. The extended period of enhanced southward magnetic field in ICME2 generates the geomagnetic storm.  This is a case where an ICME lies in a CIR formed ahead of an HSS that reached a bulk speed of 600~km~s$^{-1}$ around 16 UT on 27 August. The solar wind structures associated with this storm are similar to those of the textbook event in Figure~\ref{fig:sw4textbook}, but differences are also evident.  They include the lower speed of the magnetic cloud, the presence of an upstream transient (even though it does not appear to contribute to the generation of the storm), and the following HSS.

The solar wind speeds in the ICME structures suggest that ICME1 left the Sun at around 15~UT on 19~August 2018 and ICME2 erupted at around 20~UT on 20~August 2018.  The origin of this event has been analysed by \citet{Chen2019}, \citet{SK_Mishra2019}, \citet{Abunin2020}, and \citet{Piersanti2020}. AIA images during 18\,--\,19 August show an extended area of the quiet Sun in the northern hemisphere adjacent to coronal holes to the south and northeast.  This quiet Sun area contains a long filament channel that hosts a dark filament best seen in the AIA 193~{\AA} and 211~{\AA} channels.  The filament becomes darker on 19 August, and appears to split or disintegrate, and partly disappears early on 20 August.  These morphological changes most likely signify an eruption as both brightenings and dimmings form close to the filament channel  (indicated by the ellipse in Figure~\ref{fig:images4event34}(e)).  The entirety of the filament erupted in two stages over the course of a day.  The first stage, starting around 04~UT,  does not show an obvious eruption, but evidence of an eruption is provided by changes (brightenings and dimmings, more readily visible in difference images) primarily in the southern part of the filament channel, but also in the northern part, as indicated by the ellipses in Figures~\ref{fig:images4event34}(e) and (f)).  In the second stage, slightly different but largely overlapping areas, notably extending further northeast and south (light brown ellipses in Figure~\ref{fig:images4event34}(g) and (h)), show similar changes.  This stage includes a smaller-scale eruption in the northern part of the filament channel around 18:30 UT, identified with a ``jet’’ by \citet{SK_Mishra2019}.  As a result of the slow eruption involving the long filament channel, a slow partial halo CME is first observed at 21:24~UT by LASCO (Figure~\ref{fig:images4event34}(c) showing a later frame).  According to the CDAW catalog, which labels the CME as a ``poor’’ event, the angular width is 120$\arcdeg$, and the plane-of-the-sky speed only 126~km~s$^{-1}$ (Figure~\ref{fig:images4event34}(d)). The CME is so diffuse that difference images with long temporal separations are needed to isolate the CME front as indicated by blue arrows in Figure~\ref{fig:images4event34}(c).

At this time, STEREO-A was in near-quadrature with Earth at 108$\arcdeg$ East (Figure~\ref{fig:stereoloc}(e)), providing a better view of Earth-directed CMEs. Although LASCO observed only one CME (Figure~\ref{fig:images4event34}(c)) from the eruption on 20 August, STEREO-A COR2 revealed another CME.  This faint, narrow, Earth-directed CME (Figure~\ref{fig:images4event34}(a)) was first observed at 10:24 UT,   with a speed of 328~km~s$^{-1}$ in the CACTus CME catalog\footnote{\url{https://secchi.nrl.navy.mil/cactus/}}.  This temporally matches the first stage of the filament eruption. Since the LCSs shown in Figure~\ref{fig:images4event34}(f) appear rather evident, this is not a stealth CME.  A second, much larger and more dense, but also slow (181~km~s$^{-1}$), CME was seen in COR2 imagery from 18:54~UT (Figure~\ref{fig:images4event34}(b)). This is a side view of the diffuse partial CME observed by LASCO.  However, its origin may be problematic. Although \citet{Abunin2020} associated this CME with the smaller-scale eruption, the CME observed by COR2 may not match the spatial extent and timing (too early) of the smaller-scale eruption.  It is possible that the slow eruption of the filament channel over a day contributed to the CME, but we cannot pin down the time of liftoff, so in this sense, this CME is a stealthy event.  It is reasonable to postulate that the two CMEs observed by COR2 correspond to the two ICMEs at L1, as proposed by \citet{Abunin2020}.  As already noted, the geomagnetic storm was mostly due to the strong southward field embedded in ICME2.  In addition,  the following HSS likely inhibited expansion of the ejecta (see the mostly flat speed profile between the vertical brown lines in Figure~\ref{fig:sw4event34}), thus enhancing its geoeffectiveness. The origin of the HSS was the coronal hole northeast of the quiet-Sun region containing the filament channel, and connection of the Earth to this positive-polarity coronal hole was first established on this solar rotation.

Contrasting the coronagraph images in Figure~\ref{fig:images4event34} with those of the textbook event in Figure~\ref{fig:images4textbook}, it is clear that the partial halo CME observed in LASCO data is much weaker than that associated with the textbook event. However, since this and the preceding CME are clearly evident from the viewpoint of STEREO-A COR2, these instruments are critical in elucidating the multi-step nature of the event. Thus, this event again illustrates how coronagraph observations made away from the Sun--Earth line (e.g., from the L4 and L5 Lagrange points or in quadrature with Earth) can greatly increase the chance of detecting Earth-directed CMEs at times when observations using coronagraphs at L1 show extremely weak or no signatures. 



\subsection{Example of a Possible ``Masquerading'' Event: Event 8}
\label{sec:observations:event_15}

\begin{figure}
\includegraphics[width=0.99\textwidth,clip=]{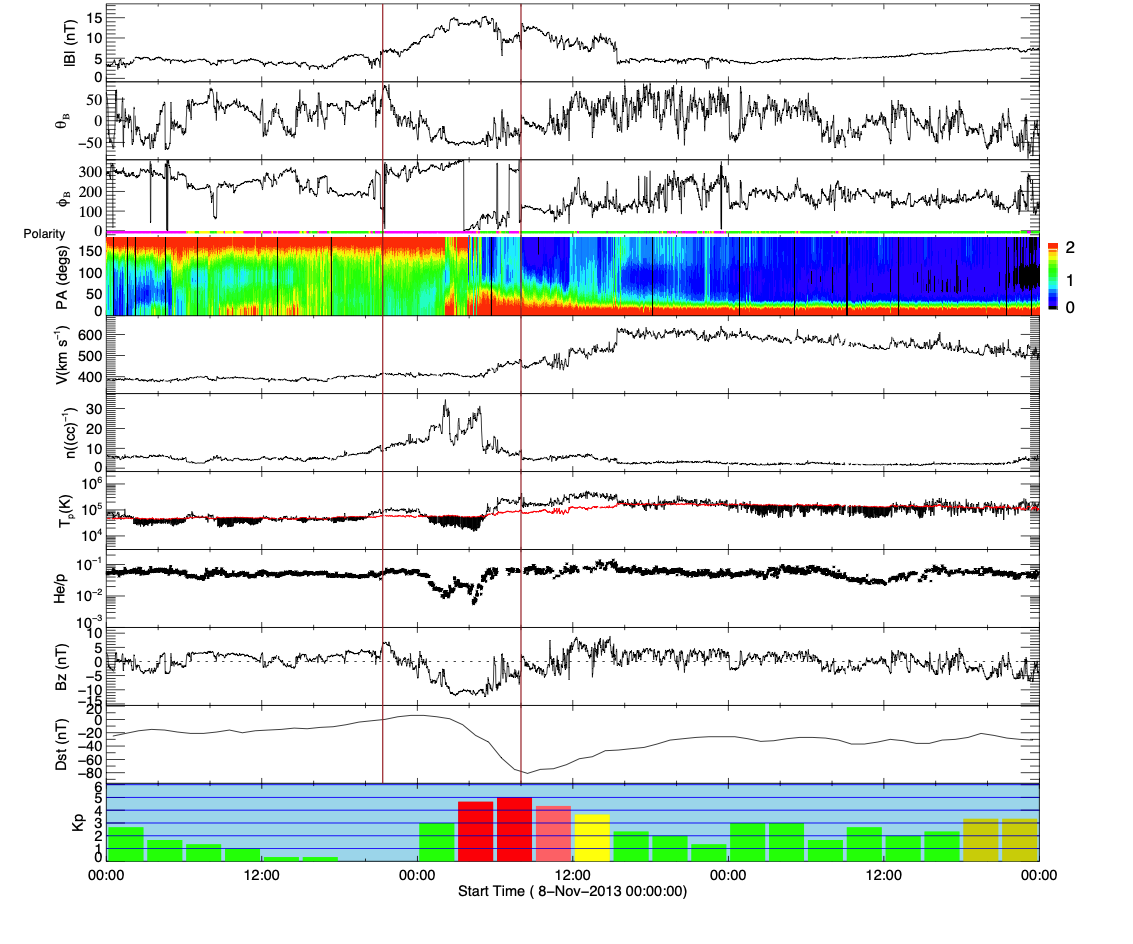}
\caption{In situ solar wind magnetic field and plasma observations at L1 and geomagnetic responses (last two rows) for Event 8 (storm of 9 November 2013) in the same format as Figure~\ref{fig:sw4textbook} except that there was no shock. }
\label{fig:sw4event15} 
\end{figure}

Our final event  illustrates another manifestation of problem geomagnetic storms. In this case, the problem is in the interpretation of the in situ solar wind observations. In particular, the question is whether or not they show an ICME or other transient embedded in a CIR ahead of a high-speed stream, as in Event 16, or if the ICME-like features are in fact associated with the CIR, and no ICME is present. 

The observations during 8\,--\,11 November 2013 in Figure~\ref{fig:sw4event15} indicate that the storm on 9 November 2013 was associated with southward magnetic fields in the solar wind region between the vertical brown lines.  This region is identified as an ICME in the Richardson \& Cane catalogue and hence is included in this study. Evidence that this region may be an ICME includes an enhanced, rotating magnetic field and depressed proton temperatures. Significant differences from the textbook event (Figure~\ref{fig:sw4textbook}) include the shorter duration of the ICME and that it is embedded in an CIR. The CIR is evidenced by the enhanced magnetic field intensities and densities ahead of a high speed stream that commences early on 9 November and extends to the end of the interval shown \citep[see][for a review of CIRs]{Richardson2018}. Thus, the ICME seems to be swept up within the slow solar wind ahead of the HSS. It also encompasses a reversal of the IMF polarity which is consistent with the change in the peak of suprathermal electron pitch angle distribution from $180^\circ$ to $0^\circ$ near the middle of this region. 

However, aspects of these in situ observations raise the question as to whether the identified region is in fact an ICME or alternatively associated with the HPS, which envelopes the HCS \citep{Winterhalter1994, Bavassano1997, Crooker2004} and is typically also characterized by high-density and low-temperature plasma. The observed depressed He/p ratio in the dense, low temperature plasma is another characteristic signature of the HPS \citep{Gosling1981, Borrini1981,Bavassano1997}, whereas ICMEs tend to be associated with an enhanced He/p ratio \citep{Hirshberg1972,Borrini1982}. The absence of bidirectional suprathermal electrons is also suggestive of the absence of an ICME in this region.  It has been proposed however, that transient structures may be present in the HPS that originate as ``blobs" formed by the pinching off of the tips of helmet streamers (e.g., \citet{Crooker2004}), so the ICME-like signatures here could be evidence of a similar structure. Indeed, \citet{Yu2018} classified the solar wind structure at 00:12\,--\,06:27~UT on 9 November as a small flux rope similar to the blobs that have been widely investigated \citep[e.g.][]{WangYM2000, Cartwright2008, Sheeley2009, Rouillard2010a,Rouillard2010b, Rouillard2011, Higginson2018}. A similar structure embedded in a CIR and interpreted as an ``ICME-like transient" is illustrated in Figure~2 of \citet{Kilpua2012}.

Despite the possibility that an ``ICME-like" structure is embedded within this CIR, the presence of an ICME is not required for a CIR to be geoeffective.  As noted in Section~\ref{sec:intro}, CIRs/HSSs alone can drive intense geomagnetic storms, such as 13\% of the intense storms discussed by \citet{Zhang2007}, while \citet{Grandin2019} included the event shown in Figure~\ref{fig:sw4event15} among their set of geomagnetic storms driven by CIRs and HSSs. In particular, the interval of persistent southward field that gives rise to the geomagnetic storm (with a minimum Dst = $-80$~nT that meets the criterion for inclusion in this study) could give the impression that a coherent magnetic field structure such as an ICME-associated flux rope is present, but this could just be an intrinsic feature of this CIR that is ``masquerading" as an ICME. Although this structure is identified as an ICME in the Richardson \& Cane catalogue, the quality of the ICME signatures is  assessed to be relatively weak (``quality"=3), suggesting some uncertainty in this identification.  

\begin{figure}
\includegraphics[width=1\textwidth]{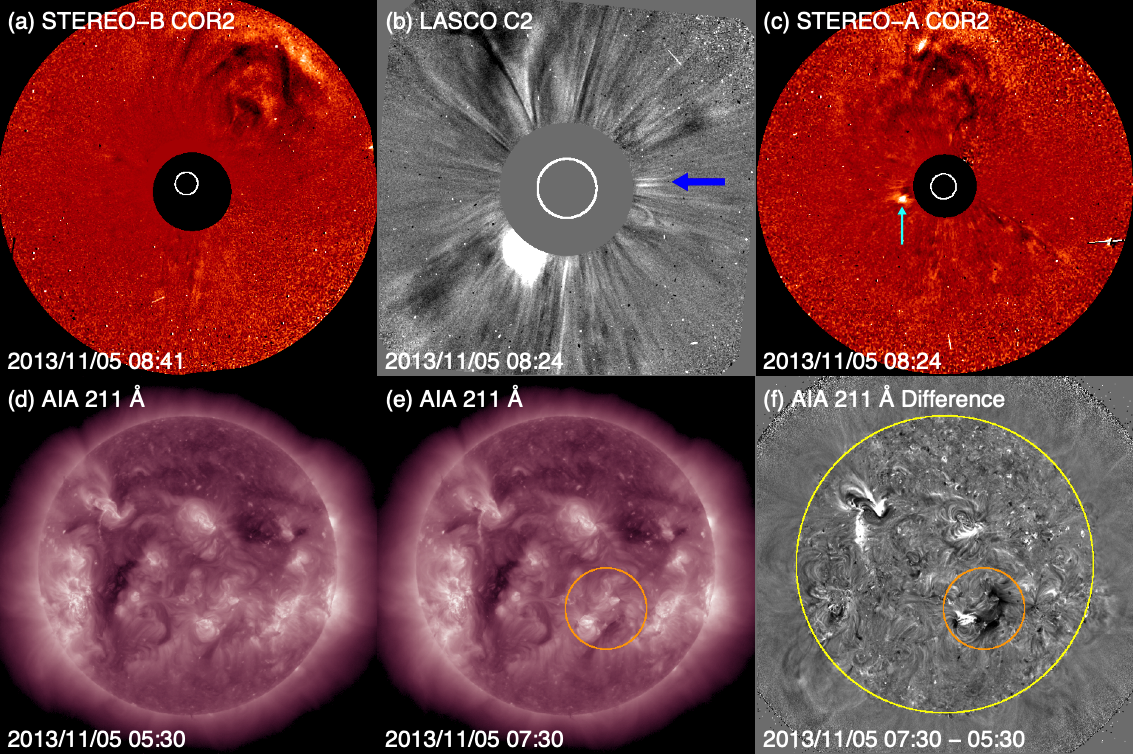}
\caption{Event 8: Coronagraph (upper row) and low coronal (lower row) information related to the event in November 2013. In the top row,  (a) and (c) are STEREO-B and -A COR2 difference images and  (b) is a LASCO C2 difference image. The arrows in (b) and (c) indicate a small CME that may be responsible for the ICME on 9~November. (d) and (e) AIA 211~\AA\ images before and after the eruption associated with this CME.  (f) The difference image formed from the images in (d) and (e) separated by 2 hours.  The possible source region for the CME in (b) and (c) is circled in (e) and (f).}
\label{fig:images4event15} 
\end{figure}

\begin{table}
\caption{Event 8: List of Partial/Full Halo CMEs}
\label{tab:cmes_201311}
\begin{tabular}{llcc}
\hline
\\
Time & Speed & Width/PA & Source \\
(UT) & (km~s$^{-1}$)  & (Degrees) & Location \\
\\ 
\hline
 2013/11/04 05:12            & 1040     & 360/67       &  N04W166 \\
 2013/11/05 02:48            & 512      & 126/24       &  N28E42 \\
 2013/11/05 08:24            & 850      & 177/136      &  S16E52 \\
 2013/11/05 22:36            & 562      & 195/160      &  S12E44 \\
\hline
\end{tabular}
\end{table}

Examining the solar observations for evidence of a CME that could account for an ICME at Earth may help decide between these possibilities. Extrapolating back to the Sun using the in situ speed in the ICME-like region suggests a time of origin at the Sun of around 4~UT on 5~November~2013. Halo CMEs (full and partial) during this period are listed in Table~\ref{tab:cmes_201311}. However, none of these CMEs are candidates. In the case of the halo CME, it was from the far side, based on STEREO observations (the spacecraft configuration is shown in Figure~\ref{fig:stereoloc}(b)). In the case of the partial halo CMEs, they were headed to the north or south (notwithstanding their modest source latitudes), appearing to move away from the ecliptic plane, which makes them unlikely candidates to be associated with an ICME at Earth. 

There are also some narrow and slow outflows evident in both LASCO C2 and STEREO COR2 data to consider. An example is shown in Figures~\ref{fig:images4event15}~(b)\,--\,(c), where a minor feature is indicated by blue arrows on C2 and COR2-A images. This is first seen in COR2-A around 07:54 UT and appears to move out following the pre-existing streamers; it is unrelated to the later partial halo CME, first observed at 08:24 UT, that moves due south in the COR2-A FOV.  It is also absent in the corresponding COR2-B image (Figure~\ref{fig:images4event15}~(a)).  It is possible that this feature may be formed from the pinching off of a helmet streamer.  However, it is hard to find the LCSs for this tiny CME or outflow.  We only tentatively find a minor eruption and accompanying dimmings around the time of this CME in AIA observations (Figures~\ref{fig:images4event15}~(d)\,--\,(f)).  Although the location appears favourable for an association with the ICME at Earth, it is not possible to demonstrate that this is the correct association. 

The identification of a possible solar source for this particular geomagnetic storm remains problematic for several reasons.  First, the nature of the interplanetary driver is not completely clear.  It remains a question of whether the geoeffective region is due to an ICME or other blob-like structure associated with the HPS embedded in a CIR, or if the geoeffective region is an inherent feature of the CIR. Second, there are partial halo and halo CMEs in a suitable time frame, but they do not appear to be Earth-directed.  Third, a tiny CME or outflow in coronagraph images with apparently frontside LCSs is identified and might be linked to the ICME-like structure, but it is not possible to demonstrate this association convincingly. 

Ultimately, cases such as this event demonstrate that there are limits to our ability to resolve stealth CMEs. Although the goal of this review is to clarify and help resolve the issue of problem geomagnetic storms, nevertheless, it remains easier to define a problem geomagnetic storm than it is to define a stealth CME. One crucial factor required in order to begin the process of identifying the existence of a stealth CME is that an unambiguous in situ signature must exist. If this is not the case, as in this ``masquerading" event, the origin of the problem geomagnetic storm caused by the geoeffective structures in the solar wind will remain elusive.


\section{Modelling the Stealth CME Drivers of Problem Geomagnetic Storms}
\label{sec:modelling}

We now turn to summarising recent progress in some of the approaches to numerical modelling of the stealth CME drivers of problem geomagnetic storms, and discuss some of the challenges that arise with these particular events.
In section~\ref{sec:modelling:mf}, we discuss magnetofrictional (MF) modelling of the global coronal magnetic field during a time period encompassing Event 12 (see section~\ref{sec:observations:event_24}). The MF evolution shows evidence for a loss-of-equilibrium and/or destabilization of the energised field structure in precisely the location of Event 12's dimming signatures adjacent to the southern coronal hole.  
In section~\ref{sec:modelling:mfmhd}, we continue by discussing some of the challenges in coupling the energised pre-eruption states obtained via global MF simulations with a full magnetohydrodynamics (MHD) treatment of the eruption process. The inclusion of plasma in the MHD treatment allows more accurate modelling of the eruption dynamics, energetics, and magnetic reconnection processes over the eruption timescales.   
In section~\ref{sec:modelling:blobs}, we discuss smaller-scale reconnection-generated flux rope transients originating high in the corona that, under certain circumstances, may result in interplanetary signatures similar to, or indistinguishable from, stealth CMEs. 
In section~\ref{sec:modelling:overview}, we summarise the \citet{Lynch2016b} MHD model of the \citet{Robbrecht2009} stealth CME event (see Figure~\ref{fig:robbrecht}) and discuss some of the physical processes governing the storage-and-release of magnetic free energy and their relationship to potential on-disc eruption signatures. 
Finally, in section~\ref{sec:modelling:lowcoronasignatures}, we discuss the synthetic observational signatures in the low corona that can be derived from MHD simulation data and the difficulties associated with attempting to compare modelling results to the naturally ambiguous on-disc signatures of stealth CME events.

\subsection{Magnetofrictional (MF) Modelling of Stealth CME Source Regions: Event 12}
\label{sec:modelling:mf}

MF modelling \citep{Yang1986, Craig1986} assumes the plasma velocity in the MHD induction equation to be proportional to the local Lorentz force. This assumption leads to a relaxation of a magnetic configuration toward a force-free state while preserving magnetic topology, except in places where current sheets form. The code solves the following form of the induction equation and auxiliary conditions:
\begin{equation}
\frac{\partial \mathbf{A}}{\partial t} = \mathbf{v} \times \mathbf{B} - \eta \mathbf{J} \; ; \;\;\;\;\; 
\mathbf{B} = \nabla \times \mathbf{A} \; ; \;\;\;\;\; 
\mathbf{J} = \nabla \times \mathbf{B} \; ; \;\;\;\;\;
\mathbf{v} = \frac{\mathbf{J} \times \mathbf{B}}{\nu}
\end{equation}
where $\mathbf{A}$ is the vector potential, $\mathbf{B}$ is the magnetic field, $\mathbf{J}$ is the electric current density, $\mathbf{v}$ is the velocity, $\nu$ is the magnetofrictional coefficient, and $\eta$ is the magnetic diffusivity. The MF approach is much more computationally efficient than MHD as the magnetic field is evolved quasi-statically.  However, full MHD simulations are required to capture the rapid evolution associated with unstable configurations.

MF modelling captures the development and persistence (``coronal memory'') of the large-scale currents. This long-term simulation capability is crucial for the accurate modelling of the coronal magnetic structure of eruptive filaments and their environments, given the long lifetimes of filament fields. 
To continuously drive the simulation, a surface flux transport model is applied to the lower photospheric boundary conditions \citep{Sheeley2005}, which consist of synoptic magnetograms. Surface flux transport includes the effects of differential rotation \citep{Snodgrass1983}, meridional flow \citep{Duvall1979}, surface diffusion and magnetic flux emergence. These effects, apart from the emergence of magnetic flux that is almost instantaneous, operate continuously over relatively long timescales, building up magnetic stress and energy in the simulated coronal magnetic field.
MF simulations performed over timescales of months and years to model filament channel formation processes and their resulting magnetic helicities compare favourably with observations \citep{Mackay2012, Yeates2014}.
In particular, MF models are good at generating sheared and twisted field over high-latitude filament channels \citep[e.g.][]{vanBallegooijen2004,yeates2012,Su2015,Jibben2016,Yardley2019} and also at reproducing the observed hemispheric pattern of filament chirality \citep{Mackay2018,Yardley2021a}.
Recently, \citet{Yeates2018} and \citet{Yardley2021b} have performed continuous MF simulations of the global photospheric and coronal magnetic field over 200 days from 1 September 2014 -- 20 March 2015 for the 2015 solar eclipse. This simulation period includes Events 11 and 12 in Tables~\ref{tab:tab_1} and \ref{tab:tab_2}. 

\begin{sidewaysfigure}
\centering
\vskip 4.80in
\includegraphics[width=0.82\textwidth]{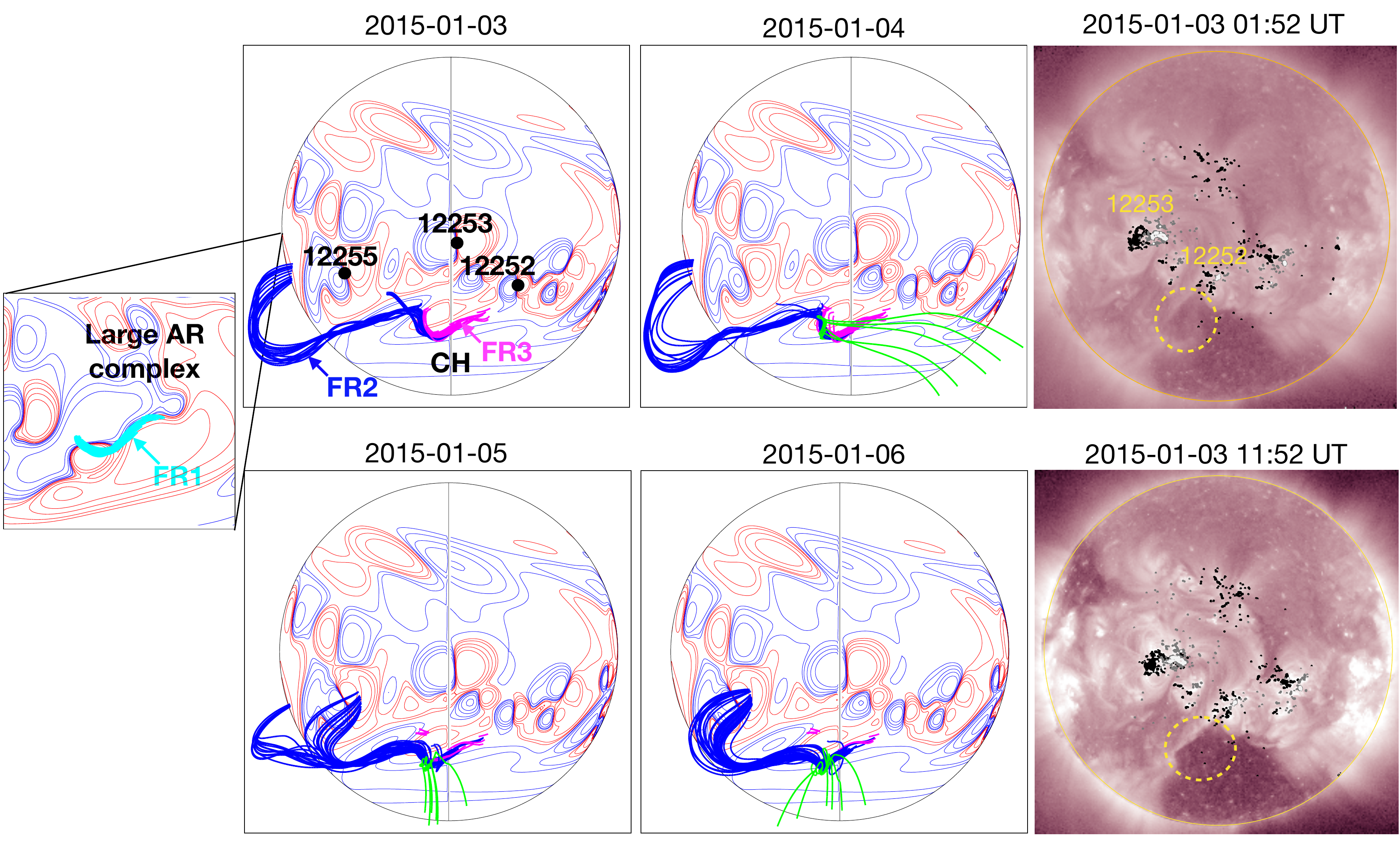}
\caption{(Left) Snapshots from the \citet{Yardley2021b} MF simulation of Event 12. The red (blue) contours represent positive (negative) radial magnetic field. The snapshots are plotted in the co-rotating Carrington frame.
Magnetic field lines corresponding to three flux ropes are shown in light blue (FR1), dark blue (FR2), and magenta (FR3). FR1 (shown in the cutout from the far side of the Sun) lies along the polarity inversion line of a large AR complex located behind the east limb. In the few days around the observed stealth eruption, reconnection occurs between FR2 and FR3 resulting in open magnetic field (green lines). The AR numbers and the coronal hole (CH) are labelled in black. (Right) The SDO/AIA 211~{\AA} observations during the eruption (top) and post-eruption (bottom). The white (black) contours represent the positive (negative) photospheric magnetic field from an SDO/HMI line-of-sight magnetogram. ARs are labelled in yellow and the dashed yellow circle shows the eruption-related dimming region.}
\label{fig:mf_fig} 
\end{sidewaysfigure}

Figure~\ref{fig:mf_fig} shows an example of the magnetic field evolution of the global corona during the time of Event 12 in our survey \citep[see][for an in-depth analysis of this event]{Yardley2021b}. The left panels show the MF model output from 3\,--\,6 January 2015, where red (blue) contours represent positive (negative) photospheric magnetic field. The light blue, dark blue, magenta and green lines are representative magnetic field lines taken from the MF model. The right column shows the SDO/AIA 211~{\AA} observations overlaid with the SDO/HMI line-of-sight magnetic field during the stealth eruption (top, 01:52~UT) and post-eruption (bottom, 11:52~UT). The only LCS observed that was associated with the eruption was a faint coronal dimming (Figure~\ref{fig:images4event24}(d)\,--\,(f)), which is most evident in difference images taken in the 211~{\AA} channel that have a long separation in temporal cadence \citep{Nitta2017}. During the eruption, the coronal dimming adjacent to the eastern periphery of AR~12252, grows and merges with the northeastern section of the extended polar coronal hole in the southern hemisphere. 

In the global model, there are three flux ropes present before the time of the observed stealth eruption: the light blue, dark blue and magenta field lines labelled as FR1, FR2, and FR3 in Figure~\ref{fig:mf_fig}, respectively. In the MF simulation all three flux ropes show signs of destabilisation around the time of the stealth eruption. FR1 lies along the polarity inversion line of a large AR complex that is located behind the East limb. FR2 has footpoints rooted in the positive polarity of this AR complex (behind the limb) and adjacent to FR3 on the disc. FR3 is located between the north east boundary of the southern polar coronal hole and the eastern periphery of AR 12252. The third flux rope (FR3) is situated at the same location that the stealth eruption-related dimming is observed.

During the few days leading up to and during the eruption, reconnection occurs between FR2 and FR3 causing the structures to merge together to form a new flux rope (as seen in Figure~\ref{fig:mf_fig}). Simultaneously, external reconnection is taking place above the newly formed flux rope, which opens up the overlying magnetic field (green field lines), resulting in the flux rope to rise in the domain. The rising motion of the flux rope along with the opening of the overlying magnetic field suggests that a loss of equilibrium has occurred in the simulation. Having successfully identified a likely source region for Event~12, and modelled the energisation of the stealth CME, the full dynamics of the eruption may be simulated by coupling the MF model with a full MHD simulation \citep[e.g.][]{Rodkin2017}, as described in the next section.

\subsection{Demonstration of MF\,--\,MHD Coupling: Event 12}
\label{sec:modelling:mfmhd}

The motivation for coupling MF models and MHD simulations is the need to understand the full life-span of eruptive structures, from their formation in the solar corona through their eruption into interplanetary space. 
The equations of ideal MHD correspond to the conservation of mass, momentum, and energy as well as the induction equation for evolution of the magnetic field:
\begin{equation}
	\frac{\partial \rho}{\partial t} + \nabla \cdot \left( \rho\mathbf{v} \right) = 0 \; ,
\end{equation}
\begin{equation}
	\frac{\partial \rho\mathbf{v}}{\partial t} + \nabla \cdot \left( \rho\mathbf{v}\mathbf{v} \right) = - \nabla p - \rho \mathbf{g} + \frac{\left( \nabla \times \mathbf{B} \right) \times \mathbf{B}}{4\pi} \; ,
\end{equation}
\begin{equation}
	\frac{\partial e}{\partial t} + \nabla \cdot \left( (e+p)\mathbf{v} \right) = 0 \; ,
\end{equation}
\begin{equation}
	\frac{\partial \mathbf{B}}{\partial t} - \nabla \times \left( \mathbf{v} \times \mathbf{B} \right) = 0 \; .
\end{equation}

\noindent Here, all the variables retain their usual meaning: mass density $\rho$, velocity $\mathbf{v}$, magnetic field $\mathbf{B}$, gas pressure $p$, and the total energy density $e$. The solar gravitational acceleration is given by $\mathbf{g} = - G M_{\odot} / ( r^2 ) \, \mathbf{\hat{r}}$ where $G$ is the gravitational constant and $M_{\odot}$ is the mass of the Sun. The total energy $e$, is then given by
\begin{equation}
e = \frac{p}{\gamma-1} + \frac{1}{2} \rho v^2 + \frac{B^2}{8\pi} \;, 
\end{equation} 

\noindent where $\gamma$ is the ratio of specific heats and the system is closed with the ideal equation of state for a hydrogen plasma, $p = (n_p + n_e) k_B T$.

The procedure for coupling MF model outputs to a full MHD model is, unfortunately, not as simple as just adopting the energised magnetic field configuration from the MF model as an initial condition to the MHD code. One of the necessary additional steps in this coupling is to construct a solar atmosphere around the imported magnetic field skeleton in order to complete the set of MHD variables. Since there are a number of issues that arise, fundamentally from the different physical systems being modeled, the technical aspects of MF\,--\,MHD model coupling remains an active area of research \citep[e.g.,][]{Pagano2018, Hoeksema2020}.

Implementation of MF\,--\,MHD model coupling can successfully describe the runaway of energised magnetic structures from equilibrium as the magnetic configurations conserve their stability properties when imported into full MHD.
%
%
At the same time however, the fate of a CME onset can be crucially sensitive to the background atmosphere. \citet{Pagano2013b} have shown that an ongoing MF eruption can, in MHD, either be quenched or continue to the outer corona, depending on the ambient plasma conditions.
%
%

For this review, we have applied the \citet{Pagano2018} MF\,--\,MHD coupling approach to study Event 12, where we couple the pre-eruption magnetic configuration obtained in the MF simulation on the same day as the stealth eruption to an MHD simulation of the subsequent eruption. 
A more detailed description can be found in \citet{Yardley2021b}; here we summarise the key steps.
We use the PLUTO code \citep{Mignone2012} and import the magnetic configuration from the MF model as the initial condition for the MHD simulation. Next, a radial hydrostatic equilibrium with a power-law density profile is constructed and perturbed with additional cool material in regions of energized magnetic fields.
The result is a heterogeneous solar corona where cold and dense flux ropes are merged with a hotter and tenuous solar corona environment. This final configuration is not in hydrostatic equilibrium, but the time scales of the plasma displacements induced by this force unbalance are much longer than those related to the Lorentz forces present in the domain that immediately lead to evolution. As this magnetic configuration and the associated reconstructed atmosphere evolve in the MHD simulation, the solar corona produces eruptions from where unstable magnetic structures are present. 

\begin{figure}
    \includegraphics[width=0.50\linewidth,trim=0.8in 0.25in 1.35in 0in,clip]{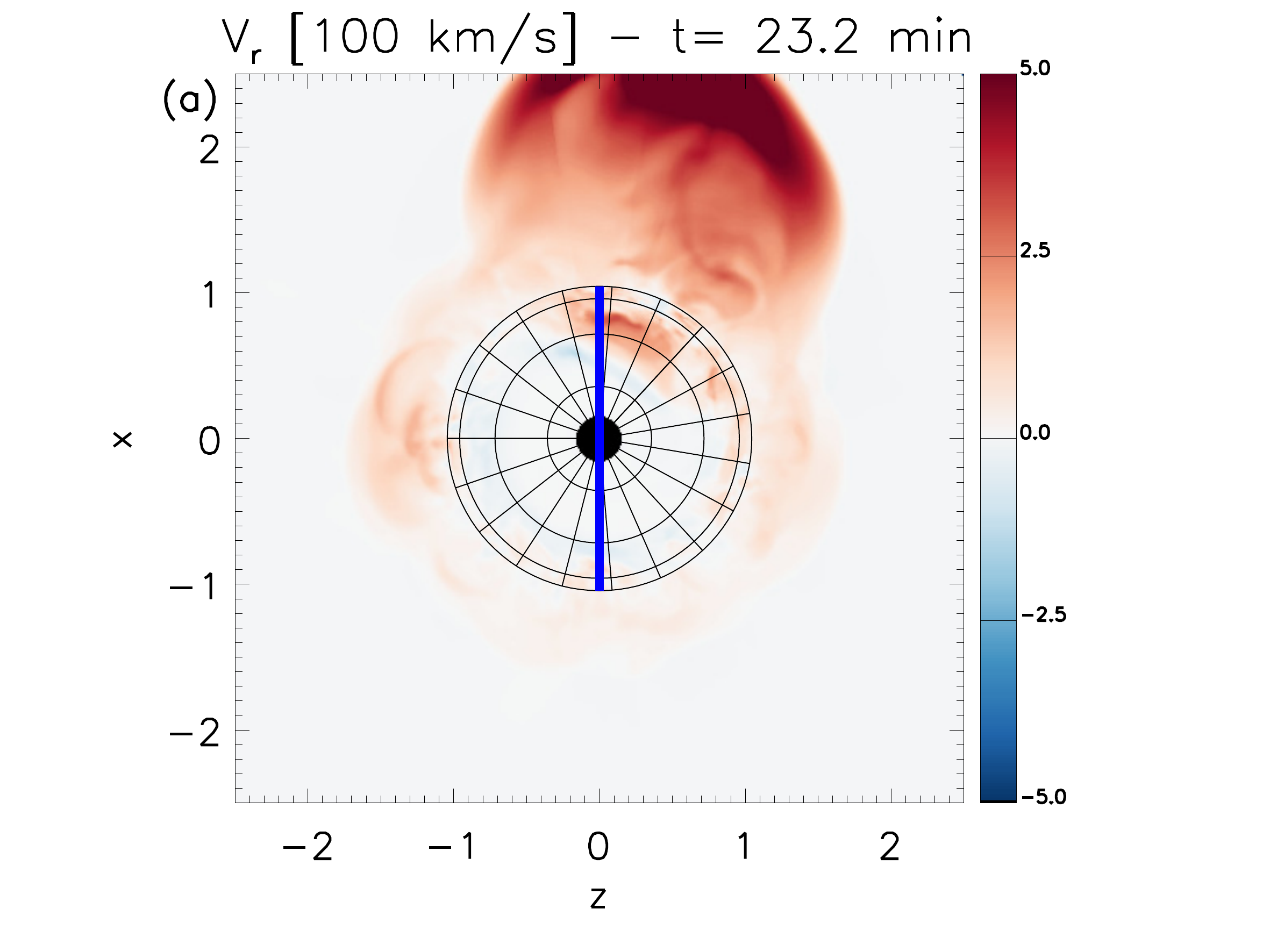}
    \includegraphics[width=0.50\linewidth,trim=0.75in 0.25in 1.40in 0in,clip]{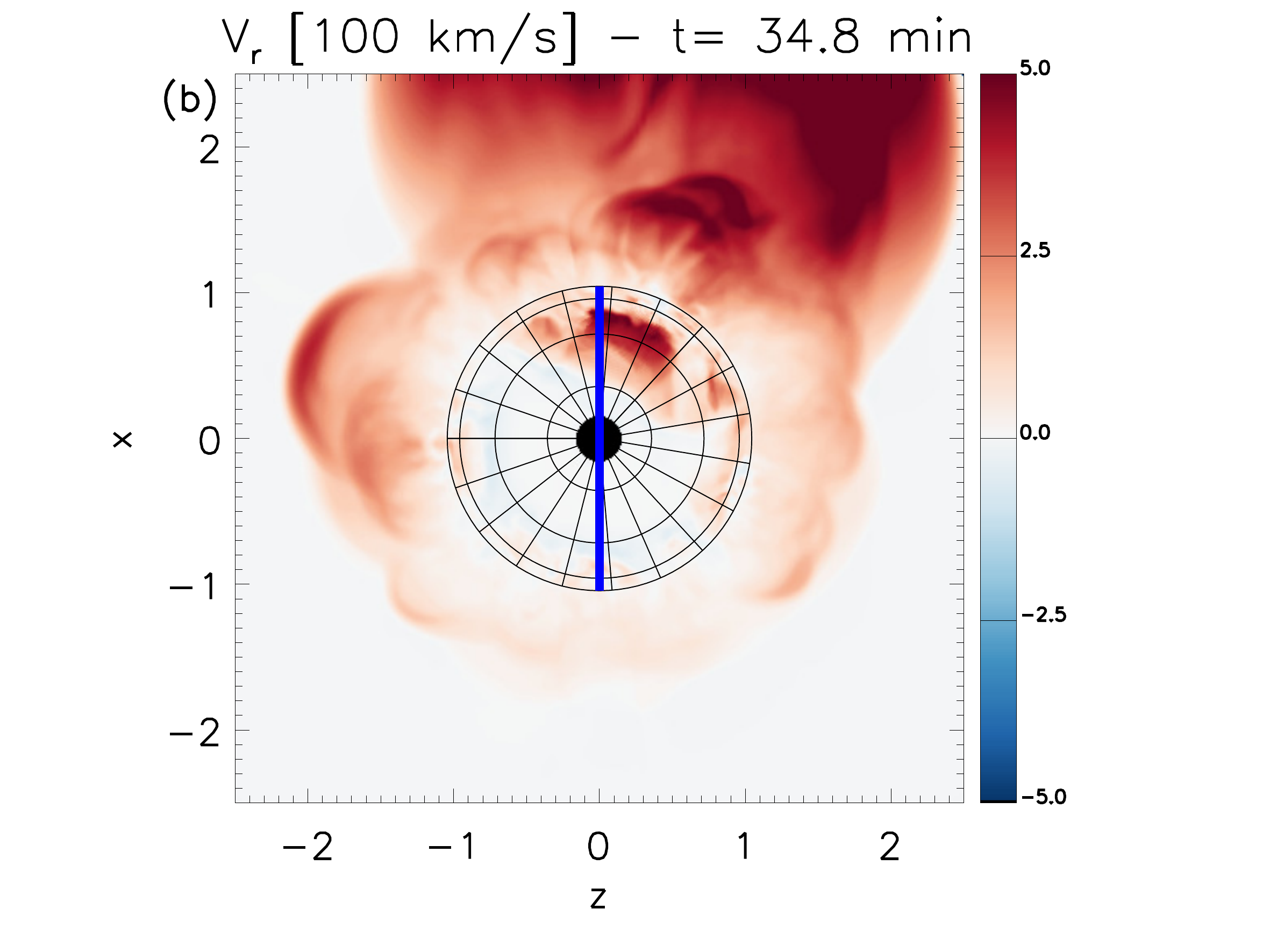}
    \caption{Two snapshots from the MHD evolution of the MF\,--\,MHD coupled simulation of Event~12. Each panel shows the ecliptic plane viewed from the solar north pole and the Earth is to the left at $(x,y,z)=(0,0,-215)$. The Sun is indicated as the spherical grid at the $r/R_\odot=1.0$ lower boundary. The large, initial eruption from just behind the east limb ($+\hat{x}$ direction) triggers subsequent radial outflows in the general direction toward Earth ($-\hat{z}$). The colour scale gives the radial plasma velocity in units of 100 km/s and the axis units are $R_\odot$. 
    }
    \label{fig:event12mhd} 
\end{figure}

Figure~\ref{fig:event12mhd} shows the evolution of the radial velocity in the MHD simulation averaged over the line of sight from a vantage point above the North pole of the Sun, at times of $t=23.2$~min and $t=34.8$~min. The thick blue line marks the plane of the sky from the point of view of Earth, which is to the left of each panel ($z < 0$) along the $x=0$ axis.
At $t=23.2$~min, the signatures of a large eruption triggered just behind the eastern limb from Earth's perspective are evident. This CME-like plasma outflow propagates towards the east ($x>0$) with peak radial velocities of the order $v_r \gtrsim 500$~km/s.  No other significant outflows are present at this time.
By t=34.8~min, additional outflows are present, resulting from the destabilisation of magnetic structures farther away by the initial, large eruption. In particular, another slower eruption lifts off near disc centre as viewed from Earth. 
%
This simulation shows that an eruption in a realistic (i.e., complex) magnetic environment with the right magnetic connectivity can lead to weaker, sympathetic eruptions some distance from the initial eruption ($\sim$90$^\circ$ longitude away) with significantly less apparent coronal signatures (e.g., $v_r \sim 250$~km/s). Relating this simulation to the observations of Event 12, \citet{Yardley2021b} suggest that a prior, larger eruption that was not Earth-directed, could have triggered a smaller, centre disc eruption from an energised structure that had not yet reached an unstable condition. Because of the lower amount of stored free magnetic energy in a such structure that is not ready to erupt, the resulting eruption releases a smaller amount of kinetic energy and thus the eruptive signatures are less evident.

\begin{figure}
    \includegraphics[width=0.50\linewidth,trim=0.8in 0.25in 1.35in 0in,clip]{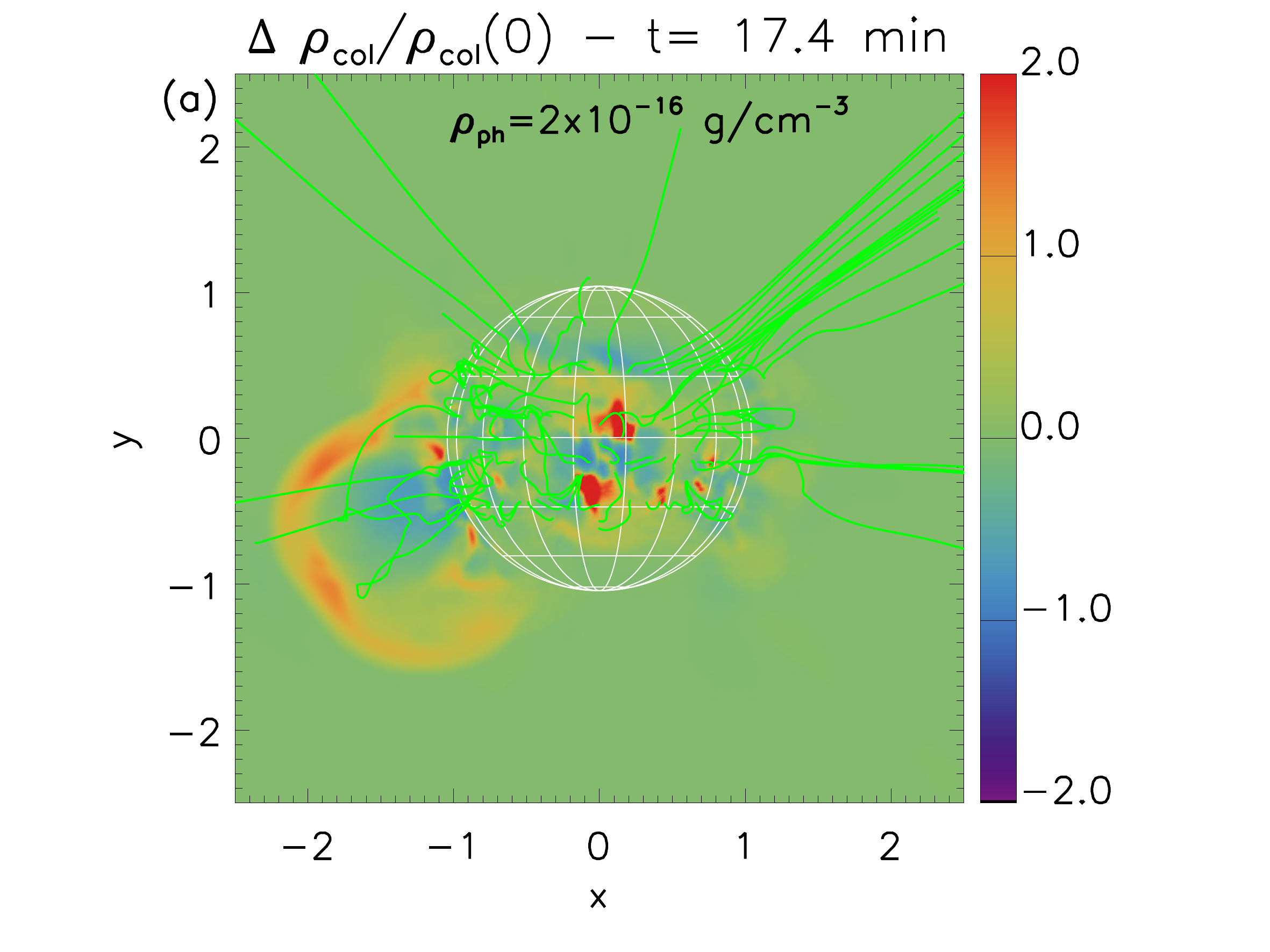}
    \includegraphics[width=0.50\linewidth,trim=0.75in 0.25in 1.40in 0in,clip]{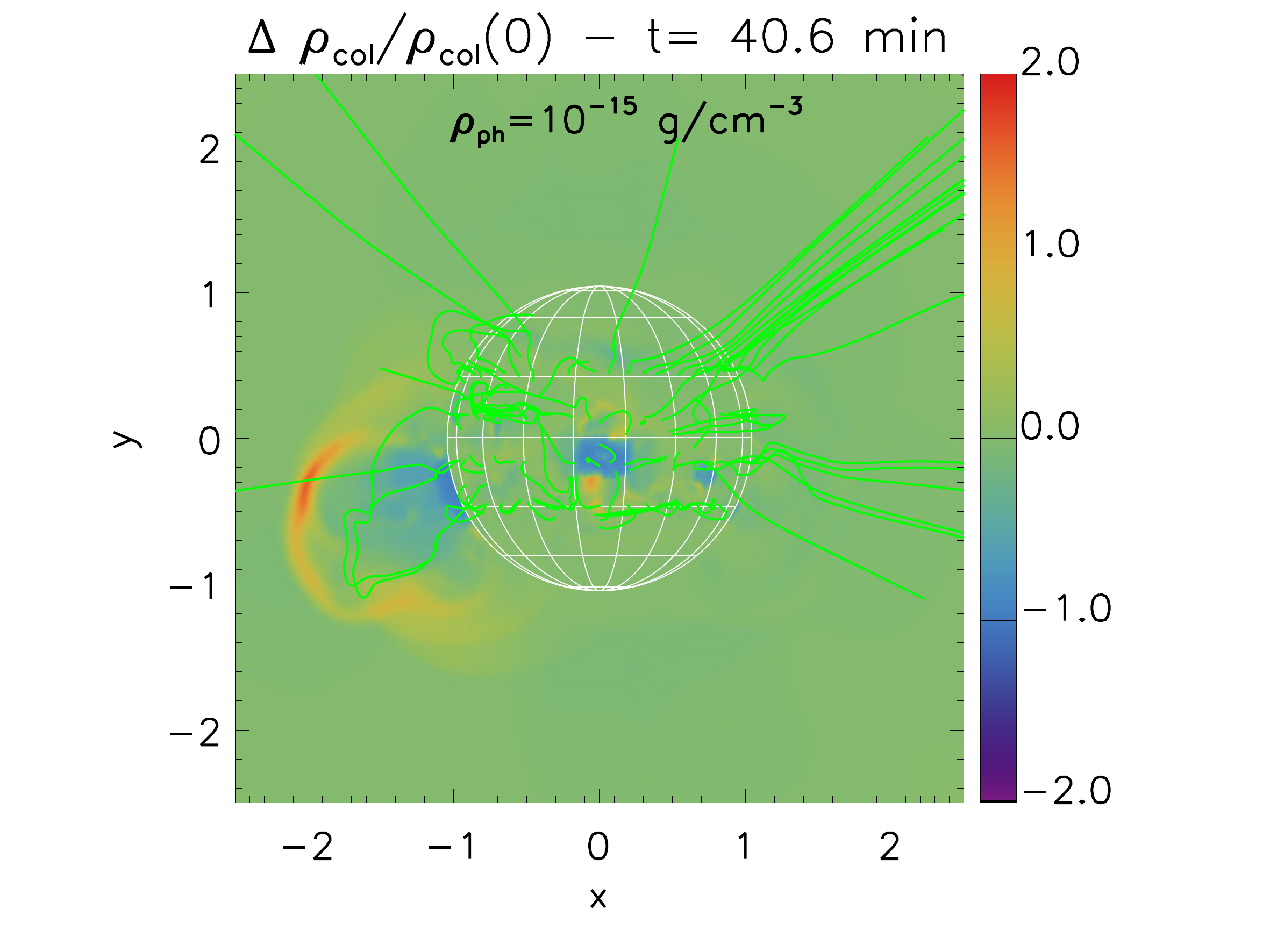}
    \caption{Demonstration of the sensitivity of eruption signatures with atmosphere parameters. Left panel: The relative variation in column density from Earth's perspective during the first behind-the-eastern-limb eruption in the MF\,--\,MHD coupled simulation of Event 12. Right panel: Variation in column density for a more subdued eruption from less intense radial outflows in an initial atmosphere with the 5 times the base mass density.}
    \label{fig:event12mhd2} 
\end{figure}

This modelling approach also allows the role of the background atmosphere in the coronal signatures of eruptions to be investigated. Figure~\ref{fig:event12mhd2} compares two simulation scenarios for Event 12 where just the initial mass density profile is changed. 
Figure~\ref{fig:event12mhd2}(a) shows the relative variations in the column density, $\Delta \rho_{\rm col}(t)/\rho_{\rm col}(0)$, from the Earth point of view at $t=17.4$~min for the simulation discussed so far.  At this time, the first, behind-the-eastern-limb, eruption is still propagating in the solar corona. In this simulation, the column density is significantly perturbed by the CME passage
%
and the column density variation is comparable to the column density itself. 
%
Figure~\ref{fig:event12mhd2}(b) shows the relative variation of the column density at time $t=40.6$~min in a simulation where the initial background density is increased by a factor of 5. At this time, the initial, eastern-limb CME has travelled a similar distance as in Figure~\ref{fig:event12mhd2}(a). 
With the increased background density, the eruption signatures both in the corona and on disc are significantly reduced. Such variations in the column density of the order of a fraction of the initial column density could make any eruption signature more arduous to detect.
This simple study highlights how the field and plasma properties of the corona may contribute to the generation of difficult to detect stealth CMEs. In particular, these may be low-energy events triggered by a previous, possibly distant eruption or the background coronal conditions may act to suppress or mask the typical signatures of a CME eruption.


\subsection{Stealth CMEs Originating from Streamer Disruptions and Reconnection Transients}
\label{sec:modelling:blobs}

Here we discuss a class of ``non-traditional'' reconnection-generated transients originating in the extended solar corona. If large enough, these may be essentially indistinguishable from CMEs or CME-like structures in interplanetary in situ measurements but only have weak or no LCSs and therefore could be a potential source of stealth CMEs and hence problematic geomagnetic impacts at Earth. In particular, we will discuss recent modelling efforts showing that streamer blobs, streamer detachments, and flare-reconnection plasmoids all produce flux rope-like transient ejecta that are carried into the solar wind. 

Streamer blobs were considered in Section~\ref{sec:observations:event_15} as a possible cause of the storm in Event~8, although we concluded that it was not possible to make a convincing association in that case.  Streamer blobs are formed by magnetic reconnection in HCSs, typically in the cusp of the helmet streamer \citep{WangYM2000} where magnetic flux reconnects to form flux rope-like plasmoid structures that incorporate coronal material.   This occurs in response to plasma perturbations at the base of the HCS in the vicinity of the Y-point or via interchange reconnection at the open--closed field boundary.  MHD modelling of the formation of blobs and their resulting propagation throughout the extended corona has been performed in 2D by \citet{Einaudi2001}, \citet{Endeve2003, Endeve2004}, \citet{Lapenta2008}, and \citet{Allred2015}, and recently in 3D by \citet{Higginson2018} and \citet{Lynch2020}.
White-light coronagraph observations of streamer blobs show they make up a significant component of the variability of the slow solar wind surrounding the HCS \citep[e.g.][]{Sheeley1997, Sheeley1999, Sheeley2007, Rouillard2010a, SanchezDiaz2017} and in situ measurements have shown these often correspond to small flux rope-like structures in the heliosphere \citep[e.g.][]{Cartwright2008, Kilpua2009, Rouillard2011, Yu2014, SanchezDiaz2019, Murphy2020}. 

Streamer detachment is similar to streamer blob formation in that the source region material and magnetic flux originate in the outer regions of the helmet streamer belt. It differs in that the detachment is much larger scale, involving more of the closed-field streamer belt flux, and is caused by the interaction of a normal (i.e., non-stealth) CME eruption with the overlying field that was not part of the stressed and twisted fields erupting from the low corona.
\begin{figure}
    \includegraphics[width=1.0\linewidth]{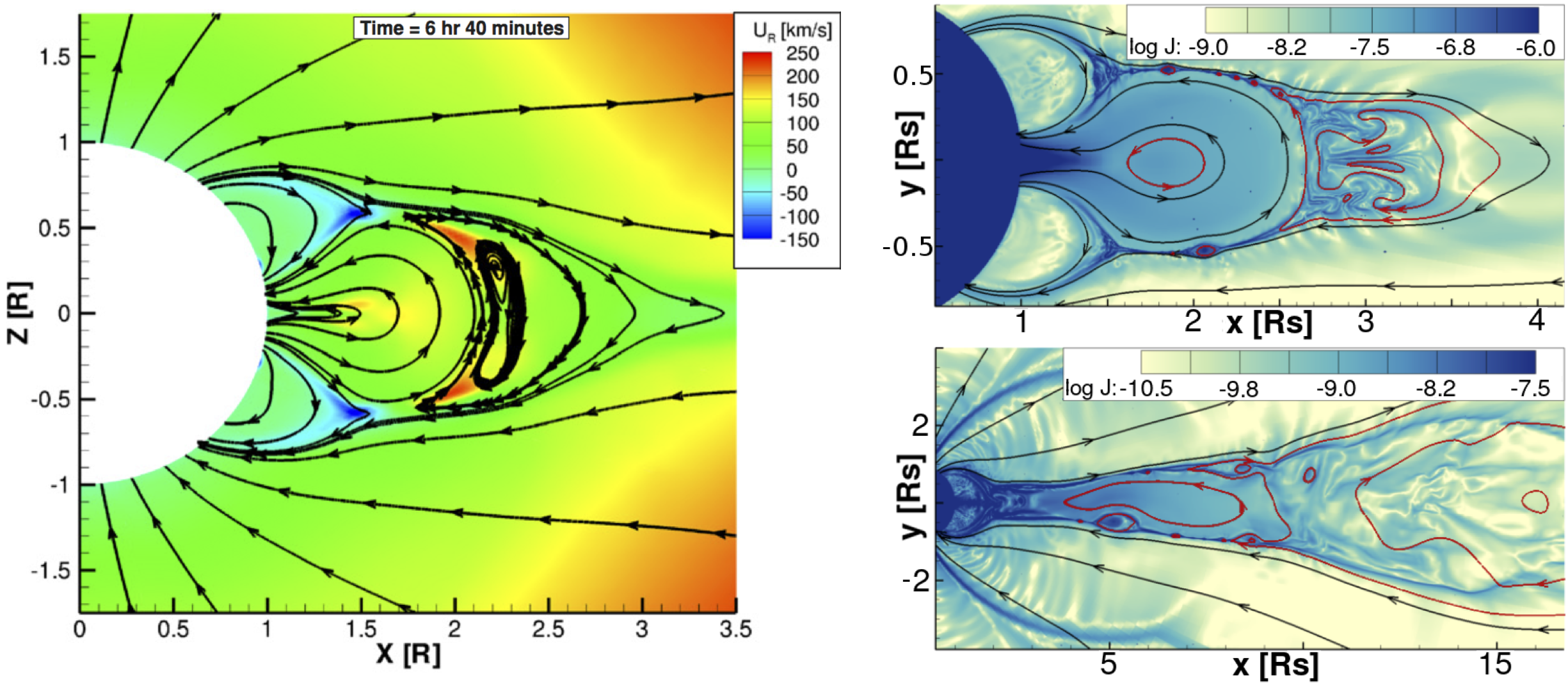}
    \caption{Examples of modelling streamer detachment via reconnection during CME eruption. Left panel shows the projection of 3D field lines  in the complex flux rope-like plasmoid structure formed above a breakout CME eruption \citep[adapted from][]{vanderHolst2009}, with the colour scale indicating the radial plasma speed. Right panel shows the same phenomena in a high-resolution 2D simulation at two different times \citep[adapted from][]{Hosteaux2018}. The colour scale indicates the logarithm of the current density magnitude in ampere~m$^{-2}$.
    }
    \label{fig:streamerdetach} 
\end{figure}
Figure~\ref{fig:streamerdetach} shows two examples of MHD simulations of streamer detachment in the context of the multipolar flux configurations associated with magnetic breakout CMEs \citep{Antiochos1999, Lynch2008}. The left panel, adapted from \citet{vanderHolst2009}, shows projections of 3D magnetic field lines onto the plane of the sky. Above the erupting central arcade, the breakout reconnection current sheet has split into two magnetic X-points, enabling the transformation of a significant portion of the overlying streamer flux system into its own flux rope-like plasmoid that is carried away with the breakout CME eruption. The right panels, adapted from \citet{Hosteaux2018}, show the continued evolution of a similar eruptive configuration in 2D, with high-resolution adaptive mesh refinement used to resolve the complex internal structure within the streamer detachment plasmoid. Here, the red magnetic field lines show that the streamer detachment flux rope can become larger than the original CME in the extended corona. While the distinction between these streamer-detachment flux ropes (that might, in their own right, give rise to a geomagnetic storm if they encounter Earth)  and their underlying non-stealth CME driver may be unnecessary in practical forecasting applications, the fact that the detachment ejecta originate at greater heights in the corona suggests that any low coronal signatures of streamer detachment are likely to be difficult to isolate, especially if they are concurrent with the much more visible signatures of the driver CME.     

The final ``alternative origin'' scenario for stealth CME-like eruptions that we discuss here is the formation of flare reconnection-generated plasmoids. These transient phenomena are also CME-related but consist of coronal plasma and magnetic flux that are swept into the eruptive flare current sheet below, and in the wake of, a preceding CME eruption. For eruptive flare reconnection scenarios that are driven by the rapid reconfiguration of the global-scale magnetic field, elongated current sheets can fragment into multiple X- and O-type null points, effectively speeding up the reconnection rate in order to facilitate the flux transfer required by the global system. The resistive tearing plasmoid instability involved is well known \citep[e.g.][]{Furth1963, Forbes1983} and has been the subject of many theoretical and simulation studies, including applications to solar flares and CMEs
\citep[e.g.][]{Uzdensky2010,ShenC2011,Loureiro2012,Karpen2012,Lynch2016a}.
In addition, \citet{Janvier2017} has discussed an interesting similarity between the size distribution of reconnection-generated plasmoids in isolated current sheet simulations and the observed distribution of heliospheric small flux rope transients measured in situ. The essentially identical power law exponents suggest that magnetic reconnection may be the common physical process underlying their origin. 

\begin{figure}[!t]
    \begin{overpic}[width=0.5\textwidth]{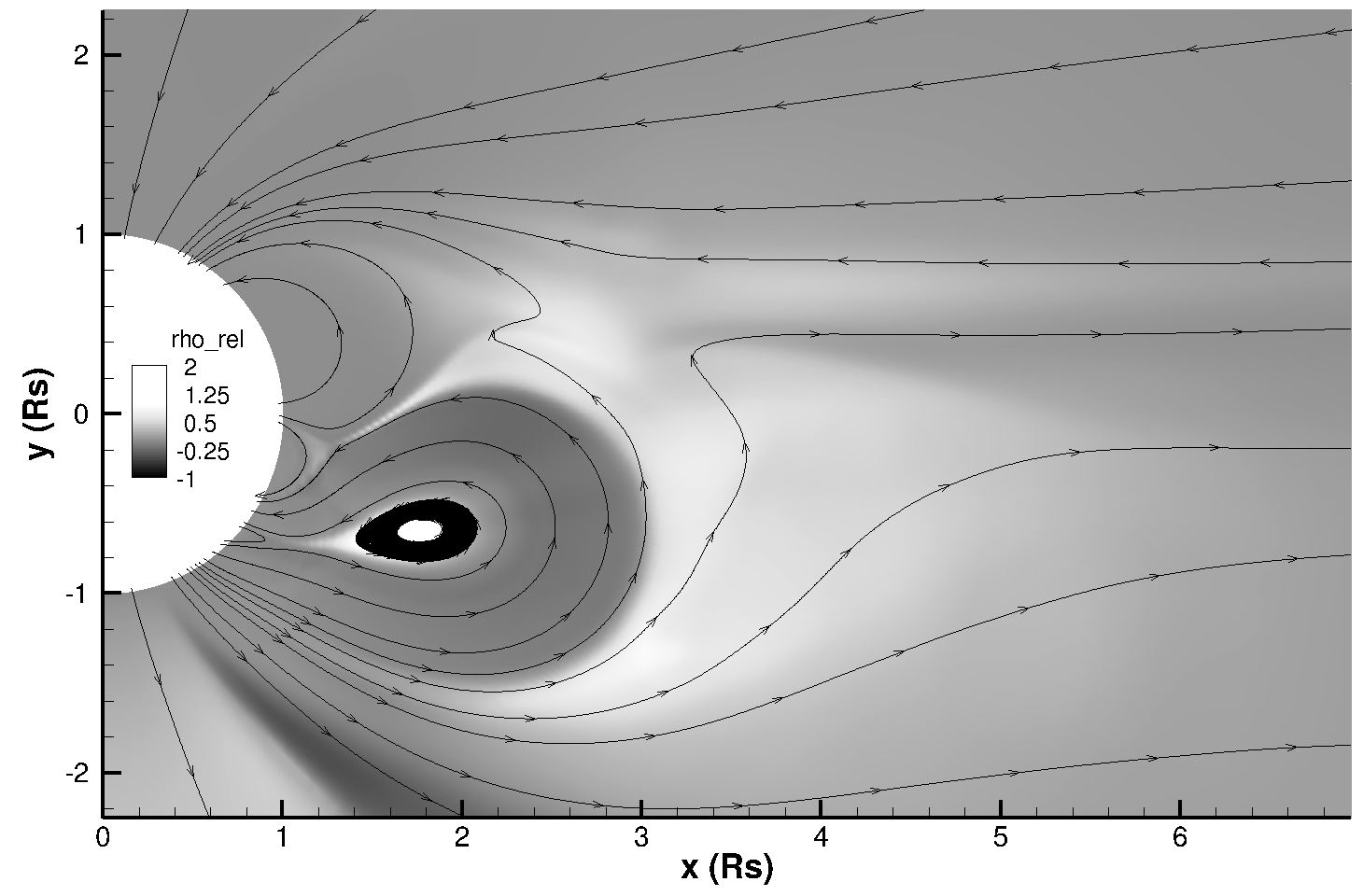}
	    \put (12,60) {\color{black} $\mathbf{time=10.33h}$}
    \end{overpic}
    \begin{overpic}[width=0.5\textwidth]{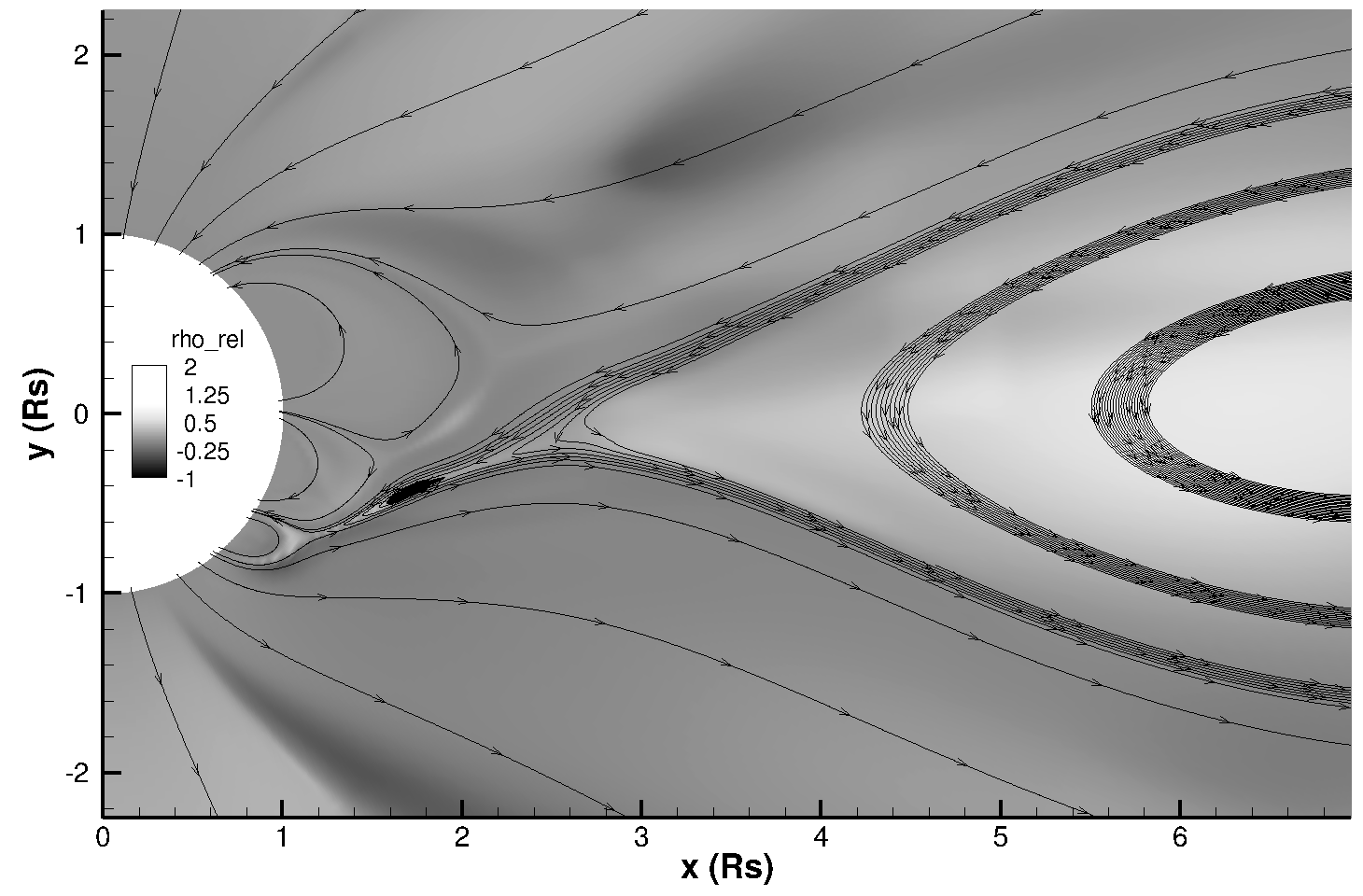}
	    \put (12,60) {\color{black} $\mathbf{time=18h}$}
    \end{overpic}

    \begin{overpic}[width=0.5\textwidth]{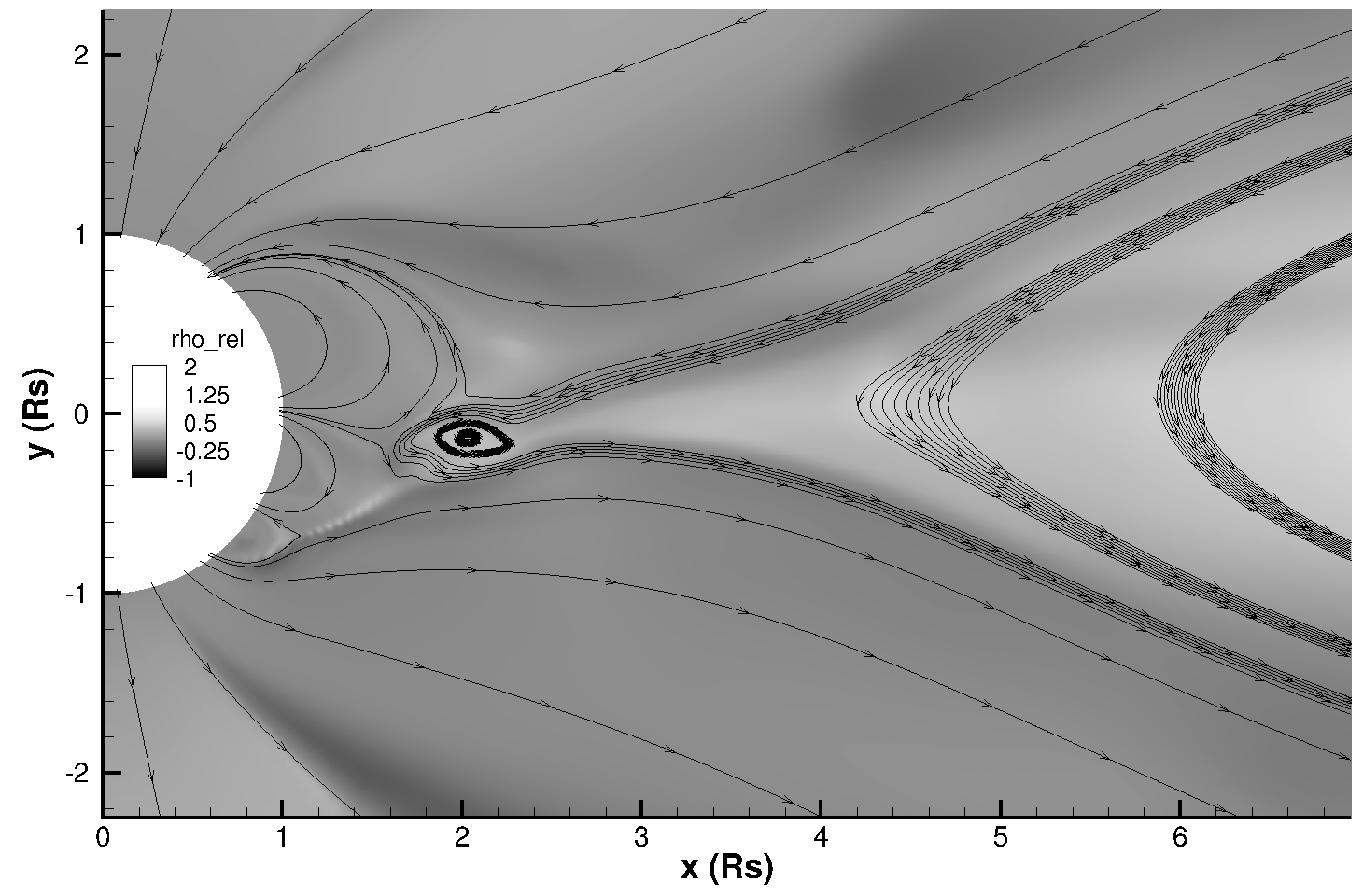}
    	\put (12,60) {\color{black} $\mathbf{time=19.33h}$}
    \end{overpic}
    \begin{overpic}[width=0.5\textwidth]{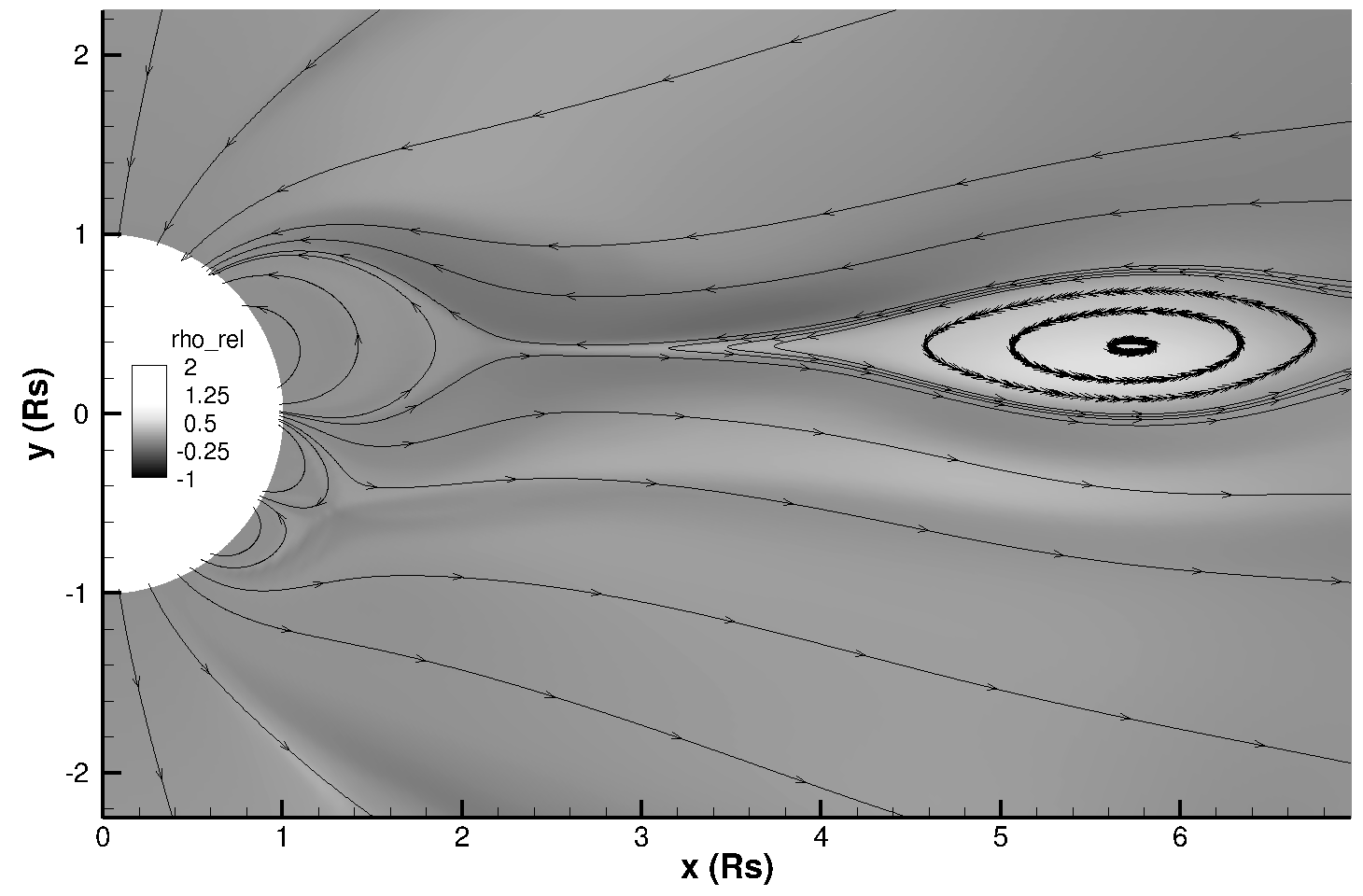}
	    \put (12,60) {\color{black} $\mathbf{time=27h}$}
    \end{overpic}\\
    \hspace*{0.015\linewidth}
    \includegraphics[width=0.5\linewidth]{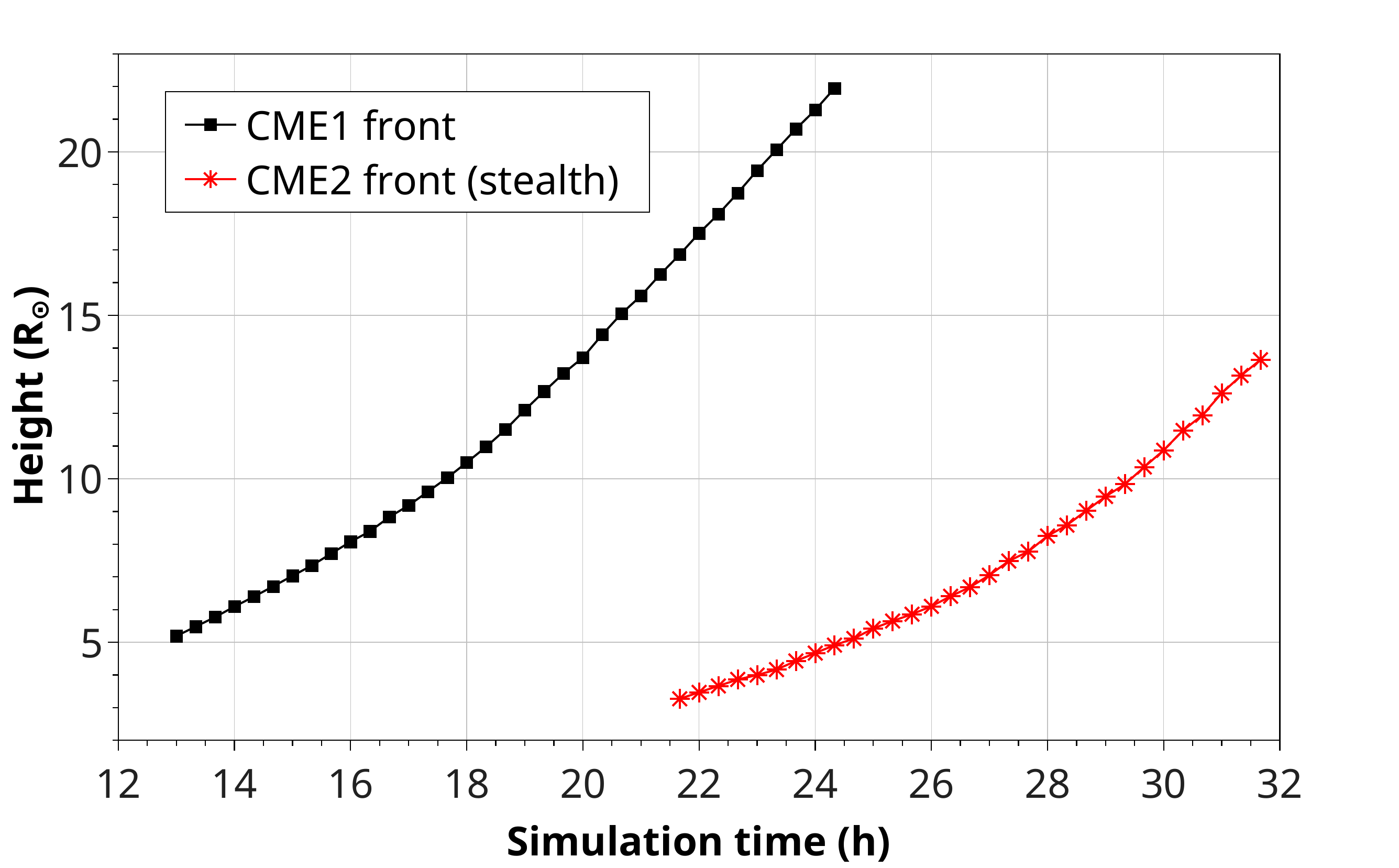}
    \includegraphics[width=0.5174\linewidth]{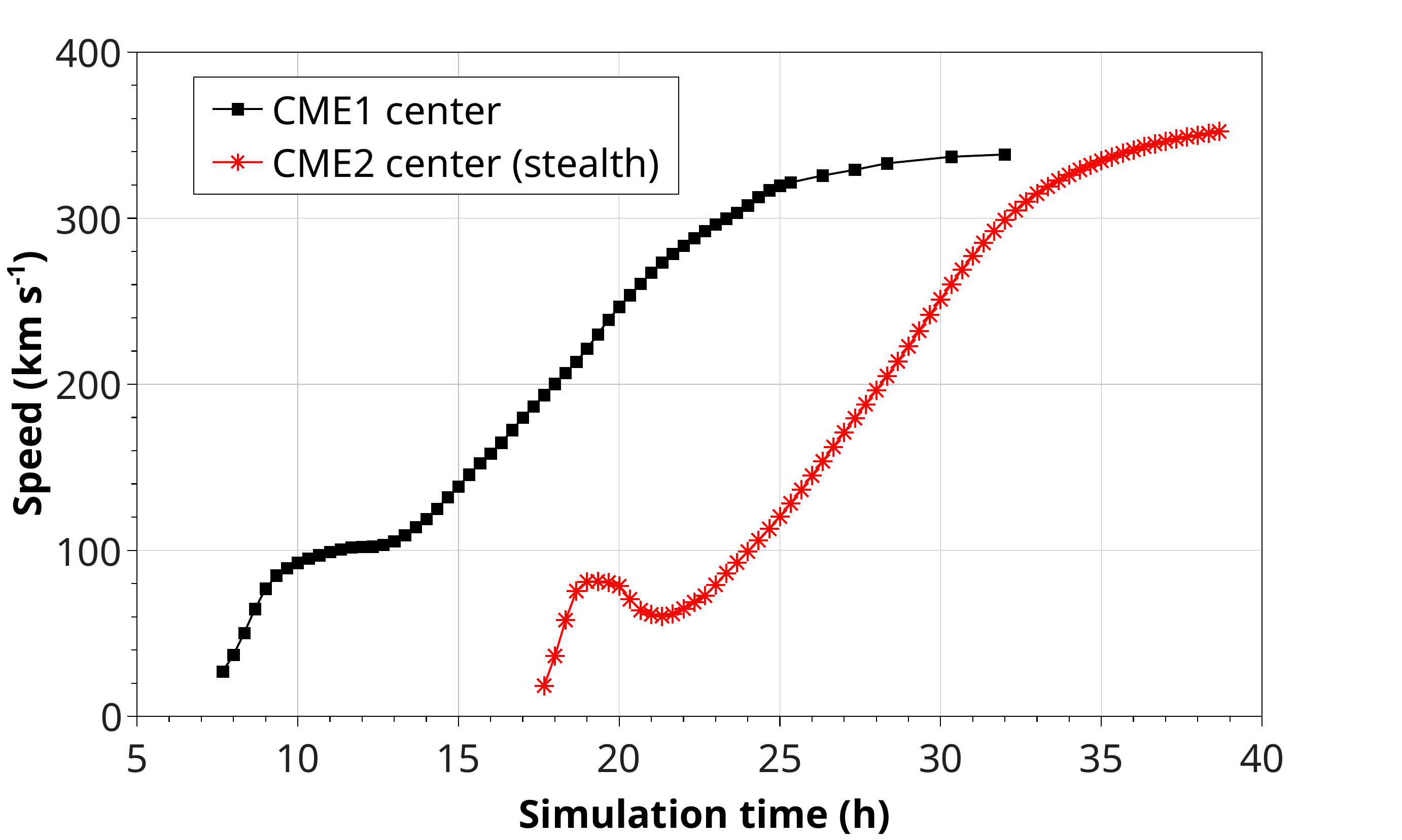}
    \caption{Top and middle panels: Simulated relative density at four times following the evolution of a large reconnection-generated plasmoid in the wake of a preceding eruption, with selected magnetic field lines shown. Bottom panels: Location of the front of the simulated CME (left side) and total speed calculated at the CME centre (right side) as a function of time for: black squares - the preceding eruption, and red stars - the stealth eruption (the reconnection-generated plasmoid). Adapted from \citet{Talpeanu2020}.}
    \label{fig:plasmoid} 
\end{figure}

Figure~\ref{fig:plasmoid} shows simulation results from \citet{Talpeanu2020} where they model the formation and eruption of a large reconnection-generated plasmoid in the flare current sheet of a CME launched from a southern hemisphere pseudostreamer. There are two interesting features to highlight: First, the preceding non-stealth CME driver, that is predominant in the top two panels, shows significant latitudinal deflection, as discussed by \citet{Zuccarello2012} and \citet{Bemporad2012} who analyse and simulate an event observed on 21\,--\,22 September 2009 that resembles this eruption scenario. Second, the plasmoid itself, first evident in the top right panel, grows to a significant size and follows the trajectory of the original CME. \citet{Webb2016} have summarised white-light coronagraph observations of these reconnection blobs, imaged as density enhancements that trail CMEs in the aftermath of streamer blowout eruptions, while \citet{Riley2007} have presented simulation analyses showing excellent agreement with such observations.  

There is indirect evidence that eruptive-flare reconnection plasmoids could be both larger and more coherent in their flux rope structure than streamer blob plasmoids due to their origin at lower coronal heights and their association with the stressed (sheared, twisted) fields of filament channels, leading to a reconnection scenario with a significantly stronger guide field component compared to the streamer blob flux ropes originating at the cusps of  helmet streamers, where the largest-scale coronal fields are much more potential and lack a strong guide field. Thus, large post-eruption reconnection plasmoids may be more likely to survive transit to 1~AU. 
Again, from a space weather forecasting perspective, large eruptive-flare plasmoids may be almost impossible to isolate from their originating CME. Similar to streamer detachment ejecta, they are not independent of their (non-stealth) CME driver, but rather arise as a natural consequence of the preceding eruption. 

It is an open area of research to identify these ``non-traditional'' CME-like transients as potential sources of problematic geomagnetic storms and quantify their impact. Streamer blob flux ropes, streamer detachment ejecta, and eruptive-flare plasmoids all originate at greater coronal heights than CMEs from ARs or even high-latitude filament channels. While it remains to be seen just how often these non-CME transients produce coherent, geoeffective interplanetary ejecta, they should be considered as potential sources of stealth CMEs, with ambiguity in the precise source region and formation/initiation mechanism, until improved observational constraints become available. 


\subsection{MHD Modelling of the Robbrecht et al. Stealth CME Event}
\label{sec:modelling:overview}

In this section we summarise the results from the \citet{Lynch2016b} MHD simulation of the \citet{Robbrecht2009} stealth CME eruption on 1\,--\,2 June 2008. The \citet{Robbrecht2009} explanation for why there were virtually no low-coronal, on-disc signatures is that the CME originated so high in the corona that it was essentially ``above'' the EUV imaging field of view.
On the other hand, \citet{Lynch2016b} argued that their simulation results supported the view that stealth CMEs are not fundamentally different from most slow, streamer-blowout CMEs; they simply represent the lowest-energy range of the slow CME distribution.

\begin{figure*}
\includegraphics[width=1.0\textwidth]{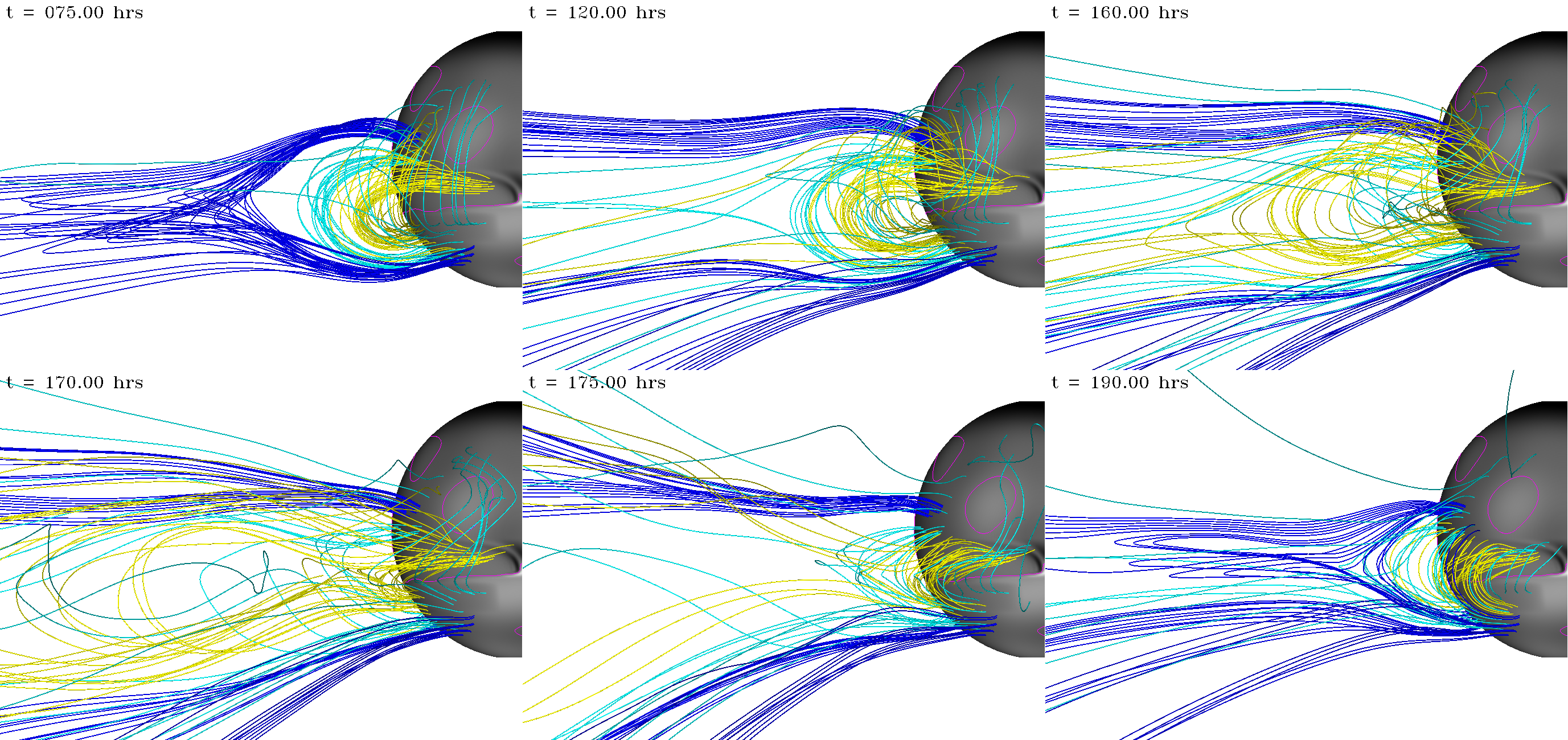}
\caption{Temporal evolution of the \citet{Lynch2016b} MHD simulation of the \citet{Robbrecht2009} stealth CME event. The representative magnetic field lines colored dark blue indicate fields rooted in the overlying streamer flux and at the original open-closed boundary, light blue indicate field lines in the streamer's restraining flux above the sheared polarity inversion line, and the yellow field lines are the innermost sheared core of the streamer flux.}
\label{fig:lynch2016remix} 
\end{figure*}

Figure~\ref{fig:lynch2016remix} shows the magnetic field evolution in the \citet{Lynch2016b} MHD simulation of the gradual eruption of the 1\,--\,2 June 2008 stealth CME. Representative field lines are coloured according to their relationship to the imposed boundary flows that energised the system: the yellow field lines are the inner-most sheared field core of the helmet streamer whereas the light blue and dark blue field lines are the unsheared overlying closed-flux, and the helmet streamer boundary and adjacent open field lines, respectively. 
The MHD simulation was initialized with a potential field extrapolation from the Michelson Doppler Imager \citep[MDI;][]{Scherrer1995} synoptic map of Carrington Rotation 2070.  The energising boundary flows were constructed to model the equivalent large-scale shear resulting from approximately two weeks of differential rotation over a 90$^\circ$ by 40$^\circ$-wide source region in longitude and latitude, respectively. 
After $t=80$~hrs, the boundary flows have ceased so the field line foot points remain the same in each of the subsequent panels.  
The sheared helmet streamer arcade expands slowly, gradually opening more of the restraining overlying flux into the solar wind---seen as the light blue field lines becoming open. This gradual opening facilitates the continued arcade expansion, which in turn, opens more closed flux in a positive-feedback loop. The expanding arcade forms a radial current sheet which allows magnetic reconnection to set in and produce the standard eruptive flare--CME flux rope eruption transition \citep[e.g.][]{Forbes1991, Gosling1995, Linker1995, Linker2003}. The flare reconnection is essential to the CME initiation because it enables the eruption of a significant amount of sheared and twisted flux from the closed field corona as a coherent structure.

\begin{figure*}
\includegraphics[width=1.0\textwidth]{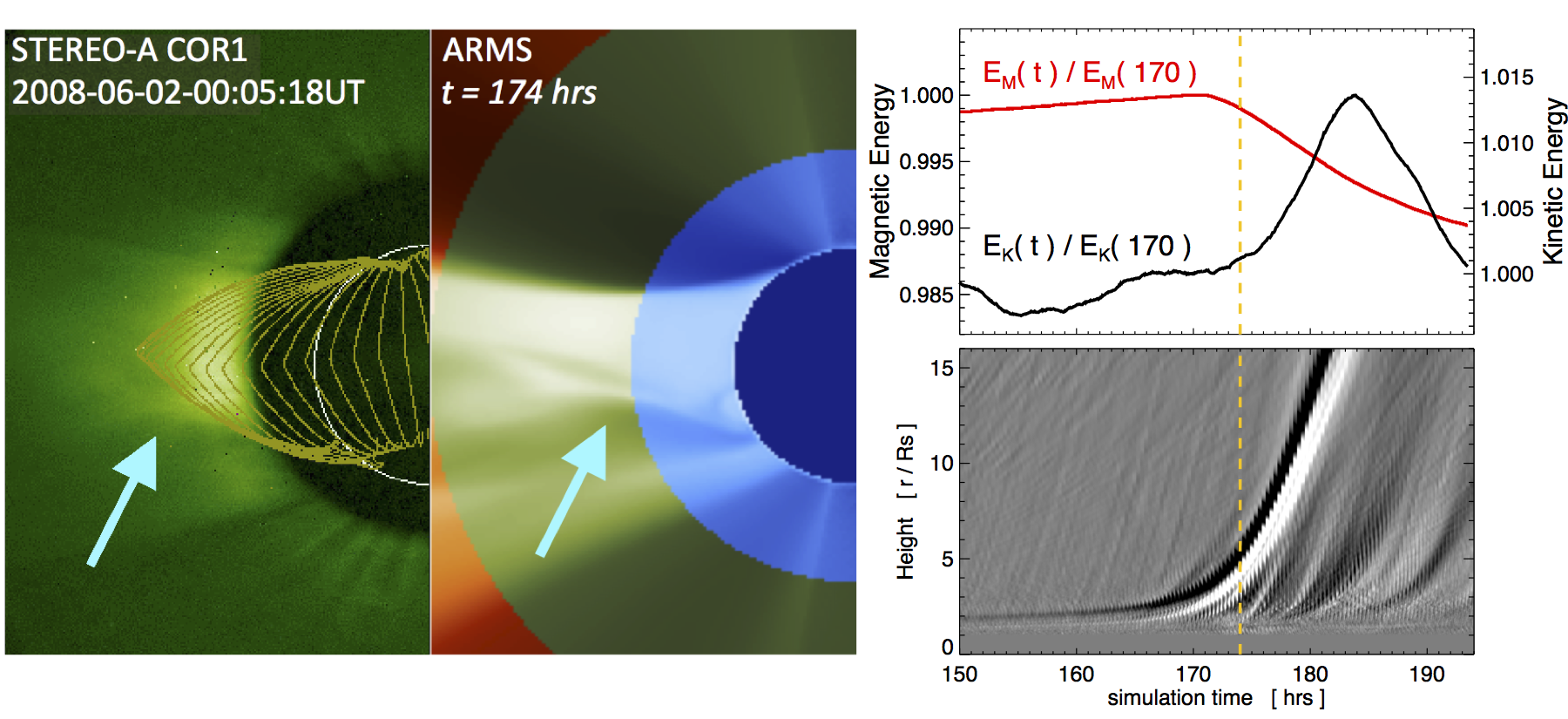}
\caption{The left panel shows the comparison between the STEREO-A COR1 coronagraph data and the synthetic white-light emission from the \citet{Lynch2016b} MHD simulation showing the high-altitude X-point indicative of magnetic reconnection beneath the erupting sheared flux. The right panel shows the temporal evolution of the magnetic and kinetic energy evolution (top) and the ``J-map'' showing the height-time evolution of the stealth CME eruption derived from running-difference synthetic white-light images (bottom).}
\label{fig:lynch2016remix2} 
\end{figure*}

The left panel of Figure~\ref{fig:lynch2016remix2} compares white-light observations from STEREO-A COR1 of the eruptive flare current sheet seen at an altitude of $\sim$1.5\,$R_{\odot}$ with the synthetic white-light structure obtained from the simulation data. The PFSS extrapolation of the helmet streamer belt is over plotted on the COR1 image as yellow dotted lines. In both panels, the cyan arrow points to the reconnection X-point structure. There is a good correspondence between the COR1-A observations and modelling results. Because of the spatial scale of the helmet streamer system and the relatively low magnetic field strength of the coronal fields at such heights, the stored magnetic energy released during the stealth CME eruptive flare reconnection was only $\sim$10$^{30}$~erg over a time of $\gtrsim$20~hrs. The upper right panel of Figure~\ref{fig:lynch2016remix2} shows the evolution of the total magnetic and kinetic energies during the stealth CME eruption normalised to their pre-eruption values. The magnetic energy released during the stealth CME eruption is only $\sim$1\% of the background magnetic energy of the system.
The lower right panel of Figure~\ref{fig:lynch2016remix2} plots a ``J-map'' constructed from radial sampling above the East limb equator of a running-difference movie of the simulation data \citep[e.g.][]{Sheeley1999}. Here, the typical three-part CME features are seen propagating as the faint leading edge enhancement, followed by the dark cavity, followed by the bright core.
The vertical yellow dashed line at $t=174$~hr corresponds to the transition to a runaway eruption, as depicted by the reconnection current sheet visualized in the left panel beneath the erupting flux rope and the rise in the global kinetic energy curve. 
The height-time profile and derived velocities during the modelled stealth CME propagation are in excellent agreement with the STA COR2 observations, reaching $\sim$350~km/s by $\sim$15\,$R_{\odot}$, indicating that the CME is passively advected with the background solar wind. 

\citet{Lynch2016b} suggested that fast CMEs, slow CMEs, stealth CMEs, and streamer blob plasmoids may well sample a continuum of spatial, temporal, and energy scales associated with a single process: the ejection of magnetic stress from the corona into the heliosphere via reconnection.  This hierarchy of eruption energetics can be organized by the relationship between the field strengths and spatial extents of the stressed field structures that eventually erupt. The largest flares that generate fast CMEs originate in very low-lying, stressed fields above polarity inversion lines of strong-field active regions \citep{Schrijver2016,Green2018,Yardley2018b}. Streamer blobs originate in the relatively weak fields at helmet streamer boundaries between the closed and open coronal flux systems \citep{WangYM2000,Higginson2018,Lynch2020}. Slow CMEs---with stealth CMEs as their low-energy subset---may span the intermediate range of field strengths and spatial scales between these two extremes; they originate at intermediate heights in closed field regions and thus have larger spatial extents than streamer blobs but are much less energetic than fast CMEs \citep{Vourlidas2000,Ma2010,Kilpua2014,Vourlidas2018}. 

\subsection{Ambiguous Low-Coronal Signatures in MHD Simulation Data}
\label{sec:modelling:lowcoronasignatures}

\begin{figure}
\includegraphics[width=1.0\linewidth]{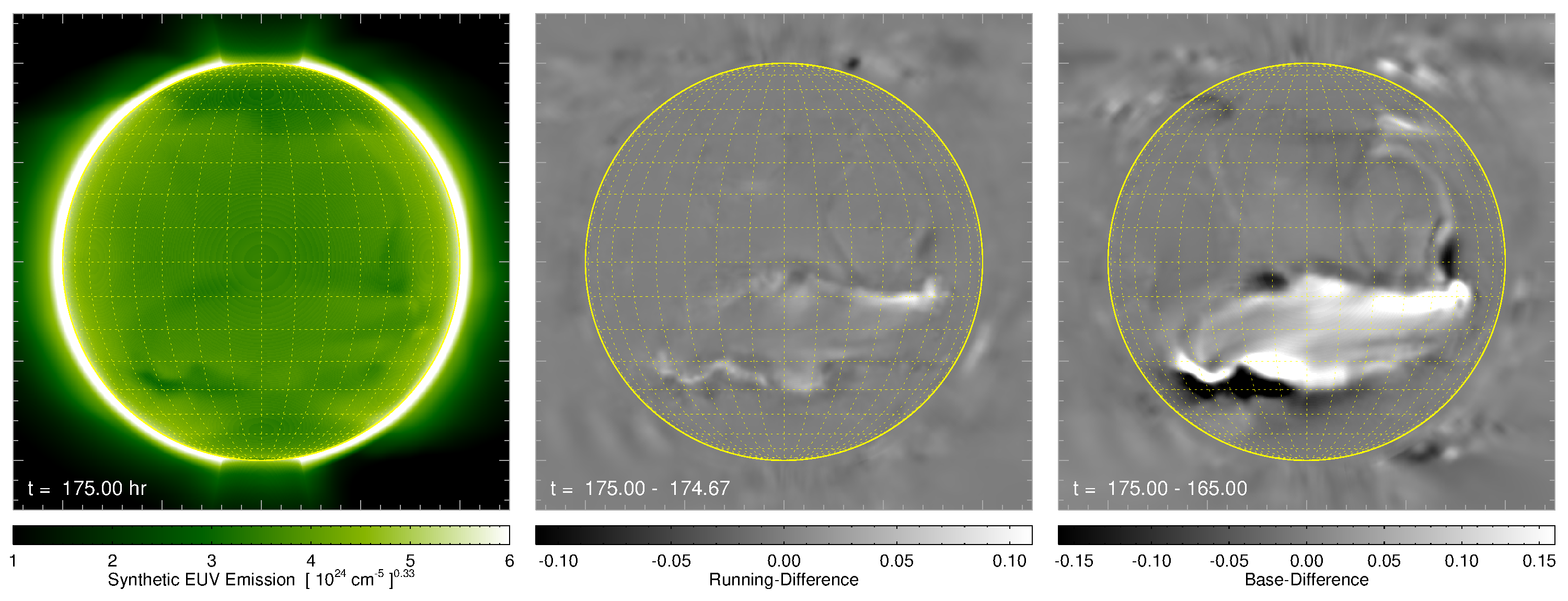}
\caption{Left panel: Synthetic EUV emission of the \citet{Lynch2016b} stealth CME eruption's on-disc low-coronal signatures. Center panel: Running-difference processing of the synthetic EUV emission showing flare ribbon-like dynamics. Right panel: Base-difference processing that highlights the CME foot points and coronal hole evolution dimmings and the post-eruption arcade brightening.}
\label{fig:syntheuv} 
\end{figure}

One of the most challenging aspects of modelling stealth CME event drivers is the lack of clear, unambiguous, observations of the low-coronal signatures of the associated eruption to  compare with simulation results. One of the usual ways to compare and validate low-coronal eruption signatures in MHD simulation results and observations is to create synthetic EUV or X-ray emission images from the simulation data \citep[e.g.][]{Reeves2010, Downs2012, Jin2017}. Here we present the \citet{Lynch2016b} stealth CME eruption as viewed from STEREO-B's perspective in synthetic EUV emission. 

As discussed earlier, large-scale EUV dimming signatures that tend to be one of the only visible indicators of stealth CMEs' source regions are most readily identified with base-difference processing rather than the usual running-difference or in the original images. This is also the case for the dimmings resulting from the large-scale magnetic field reconfiguration and eruption of stealth CME events found in simulation data. The left panel of Figure~\ref{fig:syntheuv} shows a snapshot of synthetic EUV intensity during the eruption at $t = 175$~hr (roughly 5~hr after reconnection beneath the rising magnetic field structure has begun in earnest, see Figure~\ref{fig:lynch2016remix}). 
%
%
%
The \citet{Lynch2019} procedure has been used to calculate a synthetic EUV emission proxy where 
\begin{equation}
I_{\rm EUV} \sim \int_{\rm LOS} n_e^2(\, \mathbf{r}(\ell) \, ) \, d\ell \;  
\end{equation}
and each line of sight (LOS) is perpendicular to the plane of sky viewed from STEREO-B's position at Carrington Longitude $\phi = -101.89^\circ$ so that the source region is centred on the solar disc. The middle panel of Figure~\ref{fig:syntheuv} shows the running-difference image taken between sequential frames 20~min apart, while the right panel shows the base-difference image obtained with respect to an image at $t=165$~hr (i.e., 10~hrs earlier). The EUV intensity image shows very faint dimming regions in the vicinity of the CME foot points while the running-difference EUV panel shows two brightening waves separating from each other on either side of the polarity inversion line---roughly analogous to flare ribbon signatures.  The base-difference image shows even more clearly the newly opened-flux dimming signatures of the CME foot points and the accumulated brightening of the whole region associated with the post-eruption arcade. 
%
This comparison between the original synthetic EUV intensity images and the running-difference and base-difference processed images demonstrates that the long-duration base-difference technique, employed by \citet{Nitta2017} and in Section~\ref{sec:observations}, is most effective at amplifying the subtle, large-scale, EUV coronal dimming and/or reconfiguration signatures often associated with stealth CME source regions.


\section{Discussion}
\label{sec:discussion}

This review highlights that moderate to strong geomagnetic storms may occasionally result from  stealthy solar eruptions and CMEs that occur without clear LCSs, and that the existence of such events is problematic for reliable space weather prediction. In Section~\ref{sec:observations}, we discussed several such events associated with geomagnetic storms with minimum Dst $ < -50$~nT and as a result expanded the definition of ``stealth CMEs'' to capture various aspects of the stealthiness of the CMEs in terms of white-light imagery and LCS observations. Thus, additional events are included in this review that are not discussed in previous studies of stealth CMEs \citep[e.g.,][]{DHuys2014}. In particular, we have exploited the improved capabilities of SDO/AIA with respect to its predecessors to identify weak LCSs that might be linked to the ICME that directly caused the specific problem geomagnetic storm under analysis. As in \citet{Nitta2017} and \citet{Palmerio2021a}, we found elusive LCSs to be more evident in difference images with long temporal separations that take into account the slow nature of these eruptions. We also considered multi-point coronagraph observations, when available, from the STEREO and SOHO spacecraft, to help identify any associated CME. However, in most cases, unlike the textbook case (Figure~\ref{fig:images4textbook}), the location and spatial extent of the CME itself are not well determined.

Advances in modelling the origin and eruption of the stealth-CME drivers of problem geomagnetic storms have contributed to our understanding of the local and global magnetic environments  that produce these eruptions. MF modelling enables efficient studies of the long-term evolution of the global coronal field as well as more detailed investigations into the accumulation and distribution of electric currents and their energised flux systems responsible for stealth CME eruptions (Section~\ref{sec:modelling:mf}). Progress made on MF\,--\,MHD model coupling, while still relatively early in its development, shows promise for a more fully-integrated eruption model, capable of including rapid reconnection dynamics, realistic eruption timescales, and CME plasma evolution (Section~\ref{sec:modelling:mfmhd}). Additionally, insights from the improved EUV imaging available from SDO/AIA and multi-spacecraft coronagraph imagery have motivated numerical simulation studies of the properties and consequences of coronal magnetic field connectivity. For example, prior or sympathetic CME eruptions may act as an external ``trigger'' of secondary eruptive processes that result in CME-like ejecta of lower energies, sizes, or flux content (Sections~\ref{sec:modelling:mfmhd} and \ref{sec:modelling:blobs}). Data-inspired and data-driven MF and MHD modelling of specific stealth CME events also allow quantitative analyses of the energy partition during eruptions (Section \ref{sec:modelling:overview}), and direct model-data comparisons between synthetic observations, derived from simulation data, and multi-wavelength remote-sensing and in situ plasma and field observations (Section~\ref{sec:modelling:lowcoronasignatures}).

Despite progress in observational and modelling efforts to understand the association of stealthy CMEs with geomagnetic storms, it is premature to determine whether these CMEs are caused by mechanisms different from the standard eruptive flare model \citep[e.g.,][]{Svestka1992}. The difficulty of identifying the associated LCSs appears to vary from event to event. In fact, stealthy eruptions can take place over spatial and temporal scales that range from rather small (e.g., Event~8, Section~\ref{sec:observations:event_15}) to very large (Event~14, Section~\ref{sec:observations:event_30}). In addition, there appear to be two important attributes of stealthy events. First, that stealth CMEs are faint and slow, may indicate that they involve less energy than normal CMEs. Secondly, the lack of easily detectable LCSs suggests that they start at high altitudes \citep[cf.][]{Robbrecht2009}. While the exact relationship between these attributes is unknown, they are apparently compatible when considering the rapid fall-off in magnetic field strength with height. Intuitively, the magnetic energy available for an eruption from a high-altitude region above a quiet-Sun or decayed active region configuration should be substantially less than that for an eruption from a low-lying region above a strong-field active region configuration. These are topics that have yet to be explored in detail.

Considering these open questions, it remains easier to define what is a problem geomagnetic storm than a stealth CME. In the words of \citet{McAllister1996}, a storm is problematic when ``there are no clear associations, or the associations are with solar activity that seems too insignificant to properly account for the magnitude of the geomagnetic event.'' In short, a problem geomagnetic storm has to be ``unexpected'' to be defined as such. However, such a simplified definition may not exist for stealth CMEs. Especially in the SDO era, in which the sensitivity, temperature coverage, and resolution (both temporal and spatial) of solar images all have significantly improved with respect to the past, it is rarer for an eruption to completely lack LCSs (as shown in this review). For example, when difference images can successfully reveal that ``something has happened'' on the Sun, is a CME still considered stealthy? And is there an upper limit in LCS visibility, above which a CME ceases to be stealth?

\citet{DHuys2014} have previously noted that stealth CMEs tend to originate in the proximity of polar coronal holes, while \citet{Nitta2017} found that the candidate source regions of geoeffective stealthy CMEs are often close to on-disc coronal holes. Several of the events discussed here also originated near coronal holes that ``grew'' or enlarged their area upon eruption (see the signatures in Table~\ref{tab:tab_2}). It is unlikely that a  CME, which starts from a closed structure, will be launched from inside a coronal hole, but proximity to a coronal hole may affect the trigger and subsequent evolution of the stealthy CME. However, enlargement of a coronal hole area was observed for less than half of the eruptions analysed here, suggesting that proximity to a coronal hole may not be a necessary condition for an elusive or stealthy CME. The presence of a nearby coronal hole may also contribute to the intensification of the associated geomagnetic storm, if the resulting ICME interacts with the adjacent HSS, as in the case of Event~16 (Section~\ref{sec:observations:event_34}). Event~8 (Section~\ref{sec:observations:event_15}) might be another example, although uncertainty remains as to whether the storm-driving structure was an ICME ahead of a CIR, or a CIR with HPS signatures ``masquerading'' as an ICME. Since stealth CMEs tend to be relatively modest in size and slow in speed, it is reasonable to expect their in-situ signatures to be often rather weak and indistinct, lying on a continuum of structures between those of ICMEs and blobs or other heliospheric transients.

This study once again demonstrates the importance of CME observations from far off the Sun--Earth line made possible by the STEREO mission, not only to detect minor CMEs, but also to confirm that they are earthbound. The usefulness of off-limb EUV observations from STEREO-A to identify an approximate source region on the Earth-facing disc was also demonstrated by \citet{OKane2021b} and \citet{Palmerio2021b} for stealth CMEs that erupted as recently as 2020. This is particularly crucial when multiple CMEs occur within a short period. Because the events discussed in this review occurred as the configuration of the STEREO spacecraft was changing (including the loss of contact with STEREO-B), they illustrate the effect of this changing configuration on the interpretation of the CME observations. This was a particular challenge for Event~13 (Section~\ref{sec:observations:event_26}) which took place when the STEREO spacecraft were near superior conjunction and only  observations from the Earth perspective were available. 

In the near future, with STEREO-A approaching the Sun--Earth line (and passing it in August 2023), the converging views of STEREO-A and near-Earth coronagraphs will seriously limit the ability to assess the direction of CMEs and thus identify earthbound CMEs.  In addition, the longitude range covered by EUV imagers will shrink. The recently launched Parker Solar Probe and Solar Orbiter missions can provide complementary observations from varying vantage points in the inner heliosphere, but they may not be suitably located to view specific CMEs of interest and (in the case of Solar Orbiter) their source regions. In addition, these are science missions not dedicated to space weather operations, and most data are made available only with significant delays. For space weather monitoring, a solution would be to develop a network of observatories, starting with missions to the L5 or L4 Lagrange points \citep{Vourlidas2015, Posner2021, Bemporad2021} that could monitor CMEs travelling towards Earth. More ambitious plans to improve space weather predictions might include placing satellites at L3 to monitor the far side of the Sun, and placement in high-inclination (polar) orbits to monitor CMEs emitted from all solar longitudes---the benefits of solar observations from ``unconventional'' viewpoints were reviewed by \citet{Gibson2018}. Although still in the planning stages, these distributed observatories are likely to provide simultaneous full-surface magnetograms, which should significantly improve numerical models of CMEs and their propagation in the heliosphere. We look forward to the day when what we now regard as ``stealth CMEs'' are routinely detected well before they reach Earth, and the geomagnetic storms they produce are therefore no longer considered to be ``problematic".

\begin{acknowledgements}
The authors are grateful to the International Space Science Institute (ISSI), Bern, Switzerland, for hosting two meetings in April 2018 and April 2019 of the international team on ``Problem Geomagnetic Storms'' (\url{https://www.issibern.ch/teams/geomagstorm}).
This work was supported by NASA AIA contract NNG04EA00C, and NRL STEREO/SECCHI contracts N00173-15-C-2016 and N00173-19-C-2011, all to LMSAL.
N.V.N. and T.M. were supported by NASA grant NNX17AB73G.
E.K.J.K. and J.P. acknowledge the Finnish Centre of Excellence in Research of Sustainable Space (FORESAIL; Academy of Finland grant no. 312390), and the European Union's Horizon 2020 Research and Innovation Programme Project 724391 (SolMAG). E.K.J.K. also acknowledges Academy of Finland Project 310445 (SMASH).
B.J.L. acknowledges support from NASA 80NSSC19K0088 and NSF AGS 1851945.
M.M., L.R., D.-C.T., and A.N.Z. thank the European Space Agency (ESA) and the Belgian Federal Science Policy Office (BELSPO) for their support in the framework of the PRODEX Programme.
E.P. is supported by the NASA Living With a Star (LWS) Jack Eddy Postdoctoral Fellowship Program, administered by UCAR's Cooperative Programs for the Advancement of Earth System Science (CPAESS) under award no. NNX16AK22G.
I.G.R. acknowledges support from the ACE and STEREO missions and NASA program NNH17ZDA001N-LWS.
D.-C.T. was funded by the Ph.D. fellowship of the Research Foundation – Flanders (FWO), contract no. 1118920N.
S.L.Y. would like to acknowledge STFC for support via the consolidated grant SMC1/YST037 and NERC for funding via the SWIMMR Aviation Risk and Modelling (SWARM) Project, grant no. NE/V002899/1.

\end{acknowledgements}

\bibliographystyle{spbasic}      
\bibliography{issi_stealth}   

\begin{thebibliography}{174}
\providecommand{\natexlab}[1]{#1}
\providecommand{\url}[1]{{#1}}
\providecommand{\urlprefix}{URL }
\expandafter\ifx\csname urlstyle\endcsname\relax
  \providecommand{\doi}[1]{DOI~\discretionary{}{}{}#1}\else
  \providecommand{\doi}{DOI~\discretionary{}{}{}\begingroup
  \urlstyle{rm}\Url}\fi
\providecommand{\eprint}[2][]{\url{#2}}

\bibitem[{{Abunin} et~al.(2020){Abunin}, {Abunina}, {Belov}, and
  {Chertok}}]{Abunin2020}
{Abunin} AA, {Abunina} MA, {Belov} AV, {Chertok} IM (2020) {Peculiar Solar
  Sources and Geospace Disturbances on 20-26 August 2018}. \solphys, 295(1):7,
  \doi{10.1007/s11207-019-1574-8}

\bibitem[{{Allred} and {MacNeice}(2015)}]{Allred2015}
{Allred} JC, {MacNeice} PJ (2015) {An MHD code for the study of magnetic
  structures in the solar wind}. \csd, 8(1):015002,
  \doi{10.1088/1749-4680/8/1/015002}

\bibitem[{{Alzate} and {Morgan}(2017)}]{Alzate2017}
{Alzate} N, {Morgan} H (2017) {Identification of Low Coronal Sources of
  {``}Stealth{''} Coronal Mass Ejections Using New Image Processing
  Techniques}. \apj, 840:103, \doi{10.3847/1538-4357/aa6caa}

\bibitem[{{Antiochos} et~al.(1999){Antiochos}, {DeVore}, and
  {Klimchuk}}]{Antiochos1999}
{Antiochos} SK, {DeVore} CR, {Klimchuk} JA (1999) {A Model for Solar Coronal
  Mass Ejections}. \apj, 510(1):485--493, \doi{10.1086/306563},
  \eprint{astro-ph/9807220}

\bibitem[{{Attrill} and {Wills-Davey}(2010)}]{Attrill2010}
{Attrill} GDR, {Wills-Davey} MJ (2010) {Automatic Detection and Extraction of
  Coronal Dimmings from SDO/AIA Data}. \solphys, 262(2):461--480,
  \doi{10.1007/s11207-009-9444-4}

\bibitem[{{Barnes} et~al.(2019){Barnes}, {Davies}, {Harrison}, {Byrne},
  {Perry}, {Bothmer}, {Eastwood}, {Gallagher}, {Kilpua}, {M{\"o}stl},
  {Rodriguez}, {Rouillard}, and {Odstr{\v{c}}il}}]{Barnes2019}
{Barnes} D, {Davies} JA, {Harrison} RA, {Byrne} JP, {Perry} CH, {Bothmer} V,
  {Eastwood} JP, {Gallagher} PT, {Kilpua} EKJ, {M{\"o}stl} C, {Rodriguez} L,
  {Rouillard} AP, {Odstr{\v{c}}il} D (2019) {CMEs in the Heliosphere: II. A
  Statistical Analysis of the Kinematic Properties Derived from
  Single-Spacecraft Geometrical Modelling Techniques Applied to CMEs Detected
  in the Heliosphere from 2007 to 2017 by STEREO/HI-1}. \solphys, 294(5):57,
  \doi{10.1007/s11207-019-1444-4}

\bibitem[{{Barnes} et~al.(2020){Barnes}, {Davies}, {Harrison}, {Byrne},
  {Perry}, {Bothmer}, {Eastwood}, {Gallagher}, {Kilpua}, {M{\"o}stl},
  {Rodriguez}, {Rouillard}, and {Odstr{\v{c}}il}}]{Barnes2020}
{Barnes} D, {Davies} JA, {Harrison} RA, {Byrne} JP, {Perry} CH, {Bothmer} V,
  {Eastwood} JP, {Gallagher} PT, {Kilpua} EKJ, {M{\"o}stl} C, {Rodriguez} L,
  {Rouillard} AP, {Odstr{\v{c}}il} D (2020) {CMEs in the Heliosphere: III. A
  Statistical Analysis of the Kinematic Properties Derived from Stereoscopic
  Geometrical Modelling Techniques Applied to CMEs Detected in the Heliosphere
  from 2008 to 2014 by STEREO/HI-1}. \solphys, 295(11):150,
  \doi{10.1007/s11207-020-01717-w}, \eprint{2006.14879}

\bibitem[{{Bavassano} et~al.(1997){Bavassano}, {Woo}, and
  {Bruno}}]{Bavassano1997}
{Bavassano} B, {Woo} R, {Bruno} R (1997) {Heliospheric plasma sheet and coronal
  streamers}. \grl, 24(13):1655--1658, \doi{10.1029/97GL01630}

\bibitem[{{Bemporad}(2021)}]{Bemporad2021}
{Bemporad} A (2021) {Possible advantages of a twin spacecraft Heliospheric
  mission at the Sun-Earth Lagrangian points L4 and L5}. \frass, 8:627576,
  \doi{10.3389/fspas.2021.627576}

\bibitem[{{Bemporad} et~al.(2012){Bemporad}, {Zuccarello}, {Jacobs}, {Mierla},
  and {Poedts}}]{Bemporad2012}
{Bemporad} A, {Zuccarello} FP, {Jacobs} C, {Mierla} M, {Poedts} S (2012) {Study
  of Multiple Coronal Mass Ejections at Solar Minimum Conditions}. \solphys,
  281(1):223--236, \doi{10.1007/s11207-012-9999-3}

\bibitem[{{Borrini} et~al.(1981){Borrini}, {Gosling}, {Bame}, {Feldman}, and
  {Wilcox}}]{Borrini1981}
{Borrini} G, {Gosling} JT, {Bame} SJ, {Feldman} WC, {Wilcox} JM (1981) {Solar
  wind helium and hydrogen structure near the heliospheric current sheet: A
  signal of coronal streamers at 1 AU}. \jgr, 86(A6):4565--4573,
  \doi{10.1029/JA086iA06p04565}

\bibitem[{{Borrini} et~al.(1982){Borrini}, {Gosling}, {Bame}, and
  {Feldman}}]{Borrini1982}
{Borrini} G, {Gosling} JT, {Bame} SJ, {Feldman} WC (1982) {Helium abundance
  enhancements in the solar wind}. \jgr, 87(A9):7370--7378,
  \doi{10.1029/JA087iA09p07370}

\bibitem[{{Bothmer} and {Zhukov}(2007)}]{Bothmer2007}
{Bothmer} V, {Zhukov} A (2007) {The Sun as the prime source of space weather}.
  In: {Bothmer} V, {Daglis} IA (eds) Space Weather- Physics and Effects,
  Springer Praxis Books. ISBN 978-3-540-23907-9. Praxis Publishing Ltd,
  Chichester, p~31, \doi{10.1007/978-3-540-34578-7\_3}

\bibitem[{{Brueckner} et~al.(1995){Brueckner}, {Howard}, {Koomen}, {Korendyke},
  {Michels}, {Moses}, {Socker}, {Dere}, {Lamy}, {Llebaria}, {Bout}, {Schwenn},
  {Simnett}, {Bedford}, and {Eyles}}]{Brueckner1995}
{Brueckner} GE, {Howard} RA, {Koomen} MJ, {Korendyke} CM, {Michels} DJ, {Moses}
  JD, {Socker} DG, {Dere} KP, {Lamy} PL, {Llebaria} A, {Bout} MV, {Schwenn} R,
  {Simnett} GM, {Bedford} DK, {Eyles} CJ (1995) {The Large Angle Spectroscopic
  Coronagraph (LASCO)}. \solphys, 162:357--402, \doi{10.1007/BF00733434}

\bibitem[{{Burlaga} et~al.(1981){Burlaga}, {Sittler}, {Mariani}, and
  {Schwenn}}]{Burlaga1981}
{Burlaga} L, {Sittler} E, {Mariani} F, {Schwenn} R (1981) {Magnetic loop behind
  an interplanetary shock - Voyager, Helios, and IMP 8 observations}. \jgr,
  86:6673--6684, \doi{10.1029/JA086iA08p06673}

\bibitem[{{Cane} and {Richardson}(2003)}]{Cane2003}
{Cane} HV, {Richardson} IG (2003) {Interplanetary coronal mass ejections in the
  near-Earth solar wind during 1996-2002}. \jgr, 108:1156,
  \doi{10.1029/2002JA009817}

\bibitem[{{Cartwright} and {Moldwin}(2008)}]{Cartwright2008}
{Cartwright} ML, {Moldwin} MB (2008) {Comparison of small-scale flux rope
  magnetic properties to large-scale magnetic clouds: Evidence for reconnection
  across the HCS?} \jgr, 113(A9):A09105, \doi{10.1029/2008JA013389}

\bibitem[{{Chen} et~al.(2019){Chen}, {Liu}, {Wang}, {Zhao}, {Hu}, and
  {Zhu}}]{Chen2019}
{Chen} C, {Liu} YD, {Wang} R, {Zhao} X, {Hu} H, {Zhu} B (2019) {Characteristics
  of a Gradual Filament Eruption and Subsequent CME Propagation in Relation to
  a Strong Geomagnetic Storm}. \apj, 884(1):90, \doi{10.3847/1538-4357/ab3f36}

\bibitem[{{Cid} et~al.(2016){Cid}, {Palacios}, {Saiz}, and
  {Guerrero}}]{Cid2016}
{Cid} C, {Palacios} J, {Saiz} E, {Guerrero} A (2016) {Redefining the Boundaries
  of Interplanetary Coronal Mass Ejections from Observations at the Ecliptic
  Plane}. \apj, 828(1):11, \doi{10.3847/0004-637X/828/1/11}

\bibitem[{{Craig} and {Sneyd}(1986)}]{Craig1986}
{Craig} IJD, {Sneyd} AD (1986) {A Dynamic Relaxation Technique for Determining
  the Structure and Stability of Coronal Magnetic Fields}. \apj, 311:451,
  \doi{10.1086/164785}

\bibitem[{{Crooker} et~al.(2004){Crooker}, {Huang}, {Lamassa}, {Larson},
  {Kahler}, and {Spence}}]{Crooker2004}
{Crooker} NU, {Huang} CL, {Lamassa} SM, {Larson} DE, {Kahler} SW, {Spence} HE
  (2004) {Heliospheric plasma sheets}. \jgr, 109(A3):A03107,
  \doi{10.1029/2003JA010170}

\bibitem[{{Delaboudini{\`e}re} et~al.(1995){Delaboudini{\`e}re}, {Artzner},
  {Brunaud}, {Gabriel}, {Hochedez}, {Millier}, {Song}, {Au}, {Dere}, {Howard},
  {Kreplin}, {Michels}, {Moses}, {Defise}, {Jamar}, {Rochus}, {Chauvineau},
  {Marioge}, {Catura}, {Lemen}, {Shing}, {Stern}, {Gurman}, {Neupert},
  {Maucherat}, {Clette}, {Cugnon}, and {van Dessel}}]{Delaboudiniere1995}
{Delaboudini{\`e}re} JP, {Artzner} GE, {Brunaud} J, {Gabriel} AH, {Hochedez}
  JF, {Millier} F, {Song} XY, {Au} B, {Dere} KP, {Howard} RA, {Kreplin} R,
  {Michels} DJ, {Moses} JD, {Defise} JM, {Jamar} C, {Rochus} P, {Chauvineau}
  JP, {Marioge} JP, {Catura} RC, {Lemen} JR, {Shing} L, {Stern} RA, {Gurman}
  JB, {Neupert} WM, {Maucherat} A, {Clette} F, {Cugnon} P, {van Dessel} EL
  (1995) {EIT: Extreme-Ultraviolet Imaging Telescope for the SOHO Mission}.
  \solphys, 162(1-2):291--312, \doi{10.1007/BF00733432}

\bibitem[{{D'Huys} et~al.(2014){D'Huys}, {Seaton}, {Poedts}, and
  {Berghmans}}]{DHuys2014}
{D'Huys} E, {Seaton} DB, {Poedts} S, {Berghmans} D (2014) {Observational
  Characteristics of Coronal Mass Ejections without Low-coronal Signatures}.
  \apj, 795(1):49, \doi{10.1088/0004-637X/795/1/49}, \eprint{1409.1422}

\bibitem[{{Dissauer} et~al.(2018){Dissauer}, {Veronig}, {Temmer},
  {Podladchikova}, and {Vanninathan}}]{Dissauer2018}
{Dissauer} K, {Veronig} AM, {Temmer} M, {Podladchikova} T, {Vanninathan} K
  (2018) {On the Detection of Coronal Dimmings and the Extraction of Their
  Characteristic Properties}. \apj, 855(2):137, \doi{10.3847/1538-4357/aaadb5}

\bibitem[{{Dissauer} et~al.(2019){Dissauer}, {Veronig}, {Temmer}, and
  {Podladchikova}}]{Dissauer2019}
{Dissauer} K, {Veronig} AM, {Temmer} M, {Podladchikova} T (2019) {Statistics of
  Coronal Dimmings Associated with Coronal Mass Ejections. II. Relationship
  between Coronal Dimmings and Their Associated CMEs}. \apj, 874(2):123,
  \doi{10.3847/1538-4357/ab0962}

\bibitem[{{Domingo} et~al.(1995){Domingo}, {Fleck}, and {Poland}}]{Domingo1995}
{Domingo} V, {Fleck} B, {Poland} AI (1995) {The SOHO Mission: an Overview}.
  \solphys, 162:1--37, \doi{10.1007/BF00733425}

\bibitem[{{Downs} et~al.(2012){Downs}, {Roussev}, {van der Holst}, {Lugaz}, and
  {Sokolov}}]{Downs2012}
{Downs} C, {Roussev} II, {van der Holst} B, {Lugaz} N, {Sokolov} IV (2012)
  {Understanding SDO/AIA Observations of the 2010 June 13 EUV Wave Event:
  Direct Insight from a Global Thermodynamic MHD Simulation}. \apj, 750(2):134,
  \doi{10.1088/0004-637X/750/2/134}

\bibitem[{{Duvall}(1979)}]{Duvall1979}
{Duvall} J T~L (1979) {Large-scale solar velocity fields.} \solphys,
  63(1):3--15, \doi{10.1007/BF00155690}

\bibitem[{{Echer} et~al.(2008){Echer}, {Gonzalez}, {Tsurutani}, and
  {Gonzalez}}]{Echer2008}
{Echer} E, {Gonzalez} WD, {Tsurutani} BT, {Gonzalez} ALC (2008) {Interplanetary
  conditions causing intense geomagnetic storms ($Dst \le -100$~nT) during
  solar cycle 23 (1996-2006)}. \jgr, 113(A5):A05221, \doi{10.1029/2007JA012744}

\bibitem[{{Einaudi} et~al.(2001){Einaudi}, {Chibbaro}, {Dahlburg}, and
  {Velli}}]{Einaudi2001}
{Einaudi} G, {Chibbaro} S, {Dahlburg} RB, {Velli} M (2001) {Plasmoid Formation
  and Acceleration in the Solar Streamer Belt}. \apj, 547(2):1167--1177,
  \doi{10.1086/318400}

\bibitem[{{Endeve} et~al.(2003){Endeve}, {Leer}, and {Holzer}}]{Endeve2003}
{Endeve} E, {Leer} E, {Holzer} TE (2003) {Two-dimensional Magnetohydrodynamic
  Models of the Solar Corona: Mass Loss from the Streamer Belt}. \apj,
  589(2):1040--1053, \doi{10.1086/374814}

\bibitem[{{Endeve} et~al.(2004){Endeve}, {Holzer}, and {Leer}}]{Endeve2004}
{Endeve} E, {Holzer} TE, {Leer} E (2004) {Helmet Streamers Gone Unstable:
  Two-Fluid Magnetohydrodynamic Models of the Solar Corona}. \apj,
  603(1):307--321, \doi{10.1086/381239}

\bibitem[{{Eyles} et~al.(2003){Eyles}, {Simnett}, {Cooke}, {Jackson},
  {Buffington}, {Hick}, {Waltham}, {King}, {Anderson}, and
  {Holladay}}]{Eyles2003}
{Eyles} CJ, {Simnett} GM, {Cooke} MP, {Jackson} BV, {Buffington} A, {Hick} PP,
  {Waltham} NR, {King} JM, {Anderson} PA, {Holladay} PE (2003) {The Solar Mass
  Ejection Imager (Smei)}. \solphys, 217(2):319--347,
  \doi{10.1023/B:SOLA.0000006903.75671.49}

\bibitem[{{Eyles} et~al.(2009){Eyles}, {Harrison}, {Davis}, {Waltham},
  {Shaughnessy}, {Mapson-Menard}, {Bewsher}, {Crothers}, {Davies}, {Simnett},
  {Howard}, {Moses}, {Newmark}, {Socker}, {Halain}, {Defise}, {Mazy}, and
  {Rochus}}]{Eyles2009}
{Eyles} CJ, {Harrison} RA, {Davis} CJ, {Waltham} NR, {Shaughnessy} BM,
  {Mapson-Menard} HCA, {Bewsher} D, {Crothers} SR, {Davies} JA, {Simnett} GM,
  {Howard} RA, {Moses} JD, {Newmark} JS, {Socker} DG, {Halain} JP, {Defise} JM,
  {Mazy} E, {Rochus} P (2009) {The Heliospheric Imagers Onboard the STEREO
  Mission}. \solphys, 254(2):387--445, \doi{10.1007/s11207-008-9299-0}

\bibitem[{{Forbes} and {Isenberg}(1991)}]{Forbes1991}
{Forbes} TG, {Isenberg} PA (1991) {A Catastrophe Mechanism for Coronal Mass
  Ejections}. \apj, 373:294, \doi{10.1086/170051}

\bibitem[{{Forbes} and {Priest}(1983)}]{Forbes1983}
{Forbes} TG, {Priest} ER (1983) {A numerical experiment relevant to line-tied
  reconnection in two-ribbon flares}. \solphys, 84:169--188,
  \doi{10.1007/BF00157455}

\bibitem[{{Furth} et~al.(1963){Furth}, {Killeen}, and {Rosenbluth}}]{Furth1963}
{Furth} HP, {Killeen} J, {Rosenbluth} MN (1963) {Finite-Resistivity
  Instabilities of a Sheet Pinch}. \phfl, 6:459--484, \doi{10.1063/1.1706761}

\bibitem[{{Gibson} et~al.(2018){Gibson}, {Vourlidas}, {Hassler}, {Rachmeler},
  {Thompson}, {Newmark}, {Velli}, {Title}, and {McIntosh}}]{Gibson2018}
{Gibson} SE, {Vourlidas} A, {Hassler} DM, {Rachmeler} LA, {Thompson} MJ,
  {Newmark} J, {Velli} M, {Title} A, {McIntosh} SW (2018) {Solar Physics from
  Unconventional Viewpoints}. \frass, 5:32, \doi{10.3389/fspas.2018.00032},
  \eprint{1805.09452}

\bibitem[{{Gonzalez} et~al.(1999){Gonzalez}, {Tsurutani}, and {Cl{\'u}a de
  Gonzalez}}]{Gonzalez1999}
{Gonzalez} WD, {Tsurutani} BT, {Cl{\'u}a de Gonzalez} AL (1999) {Interplanetary
  origin of geomagnetic storms}. \ssr, 88:529--562,
  \doi{10.1023/A:1005160129098}

\bibitem[{{Gopalswamy} et~al.(2009){Gopalswamy}, {Yashiro}, {Michalek},
  {Stenborg}, {Vourlidas}, {Freeland}, and {Howard}}]{Gopalswamy2009}
{Gopalswamy} N, {Yashiro} S, {Michalek} G, {Stenborg} G, {Vourlidas} A,
  {Freeland} S, {Howard} R (2009) {The SOHO/LASCO CME Catalog}. \emp,
  104(1-4):295--313, \doi{10.1007/s11038-008-9282-7}

\bibitem[{{Gosling} et~al.(1981){Gosling}, {Borrini}, {Asbridge}, {Bame},
  {Feldman}, and {Hansen}}]{Gosling1981}
{Gosling} JT, {Borrini} G, {Asbridge} JR, {Bame} SJ, {Feldman} WC, {Hansen} RT
  (1981) {Coronal streamers in the solar wind at 1 AU}. \jgr,
  86(A7):5438--5448, \doi{10.1029/JA086iA07p05438}

\bibitem[{{Gosling} et~al.(1991){Gosling}, {McComas}, {Phillips}, and
  {Bame}}]{Gosling1991}
{Gosling} JT, {McComas} DJ, {Phillips} JL, {Bame} SJ (1991) {Geomagnetic
  activity associated with earth passage of interplanetary shock disturbances
  and coronal mass ejections}. \jgr, 96(A5):7831--7839, \doi{10.1029/91JA00316}

\bibitem[{{Gosling} et~al.(1995){Gosling}, {Birn}, and {Hesse}}]{Gosling1995}
{Gosling} JT, {Birn} J, {Hesse} M (1995) {Three-dimensional magnetic
  reconnection and the magnetic topology of coronal mass ejection events}.
  \grl, 22(8):869--872, \doi{10.1029/95GL00270}

\bibitem[{{Grandin} et~al.(2019){Grandin}, {Aikio}, and
  {Kozlovsky}}]{Grandin2019}
{Grandin} M, {Aikio} AT, {Kozlovsky} A (2019) {Properties and Geoeffectiveness
  of Solar Wind High-Speed Streams and Stream Interaction Regions During Solar
  Cycles 23 and 24}. \jgr, 124(6):3871--3892, \doi{10.1029/2018JA026396}

\bibitem[{{Green} et~al.(2018){Green}, {T{\"o}r{\"o}k}, {Vr{\v{s}}nak},
  {Manchester}, and {Veronig}}]{Green2018}
{Green} LM, {T{\"o}r{\"o}k} T, {Vr{\v{s}}nak} B, {Manchester} W, {Veronig} A
  (2018) {The Origin, Early Evolution and Predictability of Solar Eruptions}.
  \ssr, 214(1):46, \doi{10.1007/s11214-017-0462-5}, \eprint{1801.04608}

\bibitem[{{Harrison} et~al.(2018){Harrison}, {Davies}, {Barnes}, {Byrne},
  {Perry}, {Bothmer}, {Eastwood}, {Gallagher}, {Kilpua}, {M{\"o}stl},
  {Rodriguez}, {Rouillard}, and {Odstr{\v{c}}il}}]{Harrison2018}
{Harrison} RA, {Davies} JA, {Barnes} D, {Byrne} JP, {Perry} CH, {Bothmer} V,
  {Eastwood} JP, {Gallagher} PT, {Kilpua} EKJ, {M{\"o}stl} C, {Rodriguez} L,
  {Rouillard} AP, {Odstr{\v{c}}il} D (2018) {CMEs in the Heliosphere: I. A
  Statistical Analysis of the Observational Properties of CMEs Detected in the
  Heliosphere from 2007 to 2017 by STEREO/HI-1}. \solphys, 293(5):77,
  \doi{10.1007/s11207-018-1297-2}, \eprint{1804.02320}

\bibitem[{{He} et~al.(2018){He}, {Liu}, {Hu}, {Wang}, and {Zhao}}]{He2018}
{He} W, {Liu} YD, {Hu} H, {Wang} R, {Zhao} X (2018) {A Stealth CME Bracketed
  between Slow and Fast Wind Producing Unexpected Geoeffectiveness}. \apj,
  860:78, \doi{10.3847/1538-4357/aac381}, \eprint{1805.03128}

\bibitem[{{Higginson} and {Lynch}(2018)}]{Higginson2018}
{Higginson} AK, {Lynch} BJ (2018) {Structured Slow Solar Wind Variability:
  Streamer-blob Flux Ropes and Torsional Alfv{\'e}n Waves}. \apj, 859(1):6,
  \doi{10.3847/1538-4357/aabc08}

\bibitem[{{Hirshberg} et~al.(1972){Hirshberg}, {Bame}, and
  {Robbins}}]{Hirshberg1972}
{Hirshberg} J, {Bame} SJ, {Robbins} DE (1972) {Solar flares and solar wind
  helium enrichments: July 1965 July 1967}. \solphys, 23(2):467--486,
  \doi{10.1007/BF00148109}

\bibitem[{{Hoeksema} et~al.(2020){Hoeksema}, {Abbett}, {Bercik}, {Cheung},
  {DeRosa}, {Fisher}, {Hayashi}, {Kazachenko}, {Liu}, {Lumme}, {Lynch}, {Sun},
  and {Welsch}}]{Hoeksema2020}
{Hoeksema} JT, {Abbett} WP, {Bercik} DJ, {Cheung} MCM, {DeRosa} ML, {Fisher}
  GH, {Hayashi} K, {Kazachenko} MD, {Liu} Y, {Lumme} E, {Lynch} BJ, {Sun} X,
  {Welsch} BT (2020) {The Coronal Global Evolutionary Model: Using HMI Vector
  Magnetogram and Doppler Data to Determine Coronal Magnetic Field Evolution}.
  \apjs, 250(2):28, \doi{10.3847/1538-4365/abb3fb}, \eprint{2006.14579}

\bibitem[{{Hosteaux} et~al.(2018){Hosteaux}, {Chan{\'e}}, {Decraemer},
  {Talpeanu}, and {Poedts}}]{Hosteaux2018}
{Hosteaux} S, {Chan{\'e}} E, {Decraemer} B, {Talpeanu} DC, {Poedts} S (2018)
  {Ultrahigh-resolution model of a breakout CME embedded in the solar wind}.
  \aap, 620:A57, \doi{10.1051/0004-6361/201832976}

\bibitem[{{House} et~al.(1981){House}, {Wagner}, {Hildner}, {Sawyer}, and
  {Schmidt}}]{House1981}
{House} LL, {Wagner} WJ, {Hildner} E, {Sawyer} C, {Schmidt} HU (1981) {Studies
  of the corona with the Solar Maximum Mission coronagraph/polarimeter}. \apjl,
  244:L117--L121, \doi{10.1086/183494}

\bibitem[{{Howard} et~al.(2008){Howard}, {Moses}, {Vourlidas}, {Newmark},
  {Socker}, {Plunkett}, {Korendyke}, {Cook}, {Hurley}, {Davila}, {Thompson},
  {St Cyr}, {Mentzell}, {Mehalick}, {Lemen}, {Wuelser}, {Duncan}, {Tarbell},
  {Wolfson}, {Moore}, {Harrison}, {Waltham}, {Lang}, {Davis}, {Eyles},
  {Mapson-Menard}, {Simnett}, {Halain}, {Defise}, {Mazy}, {Rochus}, {Mercier},
  {Ravet}, {Delmotte}, {Auchere}, {Delaboudiniere}, {Bothmer}, {Deutsch},
  {Wang}, {Rich}, {Cooper}, {Stephens}, {Maahs}, {Baugh}, {McMullin}, and
  {Carter}}]{HowardR2008}
{Howard} RA, {Moses} JD, {Vourlidas} A, {Newmark} JS, {Socker} DG, {Plunkett}
  SP, {Korendyke} CM, {Cook} JW, {Hurley} A, {Davila} JM, {Thompson} WT, {St
  Cyr} OC, {Mentzell} E, {Mehalick} K, {Lemen} JR, {Wuelser} JP, {Duncan} DW,
  {Tarbell} TD, {Wolfson} CJ, {Moore} A, {Harrison} RA, {Waltham} NR, {Lang} J,
  {Davis} CJ, {Eyles} CJ, {Mapson-Menard} H, {Simnett} GM, {Halain} JP,
  {Defise} JM, {Mazy} E, {Rochus} P, {Mercier} R, {Ravet} MF, {Delmotte} F,
  {Auchere} F, {Delaboudiniere} JP, {Bothmer} V, {Deutsch} W, {Wang} D, {Rich}
  N, {Cooper} S, {Stephens} V, {Maahs} G, {Baugh} R, {McMullin} D, {Carter} T
  (2008) {Sun Earth Connection Coronal and Heliospheric Investigation
  (SECCHI)}. \ssr, 136:67--115, \doi{10.1007/s11214-008-9341-4}

\bibitem[{{Howard} and {Harrison}(2013)}]{HowardT2013}
{Howard} TA, {Harrison} RA (2013) {Stealth Coronal Mass Ejections: A
  Perspective}. \solphys, 285:269--280, \doi{10.1007/s11207-012-0217-0}

\bibitem[{{Hudson} and {Cliver}(2001)}]{Hudson2001}
{Hudson} HS, {Cliver} EW (2001) {Observing coronal mass ejections without
  coronagraphs}. \jgr, 106(A11):25199--25214, \doi{10.1029/2000JA904026}

\bibitem[{{Janvier}(2017)}]{Janvier2017}
{Janvier} M (2017) {Three-dimensional magnetic reconnection and its application
  to solar flares}. \jplph, 83(1):535830101, \doi{10.1017/S0022377817000034},
  \eprint{1612.06513}

\bibitem[{{Jibben} et~al.(2016){Jibben}, {Reeves}, and {Su}}]{Jibben2016}
{Jibben} P, {Reeves} K, {Su} Y (2016) {Evidence for a Magnetic Flux Rope in
  Observations of a Solar Prominence-Cavity System}. \frass, 3:10,
  \doi{10.3389/fspas.2016.00010}

\bibitem[{{Jin} et~al.(2017){Jin}, {Manchester}, {van der Holst}, {Sokolov},
  {T{\'o}th}, {Vourlidas}, {de Koning}, and {Gombosi}}]{Jin2017}
{Jin} M, {Manchester} WB, {van der Holst} B, {Sokolov} I, {T{\'o}th} G,
  {Vourlidas} A, {de Koning} CA, {Gombosi} TI (2017) {Chromosphere to 1 AU
  Simulation of the 2011 March 7th Event: A Comprehensive Study of Coronal Mass
  Ejection Propagation}. \apj, 834(2):172, \doi{10.3847/1538-4357/834/2/172},
  \eprint{1611.08897}

\bibitem[{{Kaiser} et~al.(2008){Kaiser}, {Kucera}, {Davila}, {St.~Cyr},
  {Guhathakurta}, and {Christian}}]{Kaiser2008}
{Kaiser} ML, {Kucera} TA, {Davila} JM, {St~Cyr} OC, {Guhathakurta} M,
  {Christian} E (2008) {The STEREO Mission: An Introduction}. \ssr, 136:5--16,
  \doi{10.1007/s11214-007-9277-0}

\bibitem[{{Karpen} et~al.(2012){Karpen}, {Antiochos}, and
  {DeVore}}]{Karpen2012}
{Karpen} JT, {Antiochos} SK, {DeVore} CR (2012) {The Mechanisms for the Onset
  and Explosive Eruption of Coronal Mass Ejections and Eruptive Flares}. \apj,
  760:81, \doi{10.1088/0004-637X/760/1/81}

\bibitem[{{Kilpua} et~al.(2017){Kilpua}, {Koskinen}, and
  {Pulkkinen}}]{Kilpua2017a}
{Kilpua} E, {Koskinen} HEJ, {Pulkkinen} TI (2017) {Coronal mass ejections and
  their sheath regions in interplanetary space}. \lrsp, 14:5,
  \doi{10.1007/s41116-017-0009-6}

\bibitem[{{Kilpua} et~al.(2009){Kilpua}, {Luhmann}, {Gosling}, {Li}, {Elliott},
  {Russell}, {Jian}, {Galvin}, {Larson}, {Schroeder}, {Simunac}, and
  {Petrie}}]{Kilpua2009}
{Kilpua} EKJ, {Luhmann} JG, {Gosling} J, {Li} Y, {Elliott} H, {Russell} CT,
  {Jian} L, {Galvin} AB, {Larson} D, {Schroeder} P, {Simunac} K, {Petrie} G
  (2009) {Small Solar Wind Transients and Their Connection to the Large-Scale
  Coronal Structure}. \solphys, 256(1-2):327--344,
  \doi{10.1007/s11207-009-9366-1}

\bibitem[{{Kilpua} et~al.(2012){Kilpua}, {Jian}, {Li}, {Luhmann}, and
  {Russell}}]{Kilpua2012}
{Kilpua} EKJ, {Jian} LK, {Li} Y, {Luhmann} JG, {Russell} CT (2012)
  {Observations of ICMEs and ICME-like Solar Wind Structures from 2007 - 2010
  Using Near-Earth and STEREO Observations}. \solphys, 281(1):391--409,
  \doi{10.1007/s11207-012-9957-0}

\bibitem[{{Kilpua} et~al.(2014){Kilpua}, {Mierla}, {Zhukov}, {Rodriguez},
  {Vourlidas}, and {Wood}}]{Kilpua2014}
{Kilpua} EKJ, {Mierla} M, {Zhukov} AN, {Rodriguez} L, {Vourlidas} A, {Wood} B
  (2014) {Solar Sources of Interplanetary Coronal Mass Ejections During the
  Solar Cycle 23/24 Minimum}. \solphys, 289:3773--3797,
  \doi{10.1007/s11207-014-0552-4}

\bibitem[{{Klein} and {Burlaga}(1982)}]{Klein1982}
{Klein} LW, {Burlaga} LF (1982) {Interplanetary magnetic clouds at 1 AU}. \jgr,
  87(A2):613--624, \doi{10.1029/JA087iA02p00613}

\bibitem[{{Kraaikamp} and {Verbeeck}(2015)}]{Kraaikamp2015}
{Kraaikamp} E, {Verbeeck} C (2015) {Solar Demon - an approach to detecting
  flares, dimmings, and EUV waves on SDO/AIA images}. \jswsc, 5:A18,
  \doi{10.1051/swsc/2015019}

\bibitem[{{Krista} and {Reinard}(2013)}]{Krista2013}
{Krista} LD, {Reinard} A (2013) {Study of the Recurring Dimming Region Detected
  at AR 11305 Using the Coronal Dimming Tracker (CoDiT)}. \apj, 762(2):91,
  \doi{10.1088/0004-637X/762/2/91}

\bibitem[{{Lapenta} and {Restante}(2008)}]{Lapenta2008}
{Lapenta} G, {Restante} AL (2008) {Blob formation and acceleration in the solar
  wind: role of converging flows and viscosity}. \angeo, 26:3049--3060,
  \doi{10.5194/angeo-26-3049-2008}, \eprint{0710.2702}

\bibitem[{{Lemen} et~al.(2012){Lemen}, {Title}, {Akin}, {Boerner}, {Chou},
  {Drake}, {Duncan}, {Edwards}, {Friedlaender}, {Heyman}, {Hurlburt}, {Katz},
  {Kushner}, {Levay}, {Lindgren}, {Mathur}, {McFeaters}, {Mitchell}, {Rehse},
  {Schrijver}, {Springer}, {Stern}, {Tarbell}, {Wuelser}, {Wolfson}, {Yanari},
  {Bookbinder}, {Cheimets}, {Caldwell}, {Deluca}, {Gates}, {Golub}, {Park},
  {Podgorski}, {Bush}, {Scherrer}, {Gummin}, {Smith}, {Auker}, {Jerram},
  {Pool}, {Soufli}, {Windt}, {Beardsley}, {Clapp}, {Lang}, and
  {Waltham}}]{Lemen2012}
{Lemen} JR, {Title} AM, {Akin} DJ, {Boerner} PF, {Chou} C, {Drake} JF, {Duncan}
  DW, {Edwards} CG, {Friedlaender} FM, {Heyman} GF, {Hurlburt} NE, {Katz} NL,
  {Kushner} GD, {Levay} M, {Lindgren} RW, {Mathur} DP, {McFeaters} EL,
  {Mitchell} S, {Rehse} RA, {Schrijver} CJ, {Springer} LA, {Stern} RA,
  {Tarbell} TD, {Wuelser} JP, {Wolfson} CJ, {Yanari} C, {Bookbinder} JA,
  {Cheimets} PN, {Caldwell} D, {Deluca} EE, {Gates} R, {Golub} L, {Park} S,
  {Podgorski} WA, {Bush} RI, {Scherrer} PH, {Gummin} MA, {Smith} P, {Auker} G,
  {Jerram} P, {Pool} P, {Soufli} R, {Windt} DL, {Beardsley} S, {Clapp} M,
  {Lang} J, {Waltham} N (2012) {The Atmospheric Imaging Assembly (AIA) on the
  Solar Dynamics Observatory (SDO)}. \solphys, 275:17--40,
  \doi{10.1007/s11207-011-9776-8}

\bibitem[{{Lepping} et~al.(1990){Lepping}, {Jones}, and
  {Burlaga}}]{Lepping1990}
{Lepping} RP, {Jones} JA, {Burlaga} LF (1990) {Magnetic field structure of
  interplanetary magnetic clouds at 1 AU}. \jgr, 95(A8):11957--11965,
  \doi{10.1029/JA095iA08p11957}

\bibitem[{{Linker} and {Mikic}(1995)}]{Linker1995}
{Linker} JA, {Mikic} Z (1995) {Disruption of a Helmet Streamer by Photospheric
  Shear}. \apj, 438:L45, \doi{10.1086/187711}

\bibitem[{{Linker} et~al.(2003){Linker}, {Miki{\'c}}, {Lionello}, {Riley},
  {Amari}, and {Odstrcil}}]{Linker2003}
{Linker} JA, {Miki{\'c}} Z, {Lionello} R, {Riley} P, {Amari} T, {Odstrcil} D
  (2003) {Flux cancellation and coronal mass ejections}. \phpl,
  10(5):1971--1978, \doi{10.1063/1.1563668}

\bibitem[{{Loureiro} et~al.(2012){Loureiro}, {Samtaney}, {Schekochihin}, and
  {Uzdensky}}]{Loureiro2012}
{Loureiro} NF, {Samtaney} R, {Schekochihin} AA, {Uzdensky} DA (2012) {Magnetic
  reconnection and stochastic plasmoid chains in high-Lundquist-number
  plasmas}. \phpl, 19(4):042303, \doi{10.1063/1.3703318}, \eprint{1108.4040}

\bibitem[{{Lynch}(2020)}]{Lynch2020}
{Lynch} BJ (2020) {A Model for Coronal Inflows and In/Out Pairs}. \apj,
  905(2):139, \doi{10.3847/1538-4357/abc5b3}, \eprint{2010.13959}

\bibitem[{{Lynch} et~al.(2008){Lynch}, {Antiochos}, {DeVore}, {Luhmann}, and
  {Zurbuchen}}]{Lynch2008}
{Lynch} BJ, {Antiochos} SK, {DeVore} CR, {Luhmann} JG, {Zurbuchen} TH (2008)
  {Topological Evolution of a Fast Magnetic Breakout CME in Three Dimensions}.
  \apj, 683(2):1192--1206, \doi{10.1086/589738}

\bibitem[{{Lynch} et~al.(2010){Lynch}, {Li}, {Thernisien}, {Robbrecht},
  {Fisher}, {Luhmann}, and {Vourlidas}}]{Lynch2010}
{Lynch} BJ, {Li} Y, {Thernisien} AFR, {Robbrecht} E, {Fisher} GH, {Luhmann} JG,
  {Vourlidas} A (2010) {Sun to 1 AU propagation and evolution of a slow
  streamer-blowout coronal mass ejection}. \jgr, 115:A07106,
  \doi{10.1029/2009JA015099}

\bibitem[{{Lynch} et~al.(2016{\natexlab{a}}){Lynch}, {Edmondson}, {Kazachenko},
  and {Guidoni}}]{Lynch2016a}
{Lynch} BJ, {Edmondson} JK, {Kazachenko} MD, {Guidoni} SE (2016{\natexlab{a}})
  {Reconnection Properties of Large-scale Current Sheets During Coronal Mass
  Ejection Eruptions}. \apj, 826:43, \doi{10.3847/0004-637X/826/1/43},
  \eprint{1410.1089}

\bibitem[{{Lynch} et~al.(2016{\natexlab{b}}){Lynch}, {Masson}, {Li}, {Devore},
  {Luhmann}, {Antiochos}, and {Fisher}}]{Lynch2016b}
{Lynch} BJ, {Masson} S, {Li} Y, {Devore} CR, {Luhmann} JG, {Antiochos} SK,
  {Fisher} GH (2016{\natexlab{b}}) {A model for stealth coronal mass
  ejections}. \jgr, 121:10677, \doi{10.1002/2016JA023432}

\bibitem[{{Lynch} et~al.(2019){Lynch}, {Airapetian}, {DeVore}, {Kazachenko},
  {L{\"u}ftinger}, {Kochukhov}, {Ros{\'e}n}, and {Abbett}}]{Lynch2019}
{Lynch} BJ, {Airapetian} VS, {DeVore} CR, {Kazachenko} MD, {L{\"u}ftinger} T,
  {Kochukhov} O, {Ros{\'e}n} L, {Abbett} WP (2019) {Modeling a Carrington-scale
  Stellar Superflare and Coronal Mass Ejection from $\kappa^{1}Cet$}. \apj,
  880(2):97, \doi{10.3847/1538-4357/ab287e}, \eprint{1906.03189}

\bibitem[{{Ma} et~al.(2010){Ma}, {Attrill}, {Golub}, and {Lin}}]{Ma2010}
{Ma} S, {Attrill} GDR, {Golub} L, {Lin} J (2010) {Statistical Study of Coronal
  Mass Ejections With and Without Distinct Low Coronal Signatures}. \apj,
  722:289--301, \doi{10.1088/0004-637X/722/1/289}

\bibitem[{{Mackay} and {Yeates}(2012)}]{Mackay2012}
{Mackay} DH, {Yeates} AR (2012) {The Sun's Global Photospheric and Coronal
  Magnetic Fields: Observations and Models}. \lrsp, 9(1):6,
  \doi{10.12942/lrsp-2012-6}, \eprint{1211.6545}

\bibitem[{{Mackay} et~al.(2018){Mackay}, {DeVore}, {Antiochos}, and
  {Yeates}}]{Mackay2018}
{Mackay} DH, {DeVore} CR, {Antiochos} SK, {Yeates} AR (2018) {Magnetic Helicity
  Condensation and the Solar Cycle}. \apj, 869(1):62,
  \doi{10.3847/1538-4357/aaec7c}

\bibitem[{{MacQueen} et~al.(1974){MacQueen}, {Eddy}, {Gosling}, {Hildner},
  {Munro}, {Newkirk}, {Poland }, and {Ross}}]{Macqueen1974}
{MacQueen} RM, {Eddy} JA, {Gosling} JT, {Hildner} E, {Munro} RH, {Newkirk} J
  G~A, {Poland } AI, {Ross} CL (1974) {The Outer Solar Corona as Observed from
  Skylab: Preliminary Results}. \apjl, 187:L85, \doi{10.1086/181402}

\bibitem[{{Mason} et~al.(2016){Mason}, {Woods}, {Webb}, {Thompson},
  {Colaninno}, and {Vourlidas}}]{JMason2016}
{Mason} JP, {Woods} TN, {Webb} DF, {Thompson} BJ, {Colaninno} RC, {Vourlidas} A
  (2016) {Relationship of EUV Irradiance Coronal Dimming Slope and Depth to
  Coronal Mass Ejection Speed and Mass}. \apj, 830(1):20,
  \doi{10.3847/0004-637X/830/1/20}

\bibitem[{{McAllister} et~al.(1996){McAllister}, {Dryer}, {McIntosh}, {Singer},
  and {Weiss}}]{McAllister1996}
{McAllister} AH, {Dryer} M, {McIntosh} P, {Singer} H, {Weiss} L (1996) {A large
  polar crown coronal mass ejection and a ``problem'' geomagnetic storm: April
  14-23, 1994}. \jgr, 101(A6):13497--13516, \doi{10.1029/96JA00510}

\bibitem[{{Mignone} et~al.(2012){Mignone}, {Zanni}, {Tzeferacos}, {van
  Straalen}, {Colella}, and {Bodo}}]{Mignone2012}
{Mignone} A, {Zanni} C, {Tzeferacos} P, {van Straalen} B, {Colella} P, {Bodo} G
  (2012) {The PLUTO Code for Adaptive Mesh Computations in Astrophysical Fluid
  Dynamics}. \apjs, 198(1):7, \doi{10.1088/0067-0049/198/1/7},
  \eprint{1110.0740}

\bibitem[{{Mishra} and {Srivastava}(2019)}]{SK_Mishra2019}
{Mishra} SK, {Srivastava} AK (2019) {Linkage of Geoeffective Stealth CMEs
  Associated with the Eruption of Coronal Plasma Channel and Jet-Like
  Structure}. \solphys, 294(12):169, \doi{10.1007/s11207-019-1560-1}

\bibitem[{{Morgan} and {Druckm{\"u}ller}(2014)}]{Morgan2014}
{Morgan} H, {Druckm{\"u}ller} M (2014) {Multi-Scale Gaussian Normalization for
  Solar Image Processing}. \solphys, 289(8):2945--2955,
  \doi{10.1007/s11207-014-0523-9}

\bibitem[{{M{\"o}stl} et~al.(2009){M{\"o}stl}, {Farrugia}, {Temmer},
  {Miklenic}, {Veronig}, {Galvin}, {Leitner}, and {Biernat}}]{Mostl2009}
{M{\"o}stl} C, {Farrugia} CJ, {Temmer} M, {Miklenic} C, {Veronig} AM, {Galvin}
  AB, {Leitner} M, {Biernat} HK (2009) {Linking Remote Imagery of a Coronal
  Mass Ejection to Its In Situ Signatures at 1 AU}. \apjl, 705(2):L180--L185,
  \doi{10.1088/0004-637X/705/2/L180}

\bibitem[{{Murphy} et~al.(2020){Murphy}, {Winslow}, {Schwadron}, {Lugaz}, {Yu},
  {Farrugia}, and {Niehof}}]{Murphy2020}
{Murphy} AK, {Winslow} RM, {Schwadron} NA, {Lugaz} N, {Yu} W, {Farrugia} CJ,
  {Niehof} JT (2020) {A Survey of Interplanetary Small Flux Ropes at Mercury}.
  \apj, 894(2):120, \doi{10.3847/1538-4357/ab8821}

\bibitem[{{Newell} et~al.(2007){Newell}, {Sotirelis}, {Liou}, {Meng}, and
  {Rich}}]{Newell2007}
{Newell} PT, {Sotirelis} T, {Liou} K, {Meng} CI, {Rich} FJ (2007) {A nearly
  universal solar wind-magnetosphere coupling function inferred from 10
  magnetospheric state variables}. Journal of Geophysical Research (Space
  Physics), 112(A1):A01206, \doi{10.1029/2006JA012015}

\bibitem[{{Nieves-Chinchilla} et~al.(2011){Nieves-Chinchilla},
  {G{\'o}mez-Herrero}, {Vi{\~n}as}, {Malandraki}, {Dresing}, {Hidalgo},
  {Opitz}, {Sauvaud}, {Lavraud}, and {Davila}}]{Nieves-Chinchilla2011}
{Nieves-Chinchilla} T, {G{\'o}mez-Herrero} R, {Vi{\~n}as} AF, {Malandraki} O,
  {Dresing} N, {Hidalgo} MA, {Opitz} A, {Sauvaud} JA, {Lavraud} B, {Davila} JM
  (2011) {Analysis and study of the in situ observation of the June 1st 2008
  CME by STEREO}. \jastp, 73(11-12):1348--1360,
  \doi{10.1016/j.jastp.2010.09.026}

\bibitem[{{Nitta} and {Mulligan}(2017)}]{Nitta2017}
{Nitta} NV, {Mulligan} T (2017) {Earth-Affecting Coronal Mass Ejections Without
  Obvious Low Coronal Signatures}. \solphys, 292:125,
  \doi{10.1007/s11207-017-1147-7}

\bibitem[{{Nitta} et~al.(2014){Nitta}, {Aschwanden}, {Freeland}, {Lemen},
  {W{\"u}lser}, and {Zarro}}]{Nitta2014}
{Nitta} NV, {Aschwanden} MJ, {Freeland} SL, {Lemen} JR, {W{\"u}lser} JP,
  {Zarro} DM (2014) {The Association of Solar Flares with Coronal Mass
  Ejections During the Extended Solar Minimum}. \solphys, 289(4):1257--1277,
  \doi{10.1007/s11207-013-0388-3}

\bibitem[{{O'Kane} et~al.(2021{\natexlab{a}}){O'Kane}, {Green}, {Davies},
  {M{\"o}stl}, {Hinterreiter}, {von Forstner}, {Weiss}, {Long}, and
  {Amerstorfer}}]{OKane2021b}
{O'Kane} J, {Green} LM, {Davies} EE, {M{\"o}stl} C, {Hinterreiter} J, {von
  Forstner} JLF, {Weiss} AJ, {Long} DM, {Amerstorfer} T (2021{\natexlab{a}})
  {Solar origins of a strong stealth CME detected by Solar Orbiter}. \aap, in
  press, \doi{10.1051/0004-6361/202140622}, \eprint{2103.17225}

\bibitem[{{O'Kane} et~al.(2021{\natexlab{b}}){O'Kane}, {Mac Cormack},
  {Mandrini}, {D{\'e}moulin}, {Green}, {Long}, and {Valori}}]{OKane2021a}
{O'Kane} J, {Mac Cormack} C, {Mandrini} CH, {D{\'e}moulin} P, {Green} LM,
  {Long} DM, {Valori} G (2021{\natexlab{b}}) {The Magnetic Environment of a
  Stealth Coronal Mass Ejection}. \apj, 908(1):89,
  \doi{10.3847/1538-4357/abd2bf}, \eprint{2012.03757}

\bibitem[{{O{\textquoteright}Kane} et~al.(2019){O{\textquoteright}Kane},
  {Green}, {Long}, and {Reid}}]{OKane2019}
{O{\textquoteright}Kane} J, {Green} L, {Long} DM, {Reid} H (2019) {Stealth
  Coronal Mass Ejections from Active Regions}. \apj, 882(2):85,
  \doi{10.3847/1538-4357/ab371b}, \eprint{1907.12820}

\bibitem[{{Pagano} et~al.(2013){Pagano}, {Mackay}, and {Poedts}}]{Pagano2013b}
{Pagano} P, {Mackay} DH, {Poedts} S (2013) {Effect of gravitational
  stratification on the propagation of a CME}. \aap, 560:A38,
  \doi{10.1051/0004-6361/201322036}, \eprint{1310.6960}

\bibitem[{{Pagano} et~al.(2018){Pagano}, {Mackay}, and {Yeates}}]{Pagano2018}
{Pagano} P, {Mackay} DH, {Yeates} AR (2018) {A new technique for
  observationally derived boundary conditions for space weather}. \jswsc,
  8(27):A26, \doi{10.1051/swsc/2018012}, \eprint{1802.07516}

\bibitem[{{Palacios} et~al.(2017){Palacios}, {Cid}, {Saiz}, and
  {Guerrero}}]{Palacios2017}
{Palacios} J, {Cid} C, {Saiz} E, {Guerrero} A (2017) {Photospheric magnetic
  field of an eroded-by-solar-wind coronal mass ejection}. In: {Vargas
  Dom{\'\i}nguez} S, {Kosovichev} AG, {Antolin} P, {Harra} L (eds) Fine
  Structure and Dynamics of the Solar Atmosphere, IAU Symposium, vol 327, pp
  67--70, \doi{10.1017/S1743921317001077}

\bibitem[{{Palmerio} et~al.(2018){Palmerio}, {Kilpua}, {M{\"o}stl}, {Bothmer},
  {James}, {Green}, {Isavnin}, {Davies}, and {Harrison}}]{Palmerio2018}
{Palmerio} E, {Kilpua} EKJ, {M{\"o}stl} C, {Bothmer} V, {James} AW, {Green} LM,
  {Isavnin} A, {Davies} JA, {Harrison} RA (2018) {Coronal Magnetic Structure of
  Earthbound CMEs and In Situ Comparison}. \spwea, 16(5):442--460,
  \doi{10.1002/2017SW001767}, \eprint{1803.04769}

\bibitem[{{Palmerio} et~al.(2021{\natexlab{a}}){Palmerio}, {Kay}, {Al-Haddad},
  {Lynch}, {Yu}, {Stevens}, {Pal}, and {Lee}}]{Palmerio2021b}
{Palmerio} E, {Kay} C, {Al-Haddad} N, {Lynch} BJ, {Yu} W, {Stevens} ML, {Pal}
  S, {Lee} CO (2021{\natexlab{a}}) {Predicting the Magnetic Fields of a Stealth
  CME Detected by Parker Solar Probe at 0.5 AU}. \apj, 920(2):65,
  \doi{10.3847/1538-4357/ac25f4}, \eprint{2109.04933}

\bibitem[{{Palmerio} et~al.(2021{\natexlab{b}}){Palmerio}, {Nitta}, {Mulligan},
  {Mierla}, {O'Kane}, {Richardson}, {Sinha}, {Srivastava}, {Yardley}, and
  {Zhukov}}]{Palmerio2021a}
{Palmerio} E, {Nitta} NV, {Mulligan} T, {Mierla} M, {O'Kane} J, {Richardson}
  IG, {Sinha} S, {Srivastava} N, {Yardley} SL, {Zhukov} AN (2021{\natexlab{b}})
  {Investigating Remote-sensing Techniques to Reveal Stealth Coronal Mass
  Ejections}. \frass, 8:695966, \doi{10.3389/fspas.2021.695966},
  \eprint{2106.07571}

\bibitem[{{Pesnell} et~al.(2012){Pesnell}, {Thompson}, and
  {Chamberlin}}]{Pesnell2012}
{Pesnell} WD, {Thompson} BJ, {Chamberlin} PC (2012) {The Solar Dynamics
  Observatory (SDO)}. \solphys, 275:3--15, \doi{10.1007/s11207-011-9841-3}

\bibitem[{{Pevtsov} et~al.(2012){Pevtsov}, {Panasenco}, and
  {Martin}}]{Pevtsov2012}
{Pevtsov} AA, {Panasenco} O, {Martin} SF (2012) {Coronal Mass Ejections from
  Magnetic Systems Encompassing Filament Channels Without Filaments}. \solphys,
  277:185--201, \doi{10.1007/s11207-011-9881-8}

\bibitem[{{Piersanti} et~al.(2020){Piersanti}, {De Michelis}, {Del Moro},
  {Tozzi}, {Pezzopane}, {Consolini}, {Marcucci}, {Laurenza}, {Di Matteo},
  {Pignalberi}, {Quattrociocchi}, and {Diego}}]{Piersanti2020}
{Piersanti} M, {De Michelis} P, {Del Moro} D, {Tozzi} R, {Pezzopane} M,
  {Consolini} G, {Marcucci} MF, {Laurenza} M, {Di Matteo} S, {Pignalberi} A,
  {Quattrociocchi} V, {Diego} P (2020) {From the Sun to Earth: effects of the
  25 August 2018 geomagnetic storm}. \angeo, 38(3):703--724,
  \doi{10.5194/angeo-38-703-2020}

\bibitem[{{Posner} et~al.(2021){Posner}, {Arge}, {Staub}, {StCyr}, {Folta},
  {Solanki}, {Strauss}, {Effenberger}, {Gandorfer}, {Heber}, {Henney},
  {Hirzberger}, {Jones}, {K{\"u}hl}, {Malandraki}, and {Sterken}}]{Posner2021}
{Posner} A, {Arge} CN, {Staub} J, {StCyr} OC, {Folta} D, {Solanki} SK,
  {Strauss} RDT, {Effenberger} F, {Gandorfer} A, {Heber} B, {Henney} CJ,
  {Hirzberger} J, {Jones} SI, {K{\"u}hl} P, {Malandraki} O, {Sterken} VJ (2021)
  {A Multi-Purpose Heliophysics L4 Mission}. \spwea, 19(9):e2021SW002777,
  \doi{10.1029/2021SW002777}

\bibitem[{{Reeves} et~al.(2010){Reeves}, {Linker}, {Miki{\'c}}, and
  {Forbes}}]{Reeves2010}
{Reeves} KK, {Linker} JA, {Miki{\'c}} Z, {Forbes} TG (2010) {Current Sheet
  Energetics, Flare Emissions, and Energy Partition in a Simulated Solar
  Eruption}. \apj, 721(2):1547--1558, \doi{10.1088/0004-637X/721/2/1547}

\bibitem[{{Richardson}(2018)}]{Richardson2018}
{Richardson} IG (2018) {Solar wind stream interaction regions throughout the
  heliosphere}. \lrsp, 15(1):1, \doi{10.1007/s41116-017-0011-z}

\bibitem[{{Richardson} and {Cane}(1995)}]{Richardson1995}
{Richardson} IG, {Cane} HV (1995) {Regions of abnormally low proton temperature
  in the solar wind (1965--1991) and their association with ejecta}. \jgr,
  100(A12):23397--23412, \doi{10.1029/95JA02684}

\bibitem[{{Richardson} and {Cane}(2010)}]{Richardson2010}
{Richardson} IG, {Cane} HV (2010) {Near-Earth Interplanetary Coronal Mass
  Ejections During Solar Cycle 23 (1996 - 2009): Catalog and Summary of
  Properties}. \solphys, 264:189--237, \doi{10.1007/s11207-010-9568-6}

\bibitem[{{Richardson} et~al.(2001){Richardson}, {Cliver}, and
  {Cane}}]{Richardson2001}
{Richardson} IG, {Cliver} EW, {Cane} HV (2001) {Sources of geomagnetic storms
  for solar minimum and maximum conditions during 1972-2000}. \grl,
  28(13):2569--2572, \doi{10.1029/2001GL013052}

\bibitem[{{Riley} et~al.(2007){Riley}, {Lionello}, {Miki{\'c}}, {Linker},
  {Clark}, {Lin}, and {Ko}}]{Riley2007}
{Riley} P, {Lionello} R, {Miki{\'c}} Z, {Linker} J, {Clark} E, {Lin} J, {Ko} YK
  (2007) {``Bursty'' Reconnection Following Solar Eruptions: MHD Simulations
  and Comparison with Observations}. \apj, 655(1):591--597,
  \doi{10.1086/509913}

\bibitem[{{Riley} et~al.(2017){Riley}, {Ben-Nun}, {Linker}, {Owens}, and
  {Horbury}}]{Riley2017}
{Riley} P, {Ben-Nun} M, {Linker} JA, {Owens} MJ, {Horbury} TS (2017)
  {Forecasting the properties of the solar wind using simple pattern
  recognition}. \spwea, 15(3):526--540, \doi{10.1002/2016SW001589}

\bibitem[{{Riley} et~al.(2018){Riley}, {Baker}, {Liu}, {Verronen}, {Singer},
  and {G{\"u}del}}]{Riley2018}
{Riley} P, {Baker} D, {Liu} YD, {Verronen} P, {Singer} H, {G{\"u}del} M (2018)
  {Extreme Space Weather Events: From Cradle to Grave}. \ssr, 214(1):21,
  \doi{10.1007/s11214-017-0456-3}

\bibitem[{{Robbrecht} et~al.(2009){Robbrecht}, {Patsourakos}, and
  {Vourlidas}}]{Robbrecht2009}
{Robbrecht} E, {Patsourakos} S, {Vourlidas} A (2009) {No Trace Left Behind:
  STEREO Observation of a Coronal Mass Ejection Without Low Coronal
  Signatures}. \apj, 701(1):283--291, \doi{10.1088/0004-637X/701/1/283},
  \eprint{0905.2583}

\bibitem[{{Rodkin} et~al.(2017){Rodkin}, {Goryaev}, {Pagano}, {Gibb},
  {Slemzin}, {Shugay}, {Veselovsky}, and {Mackay}}]{Rodkin2017}
{Rodkin} D, {Goryaev} F, {Pagano} P, {Gibb} G, {Slemzin} V, {Shugay} Y,
  {Veselovsky} I, {Mackay} DH (2017) {Origin and Ion Charge State Evolution of
  Solar Wind Transients during 4 - 7 August 2011}. \solphys, 292(7):90,
  \doi{10.1007/s11207-017-1109-0}, \eprint{1610.05048}

\bibitem[{{Rouillard} et~al.(2010{\natexlab{a}}){Rouillard}, {Davies},
  {Lavraud}, {Forsyth}, {Savani}, {Bewsher}, {Brown}, {Sheeley}, {Davis},
  {Harrison}, {Howard}, {Vourlidas}, {Lockwood}, {Crothers}, and
  {Eyles}}]{Rouillard2010a}
{Rouillard} AP, {Davies} JA, {Lavraud} B, {Forsyth} RJ, {Savani} NP, {Bewsher}
  D, {Brown} DS, {Sheeley} NR, {Davis} CJ, {Harrison} RA, {Howard} RA,
  {Vourlidas} A, {Lockwood} M, {Crothers} SR, {Eyles} CJ (2010{\natexlab{a}})
  {Intermittent release of transients in the slow solar wind: 1. Remote sensing
  observations}. \jgr, 115:A04103, \doi{10.1029/2009JA014471}

\bibitem[{{Rouillard} et~al.(2010{\natexlab{b}}){Rouillard}, {Lavraud},
  {Davies}, {Savani}, {Burlaga}, {Forsyth}, {Sauvaud}, {Opitz}, {Lockwood},
  {Luhmann}, {Simunac}, {Galvin}, {Davis}, and {Harrison}}]{Rouillard2010b}
{Rouillard} AP, {Lavraud} B, {Davies} JA, {Savani} NP, {Burlaga} LF, {Forsyth}
  RJ, {Sauvaud} JA, {Opitz} A, {Lockwood} M, {Luhmann} JG, {Simunac} KDC,
  {Galvin} AB, {Davis} CJ, {Harrison} RA (2010{\natexlab{b}}) {Intermittent
  release of transients in the slow solar wind: 2. In situ evidence}. \jgr,
  115:A04104, \doi{10.1029/2009JA014472}

\bibitem[{{Rouillard} et~al.(2011){Rouillard}, {Sheeley}, {Cooper}, {Davies},
  {Lavraud}, {Kilpua}, {Skoug}, {Steinberg}, {Szabo}, {Opitz}, and
  {Sauvaud}}]{Rouillard2011}
{Rouillard} AP, {Sheeley} NR Jr, {Cooper} TJ, {Davies} JA, {Lavraud} B,
  {Kilpua} EKJ, {Skoug} RM, {Steinberg} JT, {Szabo} A, {Opitz} A, {Sauvaud} JA
  (2011) {The Solar Origin of Small Interplanetary Transients}. \apj, 734:7,
  \doi{10.1088/0004-637X/734/1/7}

\bibitem[{{Sanchez-Diaz} et~al.(2017){Sanchez-Diaz}, {Rouillard}, {Davies},
  {Lavraud}, {Sheeley}, {Pinto}, {Kilpua}, {Plotnikov}, and
  {Genot}}]{SanchezDiaz2017}
{Sanchez-Diaz} E, {Rouillard} AP, {Davies} JA, {Lavraud} B, {Sheeley} NR,
  {Pinto} RF, {Kilpua} E, {Plotnikov} I, {Genot} V (2017) {Observational
  Evidence for the Associated Formation of Blobs and Raining Inflows in the
  Solar Corona}. \apjl, 835:L7, \doi{10.3847/2041-8213/835/1/L7},
  \eprint{1612.05487}

\bibitem[{{Sanchez-Diaz} et~al.(2019){Sanchez-Diaz}, {Rouillard}, {Lavraud},
  {Kilpua}, and {Davies}}]{SanchezDiaz2019}
{Sanchez-Diaz} E, {Rouillard} AP, {Lavraud} B, {Kilpua} E, {Davies} JA (2019)
  {In Situ Measurements of the Variable Slow Solar Wind near Sector
  Boundaries}. \apj, 882(1):51, \doi{10.3847/1538-4357/ab341c},
  \eprint{1911.09683}

\bibitem[{{Savani} et~al.(2015){Savani}, {Vourlidas}, {Szabo}, {Mays},
  {Richardson}, {Thompson}, {Pulkkinen}, {Evans}, and
  {Nieves-Chinchilla}}]{Savani2015}
{Savani} NP, {Vourlidas} A, {Szabo} A, {Mays} ML, {Richardson} IG, {Thompson}
  BJ, {Pulkkinen} A, {Evans} R, {Nieves-Chinchilla} T (2015) {Predicting the
  magnetic vectors within coronal mass ejections arriving at Earth: 1. Initial
  architecture}. Space Weather, 13(6):374--385, \doi{10.1002/2015SW001171},
  \eprint{1502.02067}

\bibitem[{{Scherrer} et~al.(1995){Scherrer}, {Bogart}, {Bush}, {Hoeksema},
  {Kosovichev}, {Schou}, {Rosenberg}, {Springer}, {Tarbell}, {Title},
  {Wolfson}, {Zayer}, and {MDI Engineering Team}}]{Scherrer1995}
{Scherrer} PH, {Bogart} RS, {Bush} RI, {Hoeksema} JT, {Kosovichev} AG, {Schou}
  J, {Rosenberg} W, {Springer} L, {Tarbell} TD, {Title} A, {Wolfson} CJ,
  {Zayer} I, {MDI Engineering Team} (1995) {The Solar Oscillations
  Investigation - Michelson Doppler Imager}. \solphys, 162(1-2):129--188,
  \doi{10.1007/BF00733429}

\bibitem[{{Schrijver}(2016)}]{Schrijver2016}
{Schrijver} CJ (2016) {The Nonpotentiality of Coronae of Solar Active Regions,
  the Dynamics of the Surface Magnetic Field, and the Potential for Large
  Flares}. \apj, 820(2):103, \doi{10.3847/0004-637X/820/2/103},
  \eprint{1602.07244}

\bibitem[{{Schwenn} et~al.(2005){Schwenn}, {dal Lago}, {Huttunen}, and
  {Gonzalez}}]{Schwenn2005}
{Schwenn} R, {dal Lago} A, {Huttunen} E, {Gonzalez} WD (2005) {The association
  of coronal mass ejections with their effects near the Earth}. \angeo,
  23:1033--1059, \doi{10.5194/angeo-23-1033-2005}

\bibitem[{{Scolini} et~al.(2018){Scolini}, {Messerotti}, {Poedts}, and
  {Rodriguez}}]{Scolini2018}
{Scolini} C, {Messerotti} M, {Poedts} S, {Rodriguez} L (2018) {Halo Coronal
  Mass Ejections during Solar Cycle 24: reconstruction of the global scenario
  and geoeffectiveness}. \jswsc, 8:A9, \doi{10.1051/swsc/2017046}

\bibitem[{{Sheeley} and {Wang}(2007)}]{Sheeley2007}
{Sheeley} J N~R, {Wang} YM (2007) {In/Out Pairs and the Detachment of Coronal
  Streamers}. \apj, 655(2):1142--1156, \doi{10.1086/510323}

\bibitem[{{Sheeley} et~al.(1980){Sheeley}, {Michels}, {Howard}, and
  {Koomen}}]{Sheeley1980}
{Sheeley} J N~R, {Michels} DJ, {Howard} RA, {Koomen} MJ (1980) {Initial
  observations with the SOLWIND coronagraph}. \apjl, 237:L99--L101,
  \doi{10.1086/183243}

\bibitem[{{Sheeley} et~al.(2009){Sheeley}, {Lee}, {Casto}, {Wang}, and
  {Rich}}]{Sheeley2009}
{Sheeley} J N~R, {Lee} DDH, {Casto} KP, {Wang} YM, {Rich} NB (2009) {The
  Structure of Streamer Blobs}. \apj, 694(2):1471--1480,
  \doi{10.1088/0004-637X/694/2/1471}

\bibitem[{{Sheeley}(2005)}]{Sheeley2005}
{Sheeley} J Neil~R (2005) {Surface Evolution of the Sun's Magnetic Field: A
  Historical Review of the Flux-Transport Mechanism}. \lrsp, 2(1):5,
  \doi{10.12942/lrsp-2005-5}

\bibitem[{{Sheeley} et~al.(1997){Sheeley}, {Wang}, {Hawley}, {Brueckner},
  {Dere}, {Howard}, {Koomen}, {Korendyke}, {Michels}, {Paswaters}, {Socker},
  {St. Cyr}, {Wang}, {Lamy}, {Llebaria}, {Schwenn}, {Simnett}, {Plunkett}, and
  {Biesecker}}]{Sheeley1997}
{Sheeley} NR, {Wang} YM, {Hawley} SH, {Brueckner} GE, {Dere} KP, {Howard} RA,
  {Koomen} MJ, {Korendyke} CM, {Michels} DJ, {Paswaters} SE, {Socker} DG, {St
  Cyr} OC, {Wang} D, {Lamy} PL, {Llebaria} A, {Schwenn} R, {Simnett} GM,
  {Plunkett} S, {Biesecker} DA (1997) {Measurements of Flow Speeds in the
  Corona Between 2 and 30~R$_{\sun}$}. \apj, 484(1):472--478,
  \doi{10.1086/304338}

\bibitem[{{Sheeley} et~al.(1999){Sheeley}, {Walters}, {Wang}, and
  {Howard}}]{Sheeley1999}
{Sheeley} NR, {Walters} JH, {Wang} YM, {Howard} RA (1999) {Continuous tracking
  of coronal outflows: Two kinds of coronal mass ejections}. \jgr,
  104(A11):24739--24768, \doi{10.1029/1999JA900308}

\bibitem[{{Shen} et~al.(2011){Shen}, {Lin}, and {Murphy}}]{ShenC2011}
{Shen} C, {Lin} J, {Murphy} NA (2011) {Numerical Experiments on Fine Structure
  within Reconnecting Current Sheets in Solar Flares}. \apj, 737:14,
  \doi{10.1088/0004-637X/737/1/14}

\bibitem[{{Snodgrass}(1983)}]{Snodgrass1983}
{Snodgrass} HB (1983) {Magnetic rotation of the solar photosphere}. \apj,
  270:288--299, \doi{10.1086/161121}

\bibitem[{{Su} et~al.(2015){Su}, {van Ballegooijen}, {McCauley}, {Ji},
  {Reeves}, and {DeLuca}}]{Su2015}
{Su} Y, {van Ballegooijen} A, {McCauley} P, {Ji} H, {Reeves} KK, {DeLuca} EE
  (2015) {Magnetic Structure and Dynamics of the Erupting Solar Polar Crown
  Prominence on 2012 March 12}. \apj, 807(2):144,
  \doi{10.1088/0004-637X/807/2/144}, \eprint{1505.06826}

\bibitem[{{Svestka} and {Cliver}(1992)}]{Svestka1992}
{Svestka} Z, {Cliver} EW (1992) {History and Basic Characteristics of Eruptive
  Flares}. In: {Svestka} Z, {Jackson} BV, {Machado} ME (eds) IAU Colloq. 133:
  Eruptive Solar Flares, vol 399, p~1, \doi{10.1007/3-540-55246-4\_70}

\bibitem[{{Talpeanu} et~al.(2020){Talpeanu}, {Chan{\'e}}, {Poedts}, {D'Huys},
  {Mierla}, {Roussev}, and {Hosteaux}}]{Talpeanu2020}
{Talpeanu} DC, {Chan{\'e}} E, {Poedts} S, {D'Huys} E, {Mierla} M, {Roussev} I,
  {Hosteaux} S (2020) {Numerical simulations of shear-induced consecutive
  coronal mass ejections}. \aap, 637:A77, \doi{10.1051/0004-6361/202037477},
  \eprint{2004.07654}

\bibitem[{{Thernisien}(2011)}]{Thernisien2011}
{Thernisien} A (2011) {Implementation of the Graduated Cylindrical Shell Model
  for the Three-dimensional Reconstruction of Coronal Mass Ejections}. \apjs,
  194:33, \doi{10.1088/0067-0049/194/2/33}

\bibitem[{{Thernisien} et~al.(2009){Thernisien}, {Vourlidas}, and
  {Howard}}]{Thernisien2009}
{Thernisien} A, {Vourlidas} A, {Howard} RA (2009) {Forward Modeling of Coronal
  Mass Ejections Using STEREO/SECCHI Data}. \solphys, 256:111--130,
  \doi{10.1007/s11207-009-9346-5}

\bibitem[{{Thernisien} et~al.(2006){Thernisien}, {Howard}, and
  {Vourlidas}}]{Thernisien2006}
{Thernisien} AFR, {Howard} RA, {Vourlidas} A (2006) {Modeling of Flux Rope
  Coronal Mass Ejections}. \apj, 652:763--773, \doi{10.1086/508254}

\bibitem[{{Thompson} et~al.(2000){Thompson}, {Cliver}, {Nitta}, {Delann{\'e}e},
  and {Delaboudini{\`e}re}}]{Thompson2000}
{Thompson} BJ, {Cliver} EW, {Nitta} N, {Delann{\'e}e} C, {Delaboudini{\`e}re}
  JP (2000) {Coronal dimmings and energetic CMEs in April-May 1998}. \grl,
  27(10):1431--1434, \doi{10.1029/1999GL003668}

\bibitem[{{Tsurutani} and {Gonzalez}(1997)}]{Tsurutani1997}
{Tsurutani} BT, {Gonzalez} WD (1997) {The Interplanetary causes of magnetic
  storms: A review}. \gms, 98:77--89, \doi{10.1029/GM098p0077}

\bibitem[{{Tsurutani} et~al.(1992){Tsurutani}, {Gonzalez}, {Tang}, and
  {Lee}}]{Tsurutani92}
{Tsurutani} BT, {Gonzalez} WD, {Tang} F, {Lee} YT (1992) {Great magnetic
  storms}. \grl, 19(1):73--76, \doi{10.1029/91GL02783}

\bibitem[{{Uzdensky} et~al.(2010){Uzdensky}, {Loureiro}, and
  {Schekochihin}}]{Uzdensky2010}
{Uzdensky} DA, {Loureiro} NF, {Schekochihin} AA (2010) {Fast Magnetic
  Reconnection in the Plasmoid-Dominated Regime}. \prl, 105(23):235002,
  \doi{10.1103/PhysRevLett.105.235002}, \eprint{1008.3330}

\bibitem[{{van Ballegooijen}(2004)}]{vanBallegooijen2004}
{van Ballegooijen} AA (2004) {Observations and Modeling of a Filament on the
  Sun}. \apj, 612(1):519--529, \doi{10.1086/422512}

\bibitem[{{van der Holst} et~al.(2009){van der Holst}, {Manchester}, {Sokolov},
  {T{\'o}th}, {Gombosi}, {DeZeeuw}, and {Cohen}}]{vanderHolst2009}
{van der Holst} B, {Manchester} W IV, {Sokolov} IV, {T{\'o}th} G, {Gombosi} TI,
  {DeZeeuw} D, {Cohen} O (2009) {Breakout Coronal Mass Ejection or Streamer
  Blowout: The Bugle Effect}. \apj, 693:1178--1187,
  \doi{10.1088/0004-637X/693/2/1178}

\bibitem[{Vourlidas(2015)}]{Vourlidas2015}
Vourlidas A (2015) Mission to the sun-earth l5 lagrangian point: An optimal
  platform for space weather research. Space Weather, 13(4):197--201,
  \doi{10.1002/2015SW001173}

\bibitem[{{Vourlidas} and {Webb}(2018)}]{Vourlidas2018}
{Vourlidas} A, {Webb} DF (2018) {Streamer-blowout Coronal Mass Ejections: Their
  Properties and Relation to the Coronal Magnetic Field Structure}. \apj,
  861(2):103, \doi{10.3847/1538-4357/aaca3e}, \eprint{1806.00644}

\bibitem[{{Vourlidas} et~al.(2000){Vourlidas}, {Subramanian}, {Dere}, and
  {Howard}}]{Vourlidas2000}
{Vourlidas} A, {Subramanian} P, {Dere} KP, {Howard} RA (2000) {Large-Angle
  Spectrometric Coronagraph Measurements of the Energetics of Coronal Mass
  Ejections}. \apj, 534(1):456--467, \doi{10.1086/308747}

\bibitem[{{Vourlidas} et~al.(2011){Vourlidas}, {Colaninno},
  {Nieves-Chinchilla}, and {Stenborg}}]{Vourlidas2011}
{Vourlidas} A, {Colaninno} R, {Nieves-Chinchilla} T, {Stenborg} G (2011) {The
  First Observation of a Rapidly Rotating Coronal Mass Ejection in the Middle
  Corona}. \apjl, 733:L23, \doi{10.1088/2041-8205/733/2/L23}

\bibitem[{{Wang} et~al.(2000){Wang}, {Sheeley}, {Socker}, {Howard}, and
  {Rich}}]{WangYM2000}
{Wang} YM, {Sheeley} NR, {Socker} DG, {Howard} RA, {Rich} NB (2000) {The
  dynamical nature of coronal streamers}. \jgr, 105(A11):25133--25142,
  \doi{10.1029/2000JA000149}

\bibitem[{{Webb} and {Nitta}(2017)}]{Webb2017}
{Webb} D, {Nitta} N (2017) {Understanding Problem Forecasts of ISEST Campaign
  Flare-CME Events}. \solphys, 292(10):142, \doi{10.1007/s11207-017-1166-4}

\bibitem[{{Webb} and {Vourlidas}(2016)}]{Webb2016}
{Webb} DF, {Vourlidas} A (2016) {LASCO White-Light Observations of Eruptive
  Current Sheets Trailing CMEs}. \solphys, 291(12):3725--3749,
  \doi{10.1007/s11207-016-0988-9}

\bibitem[{{Webb} et~al.(2000){Webb}, {Cliver}, {Crooker}, {Cry}, and
  {Thompson}}]{Webb2000}
{Webb} DF, {Cliver} EW, {Crooker} NU, {Cry} OCS, {Thompson} BJ (2000)
  {Relationship of halo coronal mass ejections, magnetic clouds, and magnetic
  storms}. \jgr, 105:7491--7508, \doi{10.1029/1999JA000275}

\bibitem[{{Wilson} et~al.(2021){Wilson}, {Brosius}, {Gopalswamy},
  {Nieves-Chinchilla}, {Szabo}, {Hurley}, {Phan}, {Kasper}, {Lugaz},
  {Richardson}, {Chen}, {Verscharen}, {Wicks}, and {TenBarge}}]{Wilson2021}
{Wilson} I Lynn~B, {Brosius} AL, {Gopalswamy} N, {Nieves-Chinchilla} T, {Szabo}
  A, {Hurley} K, {Phan} T, {Kasper} JC, {Lugaz} N, {Richardson} IG, {Chen} CHK,
  {Verscharen} D, {Wicks} RT, {TenBarge} JM (2021) {A Quarter Century of Wind
  Spacecraft Discoveries}. \rvgeo, 59(2):e2020RG000714,
  \doi{10.1029/2020RG000714}

\bibitem[{{Winterhalter} et~al.(1994){Winterhalter}, {Smith}, {Burton},
  {Murphy}, and {McComas}}]{Winterhalter1994}
{Winterhalter} D, {Smith} EJ, {Burton} ME, {Murphy} N, {McComas} DJ (1994) {The
  heliospheric plasma sheet}. \jgr, 99(A4):6667--6680, \doi{10.1029/93JA03481}

\bibitem[{{Wuelser} et~al.(2004){Wuelser}, {Lemen}, {Tarbell}, {Wolfson},
  {Cannon}, {Carpenter}, {Duncan}, {Gradwohl}, {Meyer}, {Moore}, {Navarro},
  {Pearson}, {Rossi}, {Springer}, {Howard}, {Moses}, {Newmark},
  {Delaboudiniere}, {Artzner}, {Auchere}, {Bougnet}, {Bouyries}, {Bridou},
  {Clotaire}, {Colas}, {Delmotte}, {Jerome}, {Lamare}, {Mercier}, {Mullot},
  {Ravet}, {Song}, {Bothmer}, and {Deutsch}}]{Wuelser2004}
{Wuelser} JP, {Lemen} JR, {Tarbell} TD, {Wolfson} CJ, {Cannon} JC, {Carpenter}
  BA, {Duncan} DW, {Gradwohl} GS, {Meyer} SB, {Moore} AS, {Navarro} RL,
  {Pearson} JD, {Rossi} GR, {Springer} LA, {Howard} RA, {Moses} JD, {Newmark}
  JS, {Delaboudiniere} JP, {Artzner} GE, {Auchere} F, {Bougnet} M, {Bouyries}
  P, {Bridou} F, {Clotaire} JY, {Colas} G, {Delmotte} F, {Jerome} A, {Lamare}
  M, {Mercier} R, {Mullot} M, {Ravet} MF, {Song} X, {Bothmer} V, {Deutsch} W
  (2004) {EUVI: the STEREO-SECCHI extreme ultraviolet imager}. In: {Fineschi}
  S, {Gummin} MA (eds) Telescopes and Instrumentation for Solar Astrophysics,
  Society of Photo-Optical Instrumentation Engineers (SPIE) Conference Series,
  vol 5171, pp 111--122, \doi{10.1117/12.506877}

\bibitem[{{Yang} et~al.(1986){Yang}, {Sturrock}, and {Antiochos}}]{Yang1986}
{Yang} WH, {Sturrock} PA, {Antiochos} SK (1986) {Force-free Magnetic Fields:
  The Magneto-frictional Method}. \apj, 309:383, \doi{10.1086/164610}

\bibitem[{{Yardley} et~al.(2018){Yardley}, {Green}, {van Driel-Gesztelyi},
  {Williams}, and {Mackay}}]{Yardley2018b}
{Yardley} SL, {Green} LM, {van Driel-Gesztelyi} L, {Williams} DR, {Mackay} DH
  (2018) {The Role of Flux Cancellation in Eruptions from Bipolar ARs}. \apj,
  866(1):8, \doi{10.3847/1538-4357/aade4a}, \eprint{1808.10635}

\bibitem[{{Yardley} et~al.(2019){Yardley}, {Savcheva}, {Green}, {van
  Driel-Gesztelyi}, {Long}, {Williams}, and {Mackay}}]{Yardley2019}
{Yardley} SL, {Savcheva} A, {Green} LM, {van Driel-Gesztelyi} L, {Long} D,
  {Williams} DR, {Mackay} DH (2019) {Understanding the Plasma and Magnetic
  Field Evolution of a Filament Using Observations and Nonlinear Force-free
  Field Modeling}. \apj, 887(2):240, \doi{10.3847/1538-4357/ab54d2},
  \eprint{1911.01314}

\bibitem[{{Yardley} et~al.(2021{\natexlab{a}}){Yardley}, {Mackay}, and
  {Green}}]{Yardley2021a}
{Yardley} SL, {Mackay} DH, {Green} LM (2021{\natexlab{a}}) {Simulating the
  Coronal Evolution of Bipolar Active Regions to Investigate the Formation of
  Flux Ropes}. \solphys, 296(1):10, \doi{10.1007/s11207-020-01749-2},
  \eprint{2012.07708}

\bibitem[{{Yardley} et~al.(2021{\natexlab{b}}){Yardley}, {Pagano}, {Mackay},
  and {Upton}}]{Yardley2021b}
{Yardley} SL, {Pagano} P, {Mackay} DH, {Upton} LA (2021{\natexlab{b}})
  {Determining the source and eruption dynamics of a stealth CME using NLFFF
  modelling and MHD simulations}. \aap, 652:A160,
  \doi{10.1051/0004-6361/202141142}, \eprint{2106.14800}

\bibitem[{{Yashiro} et~al.(2004){Yashiro}, {Gopalswamy}, {Michalek}, {St. Cyr},
  {Plunkett}, {Rich}, and {Howard}}]{Yashiro2004}
{Yashiro} S, {Gopalswamy} N, {Michalek} G, {St Cyr} OC, {Plunkett} SP, {Rich}
  NB, {Howard} RA (2004) {A catalog of white light coronal mass ejections
  observed by the SOHO spacecraft}. \jgr, 109(A7):A07105,
  \doi{10.1029/2003JA010282}

\bibitem[{{Yeates}(2014)}]{Yeates2014}
{Yeates} AR (2014) {Coronal Magnetic Field Evolution from 1996 to 2012:
  Continuous Non-potential Simulations}. \solphys, 289:631--648,
  \doi{10.1007/s11207-013-0301-0}

\bibitem[{{Yeates} and {Mackay}(2012)}]{yeates2012}
{Yeates} AR, {Mackay} DH (2012) {Chirality of High-latitude Filaments over
  Solar Cycle 23}. \apj, 753(2):L34, \doi{10.1088/2041-8205/753/2/L34},
  \eprint{1206.2327}

\bibitem[{{Yeates} et~al.(2018){Yeates}, {Amari}, {Contopoulos}, {Feng},
  {Mackay}, {Miki{\'c}}, {Wiegelmann}, {Hutton}, {Lowder}, {Morgan}, {Petrie},
  {Rachmeler}, {Upton}, {Canou}, {Chopin}, {Downs}, {Druckm{\"u}ller},
  {Linker}, {Seaton}, and {T{\"o}r{\"o}k}}]{Yeates2018}
{Yeates} AR, {Amari} T, {Contopoulos} I, {Feng} X, {Mackay} DH, {Miki{\'c}} Z,
  {Wiegelmann} T, {Hutton} J, {Lowder} CA, {Morgan} H, {Petrie} G, {Rachmeler}
  LA, {Upton} LA, {Canou} A, {Chopin} P, {Downs} C, {Druckm{\"u}ller} M,
  {Linker} JA, {Seaton} DB, {T{\"o}r{\"o}k} T (2018) {Global Non-Potential
  Magnetic Models of the Solar Corona During the March 2015 Eclipse}. \ssr,
  214(5):99, \doi{10.1007/s11214-018-0534-1}, \eprint{1808.00785}

\bibitem[{{Yu} et~al.(2014){Yu}, {Farrugia}, {Lugaz}, {Galvin}, {Kilpua},
  {Kucharek}, {M{\"o}stl}, {Leitner}, {Torbert}, {Simunac}, {Luhmann}, {Szabo},
  {Wilson}, {Ogilvie}, and {Sauvaud}}]{Yu2014}
{Yu} W, {Farrugia} CJ, {Lugaz} N, {Galvin} AB, {Kilpua} EKJ, {Kucharek} H,
  {M{\"o}stl} C, {Leitner} M, {Torbert} RB, {Simunac} KDC, {Luhmann} JG,
  {Szabo} A, {Wilson} LB, {Ogilvie} KW, {Sauvaud} JA (2014) {A statistical
  analysis of properties of small transients in the solar wind 2007-2009:
  STEREO and Wind observations}. \jgr, 119(2):689--708,
  \doi{10.1002/2013JA019115}

\bibitem[{{Yu} et~al.(2018){Yu}, {Farrugia}, {Lugaz}, {Galvin}, {M{\"o}stl},
  {Paulson}, and {Vemareddy}}]{Yu2018}
{Yu} W, {Farrugia} CJ, {Lugaz} N, {Galvin} AB, {M{\"o}stl} C, {Paulson} K,
  {Vemareddy} P (2018) {The Magnetic Field Geometry of Small Solar Wind Flux
  Ropes Inferred from Their Twist Distribution}. \solphys, 293(12):165,
  \doi{10.1007/s11207-018-1385-3}

\bibitem[{{Zhang} et~al.(2007){Zhang}, {Richardson}, {Webb}, {Gopalswamy},
  {Huttunen}, {Kasper}, {Nitta}, {Poomvises}, {Thompson}, {Wu}, {Yashiro}, and
  {Zhukov}}]{Zhang2007}
{Zhang} J, {Richardson} IG, {Webb} DF, {Gopalswamy} N, {Huttunen} E, {Kasper}
  JC, {Nitta} NV, {Poomvises} W, {Thompson} BJ, {Wu} CC, {Yashiro} S, {Zhukov}
  AN (2007) {Solar and interplanetary sources of major geomagnetic storms ($Dst
  \leq$ -100 nT) during 1996-2005}. \jgr, 112:A10102,
  \doi{10.1029/2007JA012321}

\bibitem[{{Zhukov}(2007)}]{Zhukov2007}
{Zhukov} AN (2007) {Using CME Observations for Geomagnetic Storm Forecasting}.
  In: {Lilensten} J (ed) Space Weather : Research Towards Applications in
  Europe 2nd European Space Weather Week (ESWW2). Astrophysics and Space
  Science Library volume 344. Edited by Jean Lilensten. ESA / ESTEC, Noordwijk,
  The Netherlands, November 14-19, 2005. Astrophysics and space science
  library, Vol. 344. ISBN: 978-1-4020-5445-7 (e-book). Dordrecht, The
  Netherlands: Springer, 2007, p.5, Astrophysics and Space Science Library, vol
  344, p~5, \doi{10.1007/1-4020-5446-7\_2}

\bibitem[{{Zhukov} and {Auch{\`e}re}(2004)}]{Zhukov2004}
{Zhukov} AN, {Auch{\`e}re} F (2004) {On the nature of EIT waves, EUV dimmings
  and their link to CMEs}. \aap, 427:705--716, \doi{10.1051/0004-6361:20040351}

\bibitem[{{Zuccarello} et~al.(2012){Zuccarello}, {Bemporad}, {Jacobs},
  {Mierla}, {Poedts}, and {Zuccarello}}]{Zuccarello2012}
{Zuccarello} FP, {Bemporad} A, {Jacobs} C, {Mierla} M, {Poedts} S, {Zuccarello}
  F (2012) {The Role of Streamers in the Deflection of Coronal Mass Ejections:
  Comparison between STEREO Three-dimensional Reconstructions and Numerical
  Simulations}. \apj, 744(1):66, \doi{10.1088/0004-637X/744/1/66}

\bibitem[{{Zurbuchen} and {Richardson}(2006)}]{Zurbuchen2006}
{Zurbuchen} TH, {Richardson} IG (2006) {In-Situ Solar Wind and Magnetic Field
  Signatures of Interplanetary Coronal Mass Ejections}. \ssr, 123:31--43,
  \doi{10.1007/s11214-006-9010-4}

\end{thebibliography}

\end{document}